\documentclass[prd,nofootinbib,a4paper,showpacs,showkeys,preprintnumbers,twocolumn]{revtex4}

\usepackage{graphicx}
\usepackage{amsmath}
\usepackage{amssymb}
\usepackage{amsfonts}
\usepackage{latexsym}
\usepackage{mathrsfs}
\usepackage{color}
\usepackage{slashed}
\usepackage{feynmp}
\usepackage{pstricks,pst-text}
\usepackage{epsfig}
\usepackage{hyperref}
\usepackage{grffile}
\usepackage{ifpdf}
\usepackage{hyperref}
\usepackage{mathtools}
\mathtoolsset{showonlyrefs=true,showmanualtags=true}

\ifpdf
\fi

\providecommand{\unit}[1]{\ensuremath{\,\mathrm{#1}}}
\providecommand{\GeV}{\ensuremath{\,\mathrm{GeV}}}
\providecommand{\hubblerate}{H}
\providecommand{\hubbleratem}{{\cal H}}
\providecommand{\x}{z}
\providecommand{\me}{\mathcal{M}}
\providecommand{\higgs}{\phi}
\providecommand{\lepton}{{\ell}}
\providecommand{\majneutrino}{N}
\providecommand{\m}[1]{m_{#1}}
\providecommand{\width}[1]{\Gamma_{#1}}
\providecommand{\e}[2]{E_{#2}}
\providecommand{\gtwo}{g_2}
\providecommand{\gprime}{g'}
\providecommand{\gyt}{h_t}
\providecommand{\yu}{h}
\providecommand{\F}[4]{{\mathcal{F}_{#1 \leftrightarrow #2}^{#3 ; #4}}}
\providecommand{\dpi}[4]{{\dif\Pi_{#1#2}^{#3#4}}}
\providecommand{\EffAmplitude}[2]{\Xi^{#1}_{#2}}
\providecommand{\EffAmpl}[3]{\Xi^{#1}_{#2 \leftrightarrow #3}}
\providecommand{\amp}[2]{{\mathcal{M}}_{{#1}\rightarrow{#2}}}
\providecommand{\ampabs}[2]{\abs{\mathcal{M}}_{{#1}\rightarrow{#2}}}
\providecommand{\momk}{k}
\providecommand{\momp}{p}
\providecommand{\momq}{q}
\providecommand{\momr}{r}
\providecommand{\momlep}{\momp}
\providecommand{\momlep}{\momr}
\providecommand{\momhig}{\momk}
\providecommand{\mommaj}{\momq}
\providecommand{\momleploop}{\momlep}
\providecommand{\meavsqudep}[1]{\left|\me\right|_{#1}^2}
\providecommand{\lorentzdd}[1]{\dif\Pi^4_{#1}}
\providecommand{\lorentzd}[2]{\dif\Pi^{#1}_{#2}}
\providecommand{\f}[2]{f^{}_{#1}}
\providecommand{\feq}[2]{{f}_{#1}^{eq}}
\providecommand{\flong}[3]{f_{#1}({#2},{#3})}
\providecommand{\qstatf}[2]{(1-\xi^{#1}\f{#1}{#2})}
\providecommand{\qstatff}[2]{(1-\f{#1}{#2})}
\providecommand{\qstatfb}[2]{(1+\f{#1}{#2})}
\providecommand{\qstatfeq}[2]{(1-\xi^{#1}\feq{#1}{#2})}
\providecommand{\qstatfeqf}[2]{(1-\feq{#1}{#2})}
\providecommand{\qstatfeqb}[2]{(1+\feq{#1}{#2})}
\providecommand{\abundance}[1]{Y_{#1}}
\providecommand{\abundanceeq}[1]{Y^{eq}_{#1}}
\providecommand{\stat}[1]{\xi^{#1}}
\providecommand{\n}[1]{n_{#1}}
\providecommand{\nequ}[1]{n^{eq}_{#1}}
\providecommand{\dend}{{\, .}}
\providecommand{\kend}{{\, ,}}
\providecommand{\CTP}{{\cal C}}
\providecommand{\CTP}{{\cal C}}
\providecommand{\dg}{\mathscr{D}}
\providecommand{\CP}{\textit{CP}}
\providecommand{\CPT}{\textit {CPT}}
\providecommand{\ddelta}{Dirac-delta}
\providecommand{\eqn}{Eq.\,}
\providecommand{\eqns}{Eqs.\,}
\providecommand{\fig}{Fig.\,}
\providecommand{\figs}{Figs.\,}
\providecommand{\figref}[1]{\ref{fig:#1}}
\providecommand{\sect}{Sec.\,}
\providecommand{\sect}{section\,}
\providecommand{\app}{Appendix\,}
\providecommand{\abs}[1]{\left|#1\right|}
\providecommand{\sgn}{\mathrm{sgn}}
\providecommand{\sign}{\mathrm{sign}}
\providecommand{\deltafour}[1]{\delta(#1)}
\providecommand{\dif}{\,d}
\providecommand{\fermionpartial}{\displaystyle{\not}\partial}
\providecommand{\order}[1]{\mathcal{O}(#1)}
\providecommand{\Tr}{{\rm Tr}}
\providecommand{\tr}{{\rm tr}}
\providecommand{\pLept}[2]{S^{#1}_{#2}}
\providecommand{\pLeptcp}[2]{\bar{S}^{#1}_{#2}}
\providecommand{\pLeptmat}[2]{\hat{S}^{#1}_{#2}}
\providecommand{\pLeptsc}[2]{\mathbf{S}^{#1}_{#2}}
\providecommand{\sLept}[2]{\Sigma^{#1}_{#2}}
\providecommand{\sLeptcp}[2]{\bar{\Sigma}^{#1}_{#2}}
\providecommand{\sLeptmat}[2]{\hat{\Sigma}^{#1}_{#2}}
\providecommand{\pMaj}[2]{\mathscr{S}^{#1}_{#2}}
\providecommand{\pMajcp}[2]{\bar{\mathscr{S}}^{#1}_{#2}}
\providecommand{\pMajmat}[2]{\hat{\mathscr{S}}^{#1}_{#2}}
\providecommand{\pMajdiag}[2]{\mathcal{S}^{#1}_{#2}}
\providecommand{\pMajdiagmat}[2]{\hat{\mathcal{S}}^{#1}_{#2}}
\providecommand{\pMajdiagsc}[2]{\boldsymbol{\mathcal{S}}^{#1}_{#2}}
\providecommand{\pMajEQP}[2]{\tilde{\mathcal{S}}^{#1}_{#2}}
\providecommand{\pMajEQPmat}[1]{\hat{\tilde{\mathcal{S}}}_{#1}}
\providecommand{\sMaj}[2]{\Pi^{#1}_{#2}}
\providecommand{\sMajsl}[2]{\slashed{\Pi}^{#1}_{#2}}
\providecommand{\sMajsl}[2]{\slashed{\Pi}^{#1}_{#2}}
\providecommand{\sMajcp}[2]{\bar{\Pi}^{#1}_{#2}}
\providecommand{\sMajmat}[2]{\hat{\Pi}^{#1}_{#2}}
\providecommand{\MatTheta}[2]{\Theta^{#1}_{#2}}
\providecommand{\MatThetacp}[2]{\bar{\Theta}^{#1}_{#2}}
\providecommand{\HatMatTheta}[2]{\hat{\Theta}^{#1}_{#2}}
\providecommand{\pHiggs}[2]{\Delta^{#1}_{#2}}
\providecommand{\pHiggscp}[2]{\bar{\Delta}^{#1}_{#2}}
\providecommand{\OmegaNum}[2]{\Omega^{#1}_{#2}}
\providecommand{\OmegaDen}[2]{{\boldsymbol\Omega}^{#1}_{#2}}
\providecommand{\PiDen}[2]{{\boldsymbol\Pi}^{#1}_{#2}}
\providecommand{\Meff}[2]{{\boldsymbol M}^{#1}_{#2}}
\providecommand{\Gammaeff}[2]{{\boldsymbol \Gamma}^{#1}_{#2}}

\begin{document}
\pacs{11.10.Wx, 98.80.Cq}
\keywords{Leptogenesis, Kadanoff-Baym equations, Boltzmann equation, RIS subtraction}
\preprint{DESY 12-202}

\title{Systematic approach to thermal leptogenesis}

\author{T. Frossard$^{a}$}
\email[\,]{tibor.frossard@mpi-hd.mpg.de} 

\author{M. Garny$^{b}$}
\email[\,]{mathias.garny@desy.de}

\author{A. Hohenegger$^{c}$}
\email[\,]{andreas.hohenegger@epfl.ch}

\author{A. Kartavtsev$^{a}$}
\email[\,]{alexander.kartavtsev@mpi-hd.mpg.de}

\author{D. Mitrouskas$^{d}$}
\email[\,]{david.mitrouskas@math.lmu.de}

\affiliation{%
$^a$Max-Planck-Institut f\"ur Kernphysik, Saupfercheckweg 1, 69117 Heidelberg, Germany\\
$^b$Deutsches Elektronen-Synchrotron DESY, Notkestr.~85, 22607 Hamburg, Germany\\
$^c$\'Ecole Polytechnique F\'ed\'erale de Lausanne, LPPC, BSP, CH-1015 Lausanne, Switzerland\\
$^d$Ludwig-Maximilians-Universit\"at M\"unchen, Theresienstr.~39, 80333 M\"unchen, Germany}

\begin{abstract}
In this work we study thermal leptogenesis using non-equilibrium quantum field theory. 
Starting from fundamental equations for correlators of the quantum fields we describe 
the steps necessary to obtain quantum kinetic equations for quasiparticles. These can easily 
be compared to conventional results and overcome conceptional problems inherent in the 
canonical approach. Beyond \CP-violating decays we include also those scattering 
processes which are tightly related to the decays in a consistent approximation of fourth 
order in the Yukawa couplings. It is demonstrated explicitly how the S-matrix elements for 
the scattering processes in the conventional approach are related to two- and 
three-loop contributions to the effective action. We derive effective decay and scattering 
amplitudes taking medium corrections and thermal masses into account. In this context we 
also investigate \CP-violating Higgs decay within the same formalism. From the kinetic 
equations we derive rate equations for the lepton asymmetry improved in that they 
include quantum-statistical effects and medium corrections to the 
quasiparticle properties.
\end{abstract}

\maketitle

\section{\label{introduction}Introduction}

If one combines today's Standard Model of particle physics (SM) and that of cosmology, 
one finds inevitably that particles and their antiparticles annihilate at a very early 
moment in the evolution of the universe, leaving just radiation behind. The absence of a 
sizable matter-antimatter asymmetry at this epoch would imply that the universe as we 
know it could never form. The question about the origin of the observed asymmetry 
therefore represents a major challenge for modern physics. 

In the SM baryon and lepton number are (accidental) global symmetries. If baryon number 
was also conserved in the early Universe a dynamical emergence of the asymmetry would 
have been impossible. In grand-unified extensions (GUTs) of the SM baryon number (and 
also lepton number) is explicitly broken. According to past reasoning, this could provide 
a solution to the apparent discrepancy. In the class of `GUT-baryogenesis' scenarios 
the matter-antimatter imbalance is generated by asymmetric decays of new super-heavy 
bosons. 
Anomalous electroweak processes \cite{'tHooft:1976up,Klinkhamer:1984di} 
(sphalerons) which violate baryon and lepton number but conserve their difference 
essentially eliminated the prospects for GUT-baryogenesis \cite{Kuzmin:1985mm}. At the same time, it inspired 
the now widely appreciated scenarios of `electroweak baryogenesis' \cite{Kuzmin:1985mm,Morrissey:2012db} and `baryogenesis via leptogenesis' 
\cite{Fukugita:1986hr}. According to the latter scenario, 
the asymmetry is initially generated in the leptonic sector by the decay of heavy 
Majorana neutrinos at an energy scale far above the electroweak scale. 
Subsequently it is converted into the observed baryon asymmetry by sphalerons. The mass scale of the heavy Majorana neutrinos
required for leptogenesis~\cite{Davidson:2002qv,Hamaguchi:2001gw} fits together very well with the mass-differences inferred from observations of solar-, atmospheric- and reactor-neutrino oscillations. 

We focus here on the conventional, but most popular, high-energy (type-I) seesaw extension:
\begin{align} 
	\label{lagrangian}
	\mathscr{L}=\mathscr{L}_{SM}&+{\textstyle\frac12}\bar \majneutrino_i
	\bigl(i\fermionpartial - M_i\bigr)\majneutrino_i\nonumber\\
	&- \yu_{\alpha i}\bar \lepton_\alpha {\tilde \higgs} P_R \majneutrino_i
	-\yu^\dagger_{i\alpha} \bar \majneutrino_i {\tilde \higgs}^\dagger P_L \lepton_\alpha\kend
\end{align}
where $\majneutrino_i=\majneutrino^c_i$ are the heavy Majorana fields, 
$\lepton_\alpha$ are the lepton doublets, $\tilde \higgs\equiv i\sigma_2 \higgs^*$ is the conjugate 
of the Higgs doublet, and $\yu$ are the corresponding Yukawa couplings. 
The Majorana mass term violates lepton number and the Yukawa couplings can violate \CP. 
Therefore the model fulfills essential requirements for 
baryogenesis \cite{Sakharov:1967dj}. They 
can also be realized for more complicated SM extensions and a wide range of values for couplings 
and neutrino masses \cite{Buchmuller:2004nz,Giudice:2003jh,Davidson:pr2008,Blanchet:2012bk}. 
In general the right-handed neutrinos do not necessarily get into thermal 
equilibrium and \CP-violating oscillations between them can contribute to the asymmetry. 
This effect of leptogenesis through neutrino oscillations \cite{Akhmedov:1998qx} is crucial 
for neutrino-minimal extensions of the SM ($\nu$MSM) \cite{Canetti:2012kh} and poses 
interesting questions for non-equilibrium quantum field theory \cite{Garny:2011hg,Garbrecht:2011aw,Drewes:2012ma}. 
In the considered scenario of thermal leptogenesis the heavy Majorana neutrinos experience 
only a moderate deviation from thermal equilibrium at the time when the bulk of the 
asymmetry is produced. Also, for a hierarchical mass spectrum, effects related to oscillations are negligible.

The amount of the generated asymmetry is determined by the out of equilibrium 
evolution of the heavy Majorana neutrinos. Therefore, 
statistical equations for the abundance of the neutrinos and the generated asymmetry are needed. 
The conventional approach here follows the lines developed for GUT-baryogenesis \cite{Kolb:1990vq}. 
The \CP-violating amplitudes for the decay and scattering processes involving the heavy Majorana 
neutrinos are computed in terms of Feynman graphs at lowest loop order. They are used to build 
generalized Boltzmann collision terms for these processes. Each of them contributes to the 
evolution of the distributions of Majorana neutrinos and leptons or, upon momentum integration, 
their entire abundances. 

However this approach is plagued by the so-called double-counting problem 
which manifests itself in the generation of a non-vanishing asymmetry even in thermal equilibrium. 
This technical issue is expression of the fact that the `naive' generalization of the collision 
terms is quantitatively inexact, and inconsistent in the presence of \CP-violation. After a 
real intermediate state (or RIS) subtraction procedure and a number of approximations, it can be 
made consistent with fundamental requirements. Nevertheless this pragmatic solution remains unsatisfactory. The 
requirement of unitarity guarantees a consistent approximation for the amplitudes, realized by 
the RIS subtraction, if the statistical system is in thermal equilibrium. However, the deviation 
from equilibrium is a fundamental requirement for leptogenesis and it is not obvious how the 
equations have to be generalized for a system out of equilibrium.

Furthermore, the \CP-violation arises from 
one-loop contributions due to the exchange of virtual quanta. As such 
they seem to be beyond a Boltzmann approximation. But the relevant imaginary part is 
due to intermediate states in which at least some of the particles are on-shell. 
These can also be absorbed or emitted by the medium and it is not obvious how 
such contributions enter the amplitudes. It is, however, clear that the influence of medium effects 
on the one-loop contributions enters directly the \CP-violating parameter and 
therefore the source for the lepton asymmetry. Their size can be of the same 
order as that of the vacuum contributions.

Those questions can be addressed within a first-principle approach based on non-equilibrium quantum field 
theory (NEQFT). Several aspects of leptogenesis have already been investigated within this approach \cite{Buchmuller:2000nd}. 
The influence of medium effects on the generation of the asymmetry has been 
studied e.g.~in \cite{Garny:2009rv,Garny:2009qn,Garny:2010nj,Beneke:2010wd,Garny:2010nz,Garbrecht:2010sz,Drewes:2012ma}, and an analysis with special emphasis on off-shell effects was performed in~\cite{Anisimov:2010aq,Anisimov:2010dk}. The role of flavor effects as well as the range of applicability of the conventional approach to the analysis of flavored leptogenesis has been investigated in \cite{Beneke:2010dz}.
The resonant enhancement of the lepton asymmetry has been addressed within a first-principle approach in \cite{Garny:2011hg,Garbrecht:2011aw,Garbrecht:2012qv,Garbrecht:2012pq}. In addition, steps towards a consistent inclusion of gauge interactions have been taken~\cite{Giudice:2003jh,Kiessig:2011fw,Kiessig:2011ga,Besak:2010fb,Anisimov:2010gy,Laine:2011pq,Salvio:2011sf,Besak:2012qm}.

In this work we use the 2PI-formalism of NEQFT 
to derive Boltzmann-like quantum kinetic equations for the lepton asymmetry.
In particular, we show how two-body scattering processes that violate lepton number by two units and contribute to the 
washout of the asymmetry emerge within the 2PI-formalism. This approach treats 
quantum field theory and the out of equilibrium evolution on an equal footing 
and allows to overcome the conceptional difficulties inherent in the conventional 
approach. It allows us to obtain quantum-generalized Boltzmann equations which include 
medium effects and which are free of the double-counting problem. In other words, the structure 
of the obtained quantum kinetic equations automatically ensures that the asymmetry 
vanishes in thermal equilibrium and no need for RIS subtraction arises. 
The resulting equation for the lepton asymmetry $Y_L$ is given by
\begin{align}
	\frac{s\hubbleratem }{z}\frac{d Y_L}{dz} &=
	\sum_i \int \dpi{\lepton\higgs}{N_i}{\momlep\momhig}{\mommaj}
	\F{\lepton\higgs}{N_i}{\momlep\momhig}{\mommaj}\EffAmpl{}{\lepton\higgs}{\majneutrino_i}\nonumber\\
	&-\sum_i \int \dpi{\bar\lepton\bar\higgs}{N_i}{\momlep\momhig}{\mommaj}
	\F{\bar\lepton\bar\higgs}{N_i}{\momlep\momhig}{\mommaj}\EffAmpl{}{\bar\lepton\bar\higgs}{\majneutrino_i}\nonumber\\
	&-2\int \dpi{\lepton\higgs}{\bar\lepton\bar\higgs}{\momlep_1\momhig_1}{\momlep_2\momhig_2}
	\F{\bar\lepton\bar\higgs}{\lepton\higgs}{\momlep_2\momhig_2}{\momlep_1\momhig_1}
	\EffAmpl{}{\bar\lepton\bar\higgs}{\lepton\higgs}\nonumber\\
	&-\int \dpi{\lepton\lepton}{\bar\higgs\bar\higgs}{\momlep_1\momlep_2}{\momhig_1\momhig_2}
	\F{\bar\higgs\bar\higgs}{\lepton\lepton}{\momhig_1\momhig_2}{\momlep_1\momlep_2}
	\EffAmpl{}{\bar\higgs\bar\higgs}{\lepton\lepton}\nonumber\\
	&-\int \dpi{\bar\lepton\bar\lepton}{\higgs\higgs}{\momlep_1\momlep_2}{\momhig_1\momhig_2}
	\F{\bar\lepton\bar\lepton}{\higgs\higgs}{\momlep_1\momlep_2}{\momhig_1\momhig_2}
	\EffAmpl{}{\bar\lepton\bar\lepton}{\higgs\higgs}\dend
	\label{eqn:main result}
\end{align}
Together with the `effective amplitudes' $\Xi$ this is the main result of this paper. In \eqn\eqref{eqn:main result} we introduced
\begin{align}
	\label{eqn:definition of F}
	\F{ab ..}{ij ..}{p_a p_b ..}{p_i p_j ..} & \equiv 
	(2\pi)^4 \delta(p_a+p_b+.. -p_i-p_j-..)\nonumber\\
	& \times\bigl[ \f{i}{p_i} \f{j}{p_j}..\qstatf{a}{p_a}\qstatf{b}{p_b} .. \nonumber\\ 
	& - \f{a}{p_a} \f{b}{p_b} ..\qstatf{i}{p_i}\qstatf{j}{p_j} .. \bigr] \kend
\end{align}
with $\stat{a}=+(-)1$ for fermions (bosons). Note that $\F{ab ..}{ij ..}{p_a p_b ..}{p_i p_j ..}$ 
vanishes in equilibrium due to detailed balance. This ensures that the asymmetry vanishes in thermal equilibrium as mentioned before. The effective amplitudes contain medium effects ignored in the corresponding canonical expressions.

We find that, in the amplitudes of the scattering processes medium 
effects are sub-dominant and can be neglected. The total decay amplitude 
of the Majorana neutrino is barely affected as well. However, at high 
temperatures the available phase space
shrinks when taking gauge interactions in the form of effective thermal 
masses of Higgs and leptons into account. This leads to a
suppression of the decay and scattering rates. Since the \CP-violation
appears as loop effect it is more sensitive to influences of the 
surrounding medium. Even though there is a partial cancellation of the 
fermionic and bosonic contributions, the \CP-violating parameter is enhanced by medium effects.
However, the thermal masses reduce the enhancement and turn it into suppression at high 
temperatures.

We review the conventional approach to leptogenesis based on RIS subtraction in 
\sect\ref{ConventionalApproach}. In section \ref{RISQuantStat} we demonstrate explicitly 
that {\em in thermal equilibrium} the success of this procedure is guaranteed by the 
requirement of unitarity. In \sect\ref{RateEquations} we review the
derivation of rate equations for total abundances and discuss in how far quantum
statistical and medium corrections can be incorporated in the reaction densities. In \sect\ref{NEQFTapproach} we review the application of the 2PI approach of NEQFT to leptogenesis. Equation \eqref{eqn:main result} and explicit expressions for the 
effective in-medium decay and scattering amplitudes are derived within this framework in \sect\ref{Majorana}. 
We compare the results obtained within the 2PI-formalism to those of a conventional analysis with manual RIS subtraction. 
In \sect\ref{HiggsDecay} we derive rate equations and the \CP-violating amplitudes for Higgs decay within the 
framework of NEQFT. Finally, we summarize the results and present our conclusions in Sec.\,\ref{Summary}.

\section{\label{ConventionalApproach}Conventional approach}

The amount of produced asymmetry depends on the details of the non-equilibrium 
evolution of the Majorana neutrinos as well as
on the strength of \CP-violation. The latter is usually quantified by 
\CP-violating parameters\cite{Buchmuller:2004nz,Giudice:2003jh,Davidson:pr2008,Blanchet:2012bk}: 
\begin{align*}
	\epsilon_i\equiv \frac{\Gamma_{\majneutrino_i\rightarrow \lepton \higgs}
	-\Gamma_{\majneutrino_i\rightarrow \bar \lepton \bar \higgs}}{\Gamma_{\majneutrino_i\rightarrow \lepton \higgs}
	+\Gamma_{\majneutrino_i\rightarrow \bar \lepton \bar \higgs}}\kend
\end{align*}
where $\Gamma_{\majneutrino_i \rightarrow\lepton\higgs}$ and $\Gamma_{\majneutrino_i \rightarrow\bar\lepton\bar\higgs}$ are the vacuum decay rates to a particle or anti-particle pair respectively.For a hierarchical mass spectrum $\epsilon_i$ can be computed perturbatively as the interference
\begin{figure}[h!]
	\centering
	\includegraphics[width=\columnwidth]{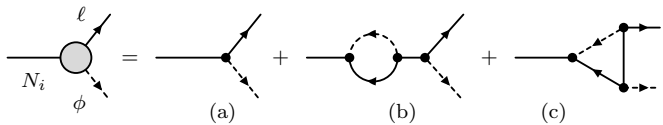}
	\caption{\label{treevertexself}Tree-level, one-loop self-energy and one-loop vertex 
	contributions to the decay of the heavy Majorana neutrino.}
\end{figure}
of the tree-level, one-loop 
vertex \cite{Fukugita:1986hr} and one-loop self-energy \cite{Flanz:1994yx,Covi:1996wh,PhysRevD.56.5431} amplitudes in \fig\ref{treevertexself}. The contribution of the loop diagrams can be accounted for by effective Yukawa couplings \cite{Pilaftsis:2003gt}:
\begin{subequations}
	\label{effective couplings}
	\noeqref{lambdaplus,lambdaminus}
	\begin{align}
		\label{lambdaplus}
		\lambda_{+,\alpha i} &\equiv h_{\alpha i} -i h_{\alpha j} (h^\dagger h)^*_{ji}\, f_{ij}\kend\\
		\label{lambdaminus}
		\lambda_{-,\alpha i} &\equiv h^*_{\alpha i} -i h^*_{\alpha j} (h^\dagger h)_{ji}\, f_{ij}\kend
	\end{align}
\end{subequations}
 where the loop function $f_{ij}$ is defined as 
\begin{align}
	\label{LoopFunction}
	f_{ij}\equiv &\frac{1}{16\pi} \frac{M_i M_j}{M^2_i-M^2_j}\\
	+ &\frac{1}{16\pi}\frac{M_j}{M_i}\biggl[1-\biggl(1+\frac{M_j^2}{M_i^2}\biggr)
	\ln \biggl(1+\frac{M_i^2}{M_j^2}\biggr)\biggr]\dend\nonumber
\end{align}
The first term in \eqn\eqref{LoopFunction} is related to the self-energy and 
the second term to the vertex contribution. The decay widths are proportional to the 
absolute values of the effective couplings, $\Gamma_{\majneutrino_i\rightarrow \lepton \higgs}
=g_w (\lambda^\dagger_{+}\lambda^{\phantom{\dagger}}_{+})_{ii}M_i/(32\pi)$ and 
$\Gamma_{\majneutrino_i\rightarrow \bar\lepton \bar\higgs}
=g_w (\lambda^\dagger_{-}\lambda^{\phantom{\dagger}}_{-})_{ii}M_i/(32\pi)$
respectively, where we have summed over flavors of the leptons and $SU(2)_L$ 
indices (hence the factor $g_w=2$) in the final state. 
Since the phase space for the decay into particles and antiparticles is the same, 
one gets for the \CP-violating parameter:
\begin{align}
	\label{epsilon vacuum}
	\epsilon^{vac}_i\approx \frac{\Im(h^\dagger h)^2_{ij}}{(h^\dagger h)_{ii}}\times 2f_{ij}\kend
	\quad j\neq i \dend
\end{align}
Let us note in passing that the divergence of the loop function for $j=i$ is not 
physical and can be removed by a resummation of the self-energy contribution 
\cite{PhysRevD.56.5431,Plumacher:1997ru,Pilaftsis:2003gt}. Here we work in a 
regime where the mass splittings $\abs{M_i -M_j}$ are large enough to render effects 
related to the enhancement of the self-energy contribution irrelevant (non-resonant 
leptogenesis). We do not require a strictly hierarchical mass-spectrum, however.

To describe the statistical evolution of the lepton asymmetry one usually employs 
generalized Boltzmann equations for the one-particle distribution functions of the 
different species \cite{Kolb:1979qa,Bernstein:1988,Kolb:1990vq}. Taking into 
account decay and inverse decay processes one writes for the distribution function 
of the leptons (for a single flavor): 
\begin{align}
	\momlep^\mu {\cal D}_\mu\f{\lepton}{\momk}&=\frac{1}{2}
	\sum_{i,s_i} \int \lorentzd{\higgs}{\momhig}\lorentzd{{\majneutrino_i}}{\momq}
	(2\pi)^4\deltafour{\momlep+\momhig-\mommaj}\nonumber\\
	&\times \bigl[\ampabs{{\majneutrino_i}}{\lepton\higgs}^2
	\qstatff{\lepton}{\momk}\qstatfb{\higgs}{\momp}\f{\majneutrino_i}{\momq} \nonumber\\
	&-	\ampabs{\lepton\higgs}{{\majneutrino_i}}^2
	\f{\lepton}{\momk}\f{\higgs}{\momp}\qstatff{\majneutrino_i}{\momq}\bigr]\kend
	\label{eqn:collision terms decay}
\end{align} 	
where $\lorentzd{a}{p}=d^3p/[(2\pi)^3 2\e{a}{p}]$ is the invariant 
phase space element, $s_i$ denotes spin degrees of freedom of $\majneutrino_i$,
and ${\cal D}_\mu$ is the covariant derivative. 
The corresponding equation for 
antileptons may be obtained by interchanging $\lepton\leftrightarrow\bar{\lepton}$ 
and $\higgs\leftrightarrow\bar{\higgs}$. \CPT-invariance implies that
$\meavsqudep{\majneutrino_i\rightarrow \lepton \higgs }=
\meavsqudep{\bar\lepton\bar\higgs \rightarrow \majneutrino_i}$ and 
$\meavsqudep{\majneutrino_i\rightarrow \bar\lepton \bar\higgs }=
\meavsqudep{\lepton\higgs \rightarrow \majneutrino_i}$.
Furthermore, in {\em thermal equilibrium} detailed balance requires that 
$(1-\feq{\lepton}{\momk})(1-\feq{\higgs}{\momp})\feq{\majneutrino_i}{\momq}
=\feq{\lepton}{\momk}\feq{\higgs}{\momp}(1-\feq{\majneutrino_i}{\momq})$. Subtracting 
the two relations we find for the contribution of the (inverse) decay terms: 
\begin{align}
	\label{eqn:asymmetry in equilibrium}
	&\momlep^\mu {\cal D}_\mu(\f{\lepton}{\momlep}-\f{\bar\lepton}{\momlep})=
 2\cdot\frac{1}{2} \sum_{i,s_i} \int 
	\lorentzd{\higgs}{\momhig}\lorentzd{{\majneutrino_i}}{\mommaj}
	(2\pi)^4\deltafour{\momlep+\momhig-\mommaj}\nonumber\\
	&\times \feq{\majneutrino_i}{\mommaj}(1-\feq{\lepton}{\momlep})(1+\feq{\higgs}{\momhig})
	\bigl[\meavsqudep{\majneutrino_i\rightarrow\lepton\higgs}-
	\meavsqudep{\majneutrino_i\rightarrow\bar\lepton\bar\higgs}\bigr]\dend
\end{align}
If the decay amplitudes in square brackets differ, the right-hand side of 
\eqn\eqref{eqn:asymmetry in equilibrium} represents the (non-zero) \CP-violating source 
term for the asymmetry generation. The total asymmetry is given by 
the sum over all flavors and $SU(2)_L$ components: $\n{L}\equiv \sum_{\alpha,a}(n_\lepton-n_{\bar\lepton})$.
Neglecting the quantum-statistical terms, $\qstatff{\lepton}{\momk}\qstatfb{\higgs}{\momp}\approx 1$, 
and integrating \eqn\eqref{eqn:asymmetry in equilibrium} over the lepton phase space we obtain for its
time derivative:
\begin{align}
	\label{eqn:decay asymmetry in equilibrium}
	\partial_t n_L&
	\approx 2\cdot\frac{g_N}{2\pi^2}\, \sum_i \epsilon_i\,\Gamma_i\, M_i^2\, 
	T\, K_1\left(\frac{M_i}{T}\right)\neq 0\kend
\end{align} 
where $K_1$ is the modified Bessel function of the second kind, 
$\Gamma_i =\Gamma_{\majneutrino_i \rightarrow\lepton\higgs}+
\Gamma_{\majneutrino_i \rightarrow\bar\lepton\bar\higgs}$ is the total tree-level decay 
width of $\majneutrino_i$, and the factor $g_\majneutrino=2$ emerges from the 
sum over the Majorana spin degrees of freedom in \eqref{eqn:asymmetry in equilibrium}, 
see \app\ref{Kinematics} for more details.
This implies that the source-term for the lepton asymmetry differs from zero even in 
equilibrium. On the other hand, combined with time translational invariance 
of an equilibrium state, \CPT-invariance requires the asymmetry to vanish 
in thermal equilibrium. Thus, we arrive at an apparent contradiction.
 
The generation of an asymmetry in equilibrium within the $S$-matrix formalism is a manifestation of the so-called 
double-counting problem. In vacuum an inverse decay immediately followed by a decay is equivalent to a scattering process 
where the intermediate particle is on the mass shell (real intermediate state or RIS). 
\begin{figure}[h!]
	\centering
	\includegraphics{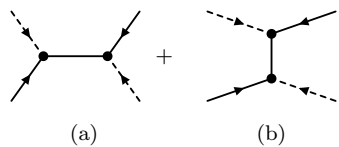}
	\caption{\label{LHbarLbarH}Two-body scattering process $\bar\lepton\bar\higgs 
\leftrightarrow \lepton\higgs$. Both graphs contribute with all $\majneutrino_i$ 
as intermediate states. Note that we read (b) as t-channel contribution.}
\end{figure}
Thus, the same contribution is taken into account twice: once by the amplitude 
for (inverse) decay processes, and once by that for the $\lepton\higgs 
\leftrightarrow \bar\lepton \bar\higgs$ scattering processes, see \fig\ref{LHbarLbarH}.(a). 

Let us convince ourselves that this is indeed the case. Including scattering 
processes we have for the distribution function of the leptons: 
\begin{align}
	\label{eqn:collision terms scattering}
	\momlep^\mu {\cal D}_\mu\f{\lepton}{\momlep}&=\ldots \nonumber\\
	&+\frac{1}{2} \sum_{\alpha,a}\int 
	\dpi{\bar\lepton\bar\higgs}{\higgs}{\momlep_2\momhig_2}{\momhig_1}
	(2\pi)^4\deltafour{\momlep+\momhig_1-\momlep_2-\momhig_2}\nonumber\\
	&\times \bigl[\ampabs{\bar\lepton\bar\higgs}{\lepton\higgs}^2
	\f{\bar\lepton}{}\f{\bar\higgs}{}\qstatff{\lepton}{\momlep}\qstatfb{\higgs}{\momhig}\nonumber\\
	&-\ampabs{\lepton\higgs}{\bar\lepton\bar\higgs}^2
	\f{\lepton}{}\f{\higgs}{}\qstatff{\bar\lepton}{\momk}\qstatfb{\bar\higgs}{\momlep}\bigr]\kend
\end{align} 
where the dots denote the contribution of the (inverse) decay processes, 
the sum is over flavors and $SU(2)_L$ components of the antileptons and 
we have introduced
\begin{align}
	\dpi{ab \dotso}{ij \dotso}{p_a p_b \dotso}{p_i p_j \dotso} 
	\equiv &\lorentzd{a}{p_a}\lorentzd{b}{p_b}\ldots\lorentzd{i}{p_i}\lorentzd{j}{p_j}\ldots \kend 
\end{align}
to shorten the notation. In the unflavored regime, to which we restrict our analysis, 
the distribution functions of leptons of all flavors are equal. If the Majorana neutrinos are 
close to equilibrium the difference between the distribution functions of the two spin 
degrees of freedom can be neglected as well. Therefore, in the expression for the total asymmetry 
$n_L$ the summation over spin, flavor and $SU(2)_L$ components reduces to summation of the 
corresponding decay and scattering amplitudes. We will denote these sums over internal 
degrees of freedom by $\EffAmplitude{}{}$ and call them effective amplitudes in the following. 
For the effective amplitude of $\bar\lepton\bar\higgs\rightarrow\lepton\higgs$ scattering
one obtains \cite{Pilaftsis:2003gt}
\begin{align}
	\EffAmplitude{}{\bar\lepton\bar\higgs\rightarrow\lepton\higgs}&=
	4(\momlep_1 \momlep_2)\textstyle\sum_{ij} M_iM_j \\
	&\times\big[2(\lambda^\dagger_{+}\lambda^{\phantom{\dagger}}_{+})^2_{ij}\,P^{*}_i(s)P_j(s)+
	2(h^\dagger h)^2_{ij}\,P^{*}_i(t)P_j(t)\nonumber\\
	&+(\lambda_+^\dagger h)^2_{ij}\,P^{*}_i(s)P_j(t)+
	(h^\dagger \lambda_+)^2_{ij}\,P^*_i(t)P_j(s)\big]\nonumber\kend
\end{align}
where $\momlep_{1,2}$ are the 
momenta of initial and final leptons respectively, and $s$ and $t$ are the usual Mandelstam variables. 
The amplitude $\EffAmplitude{}{\lepton\higgs\rightarrow\bar\lepton\bar\higgs}$ is obtained 
by interchanging $\lambda_+ \leftrightarrow\lambda_-$.
Note that the loop corrections to the Yu\-ka\-was vanish for negative momentum transfer, i.e.~in the 
$t$-channel. For this reason the above scattering amplitude contains combinations 
of the Yukawa couplings and their one-loop corrected counterparts.
The propagators $P_i$ are given by
\begin{align}
	\label{InversePropagator}
	P^{-1}_i(q^2)=q^2-M_i^2+i\theta(q^2)M_i\Gamma_i\kend
\end{align}
where $\theta$ is the Heaviside step function. The RIS contribution appears for the flavor diagonal ($i=j$) terms in the 
product of the $s$-channel amplitudes since only in this case the $s-M^2_i$ terms 
vanish simultaneously in both $P_i$ and $P_j$. In other words: 
\begin{align}
	\EffAmplitude{}{\bar\lepton\bar\higgs\rightarrow N_i \rightarrow \lepton \higgs} 
	= \frac{8 (\lambda^\dagger_{+}\lambda^{\phantom{\dagger}}_{+})^2_{ii} M_i^2
	(\momlep_1\momlep_2)}{(s-M_i^2)^2+(M_i \Gamma_i)^2} \kend
\end{align}
and a similar result for $\EffAmplitude{}{\lepton \higgs\rightarrow N_i \rightarrow \bar\lepton\bar\higgs}$.
Using the definitions of the effective couplings \eqref{effective couplings}
and the expression for the \CP-violating parameter \eqref{epsilon vacuum} we 
find that $(\lambda^\dagger_{+}\lambda^{\phantom{\dagger}}_{+})^2 \approx (h^\dagger h)^2_{ii} \, (1+\epsilon_i)^2$.
Furthermore, for a small decay width, we can approximate the Breit-Wigner propagator
by a delta-function using,
\begin{align}
	\label{deltalimit}
	\lim\limits_{\epsilon\rightarrow 0+}\frac{2\epsilon}{\omega^2+\epsilon^2}=
	\lim\limits_{\epsilon\rightarrow 0+}\frac{4\epsilon^3}{[\omega^2+\epsilon^2]^2}=
	2\pi\delta(\omega)\kend
\end{align} 
where $\omega=s-M_i^2$ and $\epsilon_i =M_i \Gamma_i$ in the considered case.
The RIS contribution to the scattering amplitude then takes the form
\begin{align}
	\label{ScatteringRISterm}
	\EffAmplitude{}{\bar\lepton\bar\higgs\rightarrow N_i \rightarrow \lepton \higgs} 
	&\approx \EffAmplitude{}{\bar \lepton\bar \higgs \rightarrow\majneutrino_i}
	\frac{\pi\delta(s-M_i^2)}{M_i \Gamma_i} \EffAmplitude{}{\majneutrino_i\rightarrow\lepton\higgs} \nonumber\\
	&\times \frac{2(\momlep_1\momlep_2)}{M_i^2 } \kend
\end{align}
where $\EffAmplitude{}{\majneutrino_i\rightarrow\lepton\higgs}=g_w (\lambda^\dagger_{+}\lambda^{\phantom{\dagger}}_{+})_{ii} 2\momlep\mommaj\approx
g_w (\lambda^\dagger_{+}\lambda^{\phantom{\dagger}}_{+})_{ii} M_i^2$ is the decay 
amplitude squared summed over all internal degrees of freedom (and a similar expression for 
anti-particles).
Just as one would expect, it is proportional to the product of the corresponding inverse 
decay and decay amplitudes. The additional momentum dependence (momenta of 
the leptons) arises because the initial and final states contain fermions. 
Close to thermal equilibrium $\f{\bar \lepton}{}\approx \f{\lepton}{}\approx\feq{\lepton}{}$ and 
$\f{\bar \higgs}{}\approx \f{\higgs}{}\approx\feq{\higgs}{}$. Neglecting the 
quantum-statistical terms we can write the RIS contribution to the 
source-term as:
\begin{align}
	\label{eqn:scattering asymmetry in equilibrium}
	\partial_t n_L & \approx 2 \int 
	\dpi{\bar\lepton\bar\higgs}{\lepton\higgs}{\momlep_1\momhig_1}{\momlep_2\momhig_2}
	(2\pi)^4\deltafour{\momlep_1+\momhig_1-\momlep_2-\momhig_2}\nonumber\\
	&\times \feq{\lepton}{}\feq{\higgs}{}\,
	\bigl[\EffAmplitude{}{\bar\lepton\bar\higgs\rightarrow N_i \rightarrow \lepton \higgs}-
	\EffAmplitude{}{\lepton \higgs\rightarrow N_i \rightarrow \bar\lepton\bar\higgs}\, 	
	\bigr]\dend
\end{align}
Taking into account that with Maxwell-Boltzmann distributions 
$\feq{\lepton}{}\feq{\higgs}{}=\feq{\majneutrino_i}{}$ in the presence of the {\ddelta}
 and performing the phase space integration 
using \eqn\eqref{eqn: scattering reaction density simplified} we obtain a result \textit{identical} to 
\eqn\eqref{eqn:decay asymmetry in equilibrium}. 

To correct the double-counting {\em in equilibrium} we may therefore subtract the RIS 
contribution from the scattering amplitude:
\begin{align} 
	\EffAmplitude{'}{\bar\lepton\bar\higgs\rightarrow \lepton \higgs}\equiv 
	\EffAmplitude{}{\bar\lepton\bar\higgs\rightarrow \lepton \higgs} - 
	\EffAmplitude{}{\bar\lepton\bar\higgs\rightarrow N_i \rightarrow \lepton \higgs} \kend
\end{align}
and similarly for the conjugate process $\lepton\higgs\rightarrow\bar\lepton\bar\higgs$. 
At first sight it might seem that the RIS subtracted scattering amplitudes 
$\EffAmplitude{'}{}$ 
do not contribute to the generation of the lepton asymmetry in equilibrium,
\begin{align}
	\partial_t n_L &\approx 2\int 
	\dpi{\bar\lepton\bar\higgs}{\lepton\higgs}{\momlep_1\momhig_1}{\momlep_2\momhig_2}
	(2\pi)^4\deltafour{\momlep_1+\momhig_1-\momlep_2-\momhig_2}\nonumber\\
	&\times \feq{\lepton}{}\feq{\higgs}{}\,\bigl[\EffAmplitude{}{\bar\lepton\bar\higgs\rightarrow \lepton \higgs}-
	\EffAmplitude{}{\bar\lepton\bar\higgs\rightarrow N_i \rightarrow \lepton \higgs}
	\nonumber\\
	&\hspace{17mm}-\EffAmplitude{}{\lepton \higgs\rightarrow \bar\lepton\bar\higgs}
	+\EffAmplitude{}{\lepton \higgs\rightarrow N_i \rightarrow \bar\lepton\bar\higgs}\,\bigr]\kend
\end{align}
but also cannot compensate the asymmetry generated in equilibrium by the 
decay processes, see \eqn\eqref{eqn:decay asymmetry in equilibrium}.
However, upon phase space integration the difference of the 
unsubtracted scattering amplitudes vanishes at leading order in $\yu$.
The remaining difference of the RIS-amplitudes precisely compensates the contribution 
of the (inverse) decay processes \eqref{eqn:decay asymmetry in equilibrium}. 

The RIS subtracted scattering amplitude can be conveniently rewritten in terms of a 
`RIS subtracted propagator' ${\cal P}_{ij}$. Motivated by \eqn\eqref{deltalimit} we 
define its diagonal components such, that they vanish upon integration 
over $s$ in the vicinity of the mass pole:
\begin{align}
	\label{RISpropagator}
	{\cal P}_{ii}(s)=\frac{(s-M^2_i)^2-(M_i\Gamma_i)^2}{[(s-M^2_i)^2+(M_i\Gamma_i)^2]^2}\dend
\end{align}
Since the second of the expressions \eqref{deltalimit} approaches the delta-function
faster than the first it is common to write \eqn\eqref{RISpropagator} in the form
\begin{align}
	\label{RIS_zero_width}
	{\cal P}_{ii}(s)&\rightarrow P^*_i(s)P_i(s)-\frac{\pi}{M_i\Gamma_i}\delta(s-M_i^2)\dend
\end{align}
For $i\neq j$ there is no need to perform 
the RIS subtraction and therefore ${\cal P}_{ij}(s)\equiv P^*_i(s)P_j(s)$. 
In the following we will also need the sum of the RIS subtracted tree-level scattering 
amplitudes. It does not contribute to the generation of the asymmetry but plays a role 
for its washout. It is defined as
\begin{align}
	\label{LHLH}
	\EffAmplitude{}{\bar\lepton\bar\higgs\leftrightarrow\lepton\higgs}&\equiv 
	{\textstyle\frac12}\bigl[\EffAmplitude{}{\bar\lepton\bar\higgs\rightarrow\lepton\higgs}+
	\EffAmplitude{}{\lepton\higgs\rightarrow\bar\lepton\bar\higgs}\bigr]\nonumber\\
	&=4(\momlep_1 \momlep_2)
	\textstyle\sum_{ij} M_i M_j\,\Re (h^\dagger h)^2_{ij} \nonumber\\
	&\times \left[\,2{\cal P}_{ij}(s)+2P^{*}_i(t)P_j(t)\right.\nonumber\\
	&+\left.P^{*}_i(s)P_j(t)+P^*_i(t)P_j(s)\,\right]\dend
\end{align}
Since it contains only the real part of $(h^\dagger h)^2_{ij}$ 
this process is \CP-conserving. 
\begin{figure}[h!]
	\centering
	\includegraphics{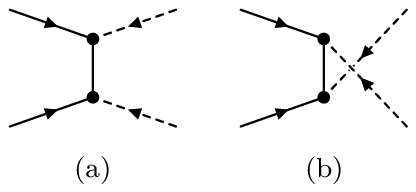}
	\caption{\label{LLbarHbarH}Two-body scattering process $\lepton\lepton \leftrightarrow \bar \higgs \bar \higgs$.}
\end{figure}
A further important washout process is $\lepton\lepton \leftrightarrow\bar\higgs\bar\higgs$ 
scattering which receives the $t$- and $u$-channel contributions, see \fig\ref{LLbarHbarH}.
By analogy with \eqn\eqref{LHLH} it is convenient to introduce
\begin{align}
	\label{LLHH}
	\EffAmplitude{}{\lepton\lepton\leftrightarrow\bar\higgs\bar\higgs}&\equiv 
	{\textstyle\frac12}\bigl[\EffAmplitude{}{\bar\higgs\bar\higgs\rightarrow\lepton\lepton}+
	\EffAmplitude{}{\higgs\higgs\rightarrow \bar\lepton\bar\lepton}\bigr]
	\nonumber\\
	&=2(\momlep_1 \momlep_2)\textstyle\sum_{ij} M_i M_j\,\Re(h^\dagger h)^2_{ij} \nonumber\\
	&\times \left[2 P^{*}_i(t)P^{*}_j(t)+2 P^{*}_i(u)P^{*}_j(u)\right.\nonumber\\
	&+\left. P^{*}_i(u) P_j(t)+P^{*}_i(t) P_j(u)\,\right]\dend
\end{align}
Since the intermediate Majorana neutrino cannot go on-shell in the $t$- and $u$-channel, 
there is no need to use the RIS subtracted propagator in \eqn\eqref{LLHH}.

Above we have briefly reviewed the canonical approach to the computation of the 
lepton asymmetry, which is based on generalized Boltzmann equations. 
Boltzmann equations, according to conventional reasoning, describe scattering 
processes of particles which propagate freely over timescales large compared to the 
duration of individual interactions. This picture seems to be consistent with the use 
of $S$-matrix elements which are intended to describe transitions between asymptotically 
free initial and final states. However, in leptogenesis the crucial processes 
({\CP}-violating decays) involve unstable particles which spoils this picture. 
In vacuum the amplitudes for such processes can be 
computed in terms of their Feynman graphs.
However the `naive' way of generalizing the Boltzmann equation by multiplying the obtained amplitudes 
by the one-particle distributions of the initial states and integrating over phase space 
leads to inconsistent equations. The origin of this problem is that the obtained collision 
terms for particle decay and inverse decay in \eqn\eqref{eqn:collision terms decay} 
miscount the rate of particle generation. In a short time-interval a finite number 
of unstable Majorana neutrinos - formed by inverse decay of particles and 
antiparticles - decays immediately back to either particles or antiparticles.
These contributions to particle generation are not included in 
\eqn\eqref{eqn:collision terms decay} where the amplitudes 
are defined in terms of Feynman graphs. For leptogenesis, in the presence of \CP-violation, it leads to inconsistent equations and must be 
corrected.
Since the missing contribution can be constructed as the 
rate of a two-body scattering process with on-shell intermediate state this issue 
can be addressed by the RIS-subtracting procedure presented above. 
It modifies the amplitudes for two-body scattering in order to cure the problem 
which appears due to the collision terms for particle decay.\footnote{The two pictures 
might seem equivalent for leptogenesis, but the first one implies that the Boltzmann equation 
for Majorana neutrino decay miscounts the rate as well. This is not corrected 
by the RIS subtraction of $\lepton\higgs\leftrightarrow\bar{\lepton}\bar{\higgs}$ 
processes. However the corresponding correction appears at order $\epsilon_i^2$,
which is usually neglected.}

\section{\label{RISQuantStat}RIS subtraction with quantum statistics}

It is well known that unitarity has important consequences for baryogenesis and 
leptogenesis \cite{Weinberg:1979bt,Kolb:1979qa,Roulet:1997xa} as it implies restrictions for 
the \CP-violating amplitudes. The issue of RIS subtraction is as well tightly related to 
unitarity as has been mentioned in e.g.~\cite{Pilaftsis:2003gt}.
As noted in \sect\ref{ConventionalApproach}, the use of `naive' Boltzmann equations of the kind 
\eqref{eqn:collision terms scattering} for unstable particles leads to problems such as 
the spurious asymmetry generation in the presence of \CP-violation in the decay of the heavy 
neutrinos. In this section we show explicitly that the success of the 
RIS subtraction \emph{in thermal equilibrium} is guaranteed by the unitarity of the $S$-matrix 
and 
how it can be generalized to include quantum-statistical terms. The approach to RIS subtraction 
differs slightly from the one discussed in the previous section.

To illustrate it we work in thermal equilibrium, $\feq{\bar{\lepton}}{}=\feq{\lepton}{}$
and $\feq{\bar{\higgs}}{}=\feq{\higgs}{}$, where
\begin{equation}
	\label{eqn:distribution function zero chemical potential}
	\feq{a}{\momp} = \bigl(e^{\e{a}{a}/T}+\stat{a}\bigr)^{-1}\dend
\end{equation}
Subtracting from the Boltzmann equation \eqref{eqn:collision terms scattering} 
 the corresponding 
equation for antiparticles, summing over internal degrees of freedom of the leptons and 
integrating with $\int \dif^3\momlep/[(2\pi)^3 \e{\lepton}{\momlep}]$
we obtain in thermal equilibrium:\footnote{In order to achieve exact thermal 
equilibrium, in this section we drop the $3\hubblerate n_L$ contribution which describes the dilution due 
to the expansion of the universe.}
\begin{align}
	\label{eqn:RIS subtraction with quantum-statistical terms}
	& \frac{\dif n_L}{\dif t} = 
	2 \sum_i \int 
	\dpi{}{\lepton\higgs\majneutrino_i}{}{\momlep\momhig\mommaj} (2\pi)^4\,\delta(\momlep +\momhig -\mommaj)\\
	&\times \bigl[ \EffAmplitude{}{\majneutrino_i\rightarrow\lepton\higgs} 
	- \EffAmplitude{}{\majneutrino_i\rightarrow\bar{\lepton}\bar{\higgs}}\bigr]
	\qstatfeqf{\lepton}{\momlep}\qstatfeqb{\higgs}{\momhig}\feq{\majneutrino_i}{\mommaj}\nonumber\\
	&+2 \int 
	\dpi{}{\lepton\higgs\lepton\higgs}{}{\momlep_1 k_1 p_2 k_2 }(2\pi)^4\,
	\delta(\momlep_1 + k_1 - p_2 - k_2 )\nonumber\\
	&\times \bigl[ \EffAmplitude{'}{\bar{\lepton}\bar{\higgs}\rightarrow\lepton\higgs} 
	- \EffAmplitude{'}{\lepton\higgs\rightarrow\bar{\lepton}\bar{\higgs}}\bigr]
	\qstatfeqf{\lepton}{\momk}\qstatfeqb{\higgs}{ k_1 }\feq{\lepton}{ p_2 }\feq{\higgs}{ k_2 }\nonumber\dend
\end{align}
We can exploit the unitarity of the 
$S$-matrix and \CPT-symmetry to obtain a requirement for a consistent approximation 
of the decay and scattering amplitudes. 
To this end we multiply \eqn\eqref{eqn:generalized optical theorem applied at order 
four difference}, which follows from the generalized optical theorem at order $h^4$, 
by $\feq{\lepton}{\momk}\feq{\higgs}{\momp}$ and integrate over 
$\lorentzd{\lepton}{\momk}\lorentzd{\higgs}{\momp}$. Assuming Maxwell-Boltzmann 
equilibrium distributions we may use $\feq{\lepton}{\momk}\feq{\higgs}{\momp}
=\feq{\majneutrino_i}{\momq}$ in the presence of the energy conserving Dirac-delta 
on the right-hand-side:
\begin{align}
	\label{eqn:generalized optical theorem applied at order four difference 1}
	&\sum_i\int \dpi{}{\lepton\higgs\majneutrino_i}{}{\momlep\momhig\mommaj} 
	(2\pi)^4\deltafour{\momlep+\momhig-\mommaj}\nonumber\\
	&\hspace{15mm}\times\big[ \EffAmplitude{}{{\majneutrino_i}\rightarrow\lepton\higgs} - 
	\EffAmplitude{}{{\majneutrino_i}\rightarrow\bar{\lepton}\bar{\higgs}}\big]\feq{\majneutrino_i}{\mommaj}\nonumber\\
	&=-\int \dpi{}{\lepton\higgs\lepton\higgs}{}{\momlep_1 k_1 p_2 k_2 }
	(2\pi)^4\deltafour{\momlep_1+ k_1 - p_2 - k_2 }\nonumber\\
	&\hspace{15mm}\times\big[ \EffAmplitude{'}{\bar{\lepton}\bar{\higgs}\rightarrow\lepton\higgs} 
	- \EffAmplitude{'}{\lepton\higgs\rightarrow\bar{\lepton}\bar{\higgs}}\big]\feq{\lepton}{ p_2 }\feq{\higgs}{ k_2 }\dend
\end{align}
We see that imposing this as a condition for the scattering amplitudes will correctly yield 
$\dif n_L/\dif t |_{\text{eq}}=0$ if we neglect the quantum-statistical 
terms in \eqref{eqn:RIS subtraction with quantum-statistical terms}. Equation 
\eqref{eqn:RIS subtraction with quantum-statistical terms} represents the zeroth-order term 
in an expansion about equilibrium. Using \eqn\eqref{eqn:generalized optical theorem applied 
at order four difference 1} we can therefore obtain consistent equations at this order
without the need to specify the detailed form of $\EffAmplitude{'}{\bar{\lepton}\bar{\higgs}
\rightarrow\lepton\higgs}$ and $\EffAmplitude{'}{\lepton\higgs\rightarrow\bar{\lepton}\bar{\higgs}}$. 
At higher order (for washout contributions) we also need to know the sum $\EffAmplitude{'}{\bar{\lepton}\bar{\higgs}
\rightarrow\lepton\higgs}+\EffAmplitude{'}{\lepton\higgs\rightarrow\bar{\lepton}\bar{\higgs}}$, 
see \sect\ref{RateEquations}. We know from \sect\ref{ConventionalApproach} that relation 
\eqref{eqn:generalized optical theorem applied at order four difference 1} can be satisfied 
by subtracting RIS contributions from the tree-level two-body scattering amplitudes and taking the zero 
width limit: 
\begin{subequations}
\label{eqn:matrix element RIS subtracted}
\noeqref{XibarLbarHLH,XiLHbarLbarH}
\begin{align}
	\label{XibarLbarHLH}
	\EffAmplitude{'}{\bar{\lepton}\bar{\higgs}\rightarrow\lepton\higgs}{}&
	=\EffAmplitude{}{\bar{\lepton}\bar{\higgs}\leftrightarrow\lepton\higgs}\nonumber\\
	&-{\textstyle\sum_i}\,\EffAmplitude{}{\bar \lepton\bar\higgs\rightarrow{\majneutrino_i}}
	\frac{\pi \delta(s-M_i^2)}{2{M_i}\width{i}}
	\EffAmplitude{}{{\majneutrino_i}\rightarrow\lepton\higgs}\kend\\
	\label{XiLHbarLbarH}
	\EffAmplitude{'}{\lepton\higgs\rightarrow\bar{\lepton}\bar{\higgs}}{}&=
	\EffAmplitude{}{{\lepton\higgs}\leftrightarrow\bar{\lepton}\bar{\higgs}}\nonumber\\
	&-{\textstyle\sum_i}\,\EffAmplitude{}{\lepton\higgs\rightarrow {\majneutrino_i}}
	\frac{\pi \delta(s-{M_i}^2)}{2{M_i}\width{i}}
	\EffAmplitude{}{{\majneutrino_i}\rightarrow\bar{\lepton}\bar{\higgs}}
	\dend
\end{align}
\end{subequations}
Note that, strictly speaking, the RIS terms in \eqn\eqref{ScatteringRISterm}
include $4(\momlep_1\momlep_2)/M_i^2$ factors. However, 
upon the phase space integration in \eqn\eqref{eqn:generalized optical theorem applied at order four difference 1} 
the two expressions give identical results and are therefore
equal in an average sense.
It is obvious from comparison of \eqns\eqref{eqn:RIS subtraction with quantum-statistical terms} 
and \eqref{eqn:generalized optical theorem applied at order four difference 1} that 
the above definition of the RIS subtracted scattering amplitudes 
is not sufficient to guarantee zero asymmetry in equilibrium if quantum-statistical 
terms are included.
However this can be achieved if we replace the vacuum decay width
in \eqn\eqref{eqn:matrix element RIS subtracted} by the thermal one \cite{Giudice:2003jh,Kiessig:2010pr}:
\begin{align}
	\label{eqn:definition thermal width}
	\Gammaeff{}{i}=\Gammaeff{}{i}(\mommaj)&\equiv
	\frac1{2g_\majneutrino M_i}\int \lorentzd{\lepton}{\momlep} \lorentzd{\higgs}{\momhig}
	(2\pi)^4\deltafour{\mommaj -\momlep -\momhig}\nonumber\\
	&\times\bigl[ \EffAmplitude{}{\majneutrino_i\rightarrow\lepton\higgs} 
	+ \EffAmplitude{}{\majneutrino_i\rightarrow\bar{\lepton}\bar{\higgs}}\bigr]
	(1-\feq{\lepton}{\momlep}+\feq{\higgs}{\momhig})\dend
\end{align}
Using the identity $1=\int ds \int d^4 q \delta_+(q^2-s)\delta(q-\momlep-\momhig)$
and the fact that the (inverse) decay amplitudes are related by \CPT-symmetry we can 
rewrite the RIS-con\-tri\-bution to the second term of \eqn\eqref{eqn:RIS subtraction with quantum-statistical terms} 
in the form:
\begin{align}
	\label{eqn:RIS subtraction with quantum-statistical terms first term}
	-&\int ds \int\frac{d^4q}{(2\pi)^3}\delta_+(q^2-s)\sum_i\frac{\delta(s-M_i^2)}{2M_i\Gammaeff{}{i}}\nonumber\\
	&\times \int \lorentzd{\lepton}{\momlep_1} \lorentzd{\higgs}{\momhig_1}(2\pi)^4\delta(q-\momlep_1-\momhig_1)\nonumber\\
	&\hspace{22mm} \times \feq{\lepton}{\momlep_1}\feq{\higgs}{\momhig_1}\bigl[ \EffAmplitude{}{\majneutrino_i\rightarrow\lepton\higgs} 
	+ \EffAmplitude{}{\majneutrino_i\rightarrow\bar{\lepton}\bar{\higgs}}\bigr]\nonumber\\
	&\times \int \lorentzd{\lepton}{\momlep_2} \lorentzd{\higgs}{\momhig_2}(2\pi)^4\delta(q-\momlep_2-\momhig_2)\nonumber\\
	&\hspace{5mm} \times (1-\feq{\lepton}{\momlep_2})(1+\feq{\higgs}{\momhig_2})\bigl[ \EffAmplitude{}{\majneutrino_i\rightarrow\lepton\higgs} 
	- \EffAmplitude{}{\majneutrino_i\rightarrow\bar{\lepton}\bar{\higgs}}\bigr]\dend
\end{align} 
The integration over $s$ is trivial. The $\delta_+(q^2-s)$ term ensures that after integration 
over $dq^0$ the intermediate Majorana neutrino is on-shell, $q^2=M_i^2$. 
Using ${\feq{\lepton}{\momq}\feq{\higgs}{\momr}}=
{\feq{\majneutrino_i}{\moml}}(1-\feq{\lepton}{\momq}+\feq{\higgs}{\momr})$ together 
with the definition \eqref{eqn:definition thermal width} we can rewrite the second term of
\eqn\eqref{eqn:RIS subtraction with quantum-statistical terms first term} as $2 g_NM_i\Gammaeff{}{i}{\feq{\majneutrino_i}{\mommaj}}$
which cancels the factors coming from RIS subtraction. The resulting 
expression reads

\begin{align} 
	-2 & \sum_i\int 
	\dpi{}{\lepton\higgs\majneutrino_i}{}{\momlep\momhig\mommaj} (2\pi)^4\,\delta(\momlep +\momhig -\mommaj)\nonumber\\
	&\times \bigl[ \EffAmplitude{}{\majneutrino_i\rightarrow\lepton\higgs} 
	- \EffAmplitude{}{\majneutrino_i\rightarrow\bar{\lepton}\bar{\higgs}}\bigr]
	\qstatfeqf{\lepton}{\momlep}\qstatfeqb{\higgs}{\momhig}\feq{\majneutrino_i}{\mommaj}\kend
\end{align}
and cancels the first term on the right-hand side of \eqn\eqref{eqn:RIS subtraction with quantum-statistical terms}.
Since $\EffAmplitude{}{\bar{\lepton}\bar{\higgs}\rightarrow\lepton\higgs}=
\EffAmplitude{}{\lepton\higgs\rightarrow\bar{\lepton}\bar{\higgs}}$ at $\order{h^4}$ the new RIS subtracted source-term for
the asymmetry vanishes in equilibrium.

The thermal width $\Gammaeff{}{i}$ defined in \eqn\eqref{eqn:definition thermal width} 
would also be obtained if one computes it using thermal cutting rules instead 
of the optical theorem (which applies in vacuum), see 
\app\ref{GeneralizedOptTh}.\footnote{Note in this context that the computation of the 
self-energy contribution to the \CP-violating parameter in thermal QFT is in effect only 
a variation of this \cite{Garny:2010nj}.} We have seen that the unitarity of the $S$-matrix 
can be employed to generalize the concept of RIS subtraction to rate equations which 
include quantum-statistical factors. As we shall see in \sect\ref{HiggsDecay}, the 
Majorana neutrino decay is at high temperature replaced by Higgs decay if the Higgs 
acquires a large effective thermal mass. In this case thermal cutting rules enforce 
relations between the amplitudes which can be used to obtain consistent equations, 
analogous to the optical theorem, see \app\ref{GeneralizedOptTh}.

Note again that in \eqn\eqref{eqn:RIS subtraction with quantum-statistical terms first term} 
we had to assume that the Majorana neutrinos are in exact thermal equilibrium. For 
leptogenesis this is an inconsistent assumption since the deviation of their distribution 
from equilibrium realizes the third Sakharov condition and drives the generation of 
the asymmetry. Not surprisingly, the NEQFT approach leads to a (slightly) 
different result for the kinetic equations. However the differences between the two 
approaches enter only at an order beyond the usual approximation as we will discuss 
in the next section.

\section{\label{RateEquations}Rate equations}

In this section we review the derivation of rate equations, discuss in how far quantum-statistical
and medium corrections can be incorporated, and compare the structure obtained when starting from
the NEQFT result \eqref{eqn:main result} with the conventional form.
 Solving a system of 
Boltzmann-like equations in general 
requires the use of numerical codes capable of treating large systems of 
stiff differential equations for the different momentum modes -- a cumbersome task 
if one wants to study a wide range of model parameters. 
In the context of baryogenesis, a commonly employed simplification is to approximate the Boltzmann equations 
by the corresponding network of `rate equations' for number densities $\n{a}$ 
or abundances $Y_a\equiv \n{a}/s$, where $s$ is the comoving entropy 
density. The resulting 
equations correspond to the hydrodynamical limit of the Boltzmann kinetic 
equations, in the comoving frame of {\em homogeneous} FRW space-time. 
To obtain evolution equations for ${Y_a}$ in the conventional approach, 
i.e.~from \eqn\eqref{eqn:collision terms scattering}, we therefore integrate 
the corresponding Boltzmann equations over $g_{a}d^3\momlep/[\e{}{\momlep} (2\pi)^3]$ to obtain, 
on the left-hand sides:
\begin{align*}
	\frac{\dif n_a}{\dif t}+3\hubblerate n_a = \frac{s\hubbleratem }{z}\frac{\dif Y_a}{dz}\kend
\end{align*}
where we have introduced the dimensionless inverse temperature $z=M_1/T$ and the 
Hubble rate $\hubbleratem =H{|_{T=M_1}}$. In the homogeneous and isotropic Universe
the derivative of the quantity 
$Y_L \equiv \n{L}/s$ can be related to the divergence of the lepton-current $j_L^\mu=(n_L,\vec{0})$ -- a quantity which is particularly easy to access in the first-principles computation -- by
\begin{align*}
	{\cal D}_{\mu} j_L^\mu(t) = \frac{s\hubbleratem }{z}\frac{d Y_L}{dz}\dend
\end{align*}
On the right-hand sides we get sums of integrated collision terms representing the 
effect of the different interactions. We separate contributions attributed to decays and scattering:
\begin{align}
	\frac{d Y_L}{dz} =\sum_i\frac{d Y_{L_i}}{dz}\bigg|_{D} + \frac{d Y_L}{dz}\bigg|_{S}\dend
\end{align}
The decay contributions ${d Y_{L_i}}/{dz}\big|_{D}$ to ${d Y_{L}}/{dz}$ are very similar 
to the decay contributions ${d Y_{\majneutrino_i}}/{dz}\big|_{D}$ to ${d Y_{\majneutrino_i}}/{dz}$ 
and we can treat them in the same way. Reordering the contributions to 
${d Y_{L_i(\majneutrino_i)}}/{dz}\big|_{D}$ we find
\begin{align}
	\label{divlepCPpred}
	\frac{s\hubbleratem }{z}&\frac{d Y_{L_i\atop\majneutrino_i}}{dz}\bigg|_{D}
	=\frac{s\hubbleratem }{z}\frac{d Y_{L_i\atop\majneutrino_i}}{dz}\bigg|_{D,\text{extra}}\\
	&+\int \dpi{\majneutrino_i\higgs\lepton}{}{\mommaj\momhig\momlep}{}
	\bigl[\pm\EffAmplitude{}{\majneutrino_i \rightleftarrows\lepton\higgs}
	\F{\lepton\higgs}{\majneutrino_i}{\momlep\momhig}{\mommaj} - 
	\EffAmplitude{}{\majneutrino_i \rightleftarrows\bar\lepton\bar\higgs}
	\F{\bar{\lepton}\bar{\higgs}}{\majneutrino_i}{\momlep\momhig}{\mommaj}\bigr]\kend\nonumber
\end{align}
where the upper (lower) signs and arrows correspond to the rate equations for $L$ 
($\majneutrino_i$) abundance and we defined 
\begin{align}
	\label{eqn:rate divlep extra term}
	\frac{s\hubbleratem }{z}&\frac{d Y_{L_i}}{dz}\bigg|_{D,\text{extra}}\equiv
	\int \dpi{\majneutrino_i\higgs\lepton}{}{\mommaj\momhig\momlep}{} 
	\, (2\pi)^4\,\deltafour{\mommaj-\momhig-\momlep}\\
	&\times\bigl(\EffAmplitude{}{\majneutrino_i \rightarrow\lepton\higgs}-
	\EffAmplitude{}{\lepton\higgs\rightarrow\majneutrino_i}\bigr)
	(1-\f{\majneutrino_i}{\mommaj})\bigl[ \f{\higgs}{\momhig} 
	\f{\lepton}{\momlep} + \f{\bar{\higgs}}{\momhig} \f{\bar{\lepton}}{\momlep}\bigr]\nonumber\kend
\end{align}
which corresponds to \eqn\eqref{eqn:decay asymmetry in equilibrium}, as well as
\begin{align}
	\label{eqn:rate Majorana extra term}
	\frac{s\hubbleratem }{z}&\frac{d Y_{\majneutrino_i}}{dz}\bigg|_{D,\text{extra}}\equiv
	\int \dpi{\majneutrino_i\higgs\lepton}{}{\mommaj\momhig\momlep}{} \, 
	(2\pi)^4\,\deltafour{\mommaj-\momhig-\momlep}\nonumber\\
	&\times \bigl(\EffAmplitude{}{\lepton\higgs \rightarrow\majneutrino_i}-
	\EffAmplitude{}{\majneutrino_i\rightarrow\lepton\higgs}\bigr)\nonumber\\
	&\times \f{\majneutrino_i}{\mommaj}\bigl[ \qstatff{\lepton}{\momlep} 
	\qstatfb{\higgs}{\momhig} - \qstatff{\bar{\lepton}}{\momlep}
	\qstatfb{\bar{\higgs}}{\momhig}\bigr]\dend
\end{align}
We used \CPT-symmetry of the amplitudes in the derivation of \eqns\eqref{eqn:rate divlep extra term} 
and \eqref{eqn:rate Majorana extra term}. Later we will see that the second term in 
\eqn\eqref{divlepCPpred} appears also in the first-principle approach, compare \eqn\eqref{eqn:main result}, while the terms 
in \eqns\eqref{eqn:rate divlep extra term} and \eqref{eqn:rate Majorana extra term} are absent. 
This motivates the separation into `regular' and `extra' terms performed in 
\eqn\eqref{divlepCPpred}. For the contributions attributed to scattering we get:
\begin{align}
	\label{divlepCPscatt}
	\frac{s\hubbleratem }{z}\frac{d Y_L}{dz}\bigg|_{S}&=
	\frac{s\hubbleratem }{z}\frac{d Y_L}{dz}\bigg|_{S,\text{extra}}\\
	+\int & \dpi{\lepton\higgs\lepton\higgs}{}{\momk\momp\momq\momr}{}\,
	\textstyle \bigl(\EffAmplitude{'}{\bar{\lepton}\bar{\higgs}
	\rightarrow\lepton\higgs}+\EffAmplitude{'}{\lepton\higgs\rightarrow
	\bar{\lepton}\bar{\higgs}}\bigr)\F{\lepton\higgs}{\bar{\lepton}
	\bar{\higgs}}{\momk\momp}{\momq\momr}\nonumber\\
	+ \int & \dpi{\lepton\lepton\higgs\higgs}{}{\momk\momp\momq\momr}{}
	\bigl[\EffAmplitude{}{{\lepton}{\lepton}\rightarrow\bar{\higgs}\bar{\higgs}}
	\F{\lepton\lepton}{\bar{\higgs}\bar{\higgs}}{\momk\momp}{\momq\momr}
	-\EffAmplitude{}{\bar{\lepton}\bar{\lepton}\rightarrow{\higgs}{\higgs}} 
	\F{\bar{\lepton}\bar{\lepton}}{{\higgs}{\higgs}}{\momk\momp}{\momq\momr}\bigr]\kend\nonumber
\end{align}
with
\begin{align}
	\label{divlepCPscatt extra term}
	&\frac{s\hubbleratem }{z}\frac{d Y_L}{dz}\bigg|_{S,\text{extra}}=\\
	&+ \int \dpi{\lepton\higgs\lepton\higgs}{}{\momk\momp\momq\momr}{}\, 
	(2\pi)^4\deltafour{\momk+\momp-\momq-\momr} \bigl(\EffAmplitude{'}{\bar{\lepton}\bar{\higgs}
	\rightarrow\lepton\higgs}-\EffAmplitude{'}{\lepton\higgs\rightarrow\bar{\lepton}\bar{\higgs}}\bigr)\nonumber\\
	&\times \bigl[ \qstatff{\lepton}{\momk}\qstatfb{\higgs}{\momp}\f{\bar{\lepton}}{\momq}\f{\bar{\higgs}}{\momr}
	+ \qstatff{\bar{\lepton}}{\momk}\qstatfb{\bar{\higgs}}{\bar{\momp}}
	\f{\lepton}{\momq}\f{\higgs}{\momr}\bigr]\kend\nonumber
\end{align}
corresponding to \eqn\eqref{eqn:scattering asymmetry in equilibrium}.
Again, \eqn\eqref{divlepCPscatt extra term} 
does not appear in the first-principle approach.
Since in equilibrium the regular terms in each of \eqns\eqref{divlepCPpred} and 
\eqref{divlepCPscatt} vanish by detailed balance we retain 
\eqn\eqref{eqn:RIS subtraction with quantum-statistical terms} in the sum of decay and 
scattering contributions. The latter vanishes as well in equilibrium if we adopt 
e.g.~\eqn\eqref{eqn:matrix element RIS subtracted} with thermal width for the RIS subtracted 
amplitudes $\EffAmplitude{'}{\bar{\lepton}\bar{\higgs}\rightarrow\lepton\higgs},\,
\EffAmplitude{'}{\lepton\higgs\rightarrow\bar{\lepton}\bar{\higgs}}$. Out of equilibrium 
the last terms constitute a structural difference compared to the results obtained from 
first-principles. This difference carries over to the rate equations. We will therefore 
analyze these contributions separately.

The computational advantage of rate equations over full Boltzmann equations is maximized by a
number of common approximations. In particular, assuming that all species are close to 
equilibrium and that the Majorana neutrino distribution 
function $\flong{\majneutrino_i}{t}{\abs{{\vec \mommaj}}}$ is proportional to its equilibrium 
distribution for all values of the momentum 
$\abs{\vec \mommaj}$. The temperature for all 
kinetic equilibrium distributions is set to a common value $T$ while finite deviations of 
the chemical potential with small $\mu/T$ are permitted. These approximations result for 
$\mu_\lepton /T, \mu_\higgs /T, \mu_{\majneutrino_i} /T \ll 1$ in a closed network of rate 
equations for the abundances of the form
(compare with \cite{Kolb:1979qa,Buchmuller:2004nz,Pilaftsis:2003gt,Giudice:2003jh}):
\begin{subequations}
\label{RateEquations1}
\begin{align}
	\label{RateEquationsa}
	&\frac{s\hubbleratem }{z}\frac{d Y_L}{dz}= \textstyle\sum_{i}\langle \epsilon_i\, 
	\gamma^D_{\majneutrino_i}\rangle\bigg(\frac{Y_{\majneutrino_i}}{Y_{\majneutrino_i}^{eq}}-1\bigg)\\
	&-\frac{Y_L}{2\,Y_\lepton^{eq}}(1+c_{\higgs\lepton}) c_\lepton\bigl(\textstyle\sum_{i}\langle 
	\gamma^W_{\majneutrino_i}\rangle + 4\langle \gamma'^{\lepton\higgs}_{\bar \lepton \bar \higgs} 
	\rangle +4\langle \gamma^{\bar\lepton \bar\lepton}_{ \higgs \higgs}\rangle \bigr)\nonumber \kend\\
	\label{RateEquationsb}
	&\frac{s\hubbleratem }{z}\frac{d Y_{\majneutrino_i}}{dz}=-\langle\gamma^D_{\majneutrino_i}
	\rangle\bigg(\frac{Y_{\majneutrino_i}}{Y_{\majneutrino_i}^{eq}}-1\bigg)\kend
\end{align}
\end{subequations}
where we have introduced 
\begin{subequations}
	\label{AmplsqAndEpsDef}
	\noeqref{EpsilonUnintegrated,XiUnintegrated}
	\begin{align}
		\label{EpsilonUnintegrated}
		\epsilon_i & \equiv
		\frac{\EffAmplitude{}{\majneutrino_i \rightarrow \lepton\higgs}-
		\EffAmplitude{}{\majneutrino_i \rightarrow \bar\lepton\bar\higgs}}{
		\EffAmplitude{}{\majneutrino_i}}\kend\\	
		\label{XiUnintegrated}
		\EffAmplitude{}{\majneutrino_i} & \equiv 
		\EffAmplitude{}{\majneutrino_i \rightarrow \lepton\higgs}+
		\EffAmplitude{}{\majneutrino_i \rightarrow \bar\lepton\bar\higgs}\dend
	\end{align}
\end{subequations}
The factor $c_\lepton\equiv {9\zeta(3)}/{\pi^2}\approx 1.1$ (we neglect the thermal lepton masses here)
relates the chemical potential of the leptons to their number density,
\begin{align*}
	\frac{\mu_\lepton}{T}\approx c_\lepton\cdot \frac{\abundance{L}}{2\abundance{\lepton}}\kend
\end{align*}
and the coefficient $c_{\higgs\lepton}$ takes into account that in the SM the chemical potentials of leptons 
and Higgs are related by $\mu_\higgs=c_{\higgs\lepton}\cdot \mu_\lepton$ with $c_{\higgs\lepton}=4/7$ 
through equilibrium gauge, Yukawa and sphaleron interactions \cite{Buchmuller:2005eh,Kartavtsev:2005rs,PhysRevD.42.3344}. 

Hence, the evolution of the abundances close to equilibrium is roughly governed by a few average 
quantities called \textit{reaction densities} which describe decay and scattering processes. We 
will refer to $\langle \epsilon_i\, \gamma^D_{\majneutrino_i}\rangle$, $\langle \gamma^D_{\majneutrino_i}
\rangle$, $\langle \gamma^W_{\majneutrino_i}\rangle$ as {\CP}-violating decay reaction density, 
decay reaction density and washout reaction density respectively.
For comparison with standard results we want to maintain the form of 
\eqns\eqref{RateEquations1} and repeat their derivation from \eqns\eqref{divlepCPpred} 
and \eqref{divlepCPscatt} to obtain expressions for the reaction densities which take the quantum 
statistical factors of the Boltzmann equation into account. This is important in the present context 
because the thermal corrections to the {\CP}-violating parameter, to be derived later, are of a 
similar kind. To this end we use that the SM gauge and Yukawa interactions keep the Higgs and 
leptons very close to kinetic equilibrium:
\begin{equation*}
	\label{eqn:kinetic equilibrium distribution general}
	\f{a}{\momp} = \bigl(e^{(\e{a}{\momp}-\mu_a)/T}+\stat{a}\bigr)^{-1}\kend
\end{equation*}
with a common temperature $T_\lepton =T_\higgs =T$ and chemical potentials 
$\mu_\lepton =-\mu_{\bar{\lepton}}$, $\mu_\higgs =-\mu_{\bar{\higgs}}$. 
We shall also use $\feq{a}{p}$ for the equilibrium distribution functions with zero 
chemical potential defined in \eqn\eqref{eqn:distribution function zero chemical potential}.

Since a chemical potential with positive sign will appear for either the Higgs or its 
antiparticle, we need to include at least the thermal mass of the Higgs to be consistent. 
In the dense plasma gauge-, Yukawa- and Higgs self-interactions induce a large thermal 
Higgs mass of about $0.4\,T$. With 
$\mu_\higgs/T\sim\mu_\lepton/T\sim\epsilon_i^{vac}\ll \m{\higgs}/T\sim 0.4$, the Higgs cannot 
acquire a condensate component. It is then safe to use a Bose-Einstein equilibrium 
distribution function to describe the distribution of the Higgs particles. Using that 
$\qstatf{a}{\momk}=\exp((\e{a}{\momk}-\mu_a)/T)\f{a}{\momk}$ and hence, for a general decay 
collision term $N\leftrightarrow a b$ in the presence of the energy conserving 
{\ddelta},\footnote{$N$, $a$ and $b$ can be any species for which the above conditions apply. 
Here we identify $N=\majneutrino_i$, $a\in\{\lepton ,\bar{\lepton}\}$, $b\in\{\higgs ,\bar{\higgs}\}$.} 
$\f{a}{\momlep}\f{b}{\momhig}=\exp ((\mu_a +\mu_b)/{T})\qstatf{a}{\momlep}\qstatf{b}{\momhig}\
\feq{N}{\mommaj}/\qstatfeqf{N}{\mommaj}$, we may write:
\begin{align}
	\label{eqn:general decay collision term}
	&\int \dpi{N a b}{}{\mommaj\momlep\momhig}{}\EffAmplitude{}{a b\leftrightarrows N}
	\F{a b}{N}{\momlep\momhig}{\mommaj}=\nonumber\\
	&\int \dpi{N a b}{}{\mommaj\momlep\momhig}{} \, (2\pi)^4\,\deltafour{\mommaj-\momhig-\momlep} 
	\EffAmplitude{}{a b\leftrightarrows N}\nonumber\\
	&\times\qstatf{a}{\momlep}\qstatf{b}{\momhig}\left[ \frac{\f{N}{\mommaj}
	-\feq{N}{\mommaj}}{\qstatf{N}{\mommaj}\feq{N}{\mommaj}} - (e^{\frac{\mu_{a}+\mu_{b}}{T}}-1)\right]\nonumber\\
	&\times\qstatf{N}{\mommaj}\frac{\feq{N}{\mommaj}}{\qstatfeq{N}{\mommaj}}\dend
\end{align}
We can now expand the exponential in square brackets in the small quantity $(\mu_{a}+\mu_{b})/{T}$. 
If this quantity is tiny at all times the integral \eqref{eqn:general decay collision term} will not 
change much if we neglect quadratic and higher order terms.\footnote{By inserting equilibrium 
distribution functions for leptons and Higgs in the derivation of the {\CP}-violating parameter 
we will neglect terms of the order $\epsilon_i^{vac} ({\mu_{{\lepton}}+\mu_{{\higgs}}})/{T}$ as well.} 
For the zeroth-order (first) term in square brackets we use the linear expansion 
$\qstatf{a}{\momlep}\qstatf{b}{\momhig}\approx \qstatfeq{a}{\momlep}\qstatfeq{b}{\momhig}
\bigl[1-(\xi_a\frac{\mu_a}{T}\feq{a}{\momlep}+\xi_b\frac{\mu_b}{T}\feq{b}{\momhig})\bigr]$ of 
the prefactor. The linear order (second) term in square brackets will appear preceded by just 
the zeroth-order factor $\qstatfeq{a}{\momlep}\qstatfeq{b}{\momhig}=f_{a b}\qstatfeq{N}{\mommaj}$ with
\begin{align}
	\label{eqn:linear expansion of product of quantum-statistical terms}
	f_{a b}\equiv(1-\xi_a\feq{a}{\momlep}-\xi_b\feq{b}{\momhig})\dend
\end{align}
To write the results in a compact form we introduce decay reaction densities with 
quantum-statistical factors included: 
\begin{align}
	\label{eqn:reaction density decays}
	\hskip -5mm
	\langle X \gamma^D_{\majneutrino_i}\rangle \equiv \int \hspace{-1mm} &
	\dpi{{\lepton}{\higgs}{\majneutrino_i}}{}{{\momlep}{\momhig}{\mommaj}}{}\, 
	(2\pi)^4 \deltafour{\mommaj-\momk-\momlep} X\EffAmplitude{}{\majneutrino_i} 
	\feq{\majneutrino_i}{q} f_{\lepton\higgs},
\end{align}
and
\begin{align}
	\label{eqn:reaction density washout}
	& \langle X\gamma^W_{\majneutrino_i}\rangle \equiv \big< X\qstatfeqf{\majneutrino_i}{}
	\gamma^D_{\majneutrino_i}\big>\\
	& = \int \dpi{{\lepton}{\higgs}{\majneutrino_i}}{}{{\momlep}{\momhig}{\mommaj}}{}\, 
	(2\pi)^4 \deltafour{\mommaj-\momk-\momlep} X\EffAmplitude{}{\majneutrino_i} 
	\qstatfeqf{\majneutrino_i}{q} \feq{\majneutrino_i}{q} f_{\lepton\higgs}\kend\nonumber
\end{align}
where $\EffAmplitude{}{\majneutrino_i}$ is the total Majorana decay amplitude. Similarly 
we define the scattering reaction densities as 
\begin{align}
\label{ReactDensExact}
\langle X\gamma^{ab}_{ij}\rangle \equiv\int &\dpi{{a}{b}{i}{j}}{}{{\momk}{\momp}{\momq}{\momr}}{} 
(2\pi)^4 \delta(\momk +\momp -\momq -\momr) X\EffAmplitude{}{ab\leftrightarrow ij}\nonumber\\
\times & \qstatfeq{a}{\momk}\qstatfeq{b}{\momp}\feq{i}{\momq}\feq{j}{\momr}\dend
\end{align}
Since $\EffAmplitude{}{ab\leftrightarrow ij}$ refers here to a CP-symmetric (tree-level) 
amplitude squared we have $\langle X\gamma^{ab}_{ij}\rangle=\langle X\gamma^{ij}_{ab}\rangle$ 
if $X$ is symmetric as well.

With help of \eqn\eqref{eqn:general decay collision term expanded} we may separate the contributions 
to ${d Y_{L_i}}/{dz}\big|_D$ into terms proportional to $\Delta\f{\majneutrino_i}{\mommaj}
\equiv({\f{\majneutrino_i}{\mommaj}-\feq{\majneutrino_i}{\mommaj}})$, terms proportional to 
$\Delta\f{\majneutrino_i}{\mommaj}\cdot {\frac{\mu_\lepton}{T}}$, or just proportional to 
$\mu_\lepton/T$ (see \app\ref{RateEquationsApp} for details):
\begin{subequations}
	\label{eqn:quantum corrected Boltzmann equation Li result} 
	\begin{align}
		\label{eqn:quantum corrected Boltzmann equation Delta f Li result}
		&\frac{s\hubbleratem }{z}\frac{d Y_{L_i}}{dz}\Big|_{\Delta\f{\majneutrino_i}{}} 
		=\biggl\langle \epsilon_i\frac{\Delta\f{\majneutrino_i}{\mommaj}}{\feq{\majneutrino_i}{}}
		\gamma^D_{\majneutrino_i}\biggr\rangle\kend\\
		\label{eqn:quantum corrected Boltzmann equation mu by T times Delta f Li result}
		&\frac{s\hubbleratem }{z}\frac{d Y_{L_i}}{dz} \Big|_{\Delta\f{\majneutrino_i}{}{\frac{\mu_\lepton}{T}}} 
		=\frac{\mu_\lepton}{T}\bigl\langle(c_{\higgs\lepton}\feq{\higgs}{\momhig}-\feq{\lepton}{\momlep}) 
		\frac{\Delta\f{\majneutrino_i}{\mommaj}}{\feq{\majneutrino_i}{}}\gamma^D_{\majneutrino_i}\bigr\rangle
		\kend\hphantom{aaaa}\\
		\label{eqn:quantum corrected Boltzmann equation mu by T Li result}
		&\frac{s\hubbleratem }{z}\frac{d Y_{L_i}}{dz}\Big|_{\frac{\mu_\lepton}{T}}=
		-\frac{\mu_\lepton}{T}(1+c_{\higgs\lepton})\biggl\langle\frac{(1-\f{\majneutrino_i}{})}{(1-\feq{\majneutrino_i}{})}
		\gamma^W_{\majneutrino_i}\biggr\rangle\dend
	\end{align}
\end{subequations}
In addition we get with \eqn\eqref{eqn:quantum corrected Boltzmann equation decay extra term} 
for the extra term in \eqn\eqref{divlepCPpred}:
\begin{align}
	\label{eqn:quantum corrected Boltzmann equation decay extra term Li result}
	&\frac{s\hubbleratem }{z}\frac{d Y_{L_i}}{dz}\Big|_{D,\text{extra}}= 2
	\bigl\langle\qstatff{\majneutrino_i}{}\epsilon_i\gamma^D_{\majneutrino_i}\bigr\rangle\dend
\end{align}
Equations \eqref{eqn:quantum corrected Boltzmann equation Li result} describe 
the generation of a net asymmetry due to out of equilibrium decays of heavy Majorana neutrinos. 
Once $\mu_\lepton/T$ has a non-zero value, there will be a slight difference in the decay rates 
to particles and antiparticles respectively which is not due to {\CP}-violation in the decay amplitude, 
but due to the presence of slightly different occupation numbers of leptons and Higgs in the final 
states of the decays. At linear order this combined effect of blocking and stimulated emission is 
accounted for by \eqn\eqref{eqn:quantum corrected Boltzmann equation mu by T times Delta f Li result}. 
Depending on the `typical' sign of $(c_{\higgs\lepton}\feq{\higgs}{\momhig}-\feq{\lepton}{\momlep})$ 
it can add to or diminish an existing asymmetry. Finally, \eqn\eqref{eqn:quantum corrected Boltzmann 
equation mu by T Li result} describes washout due to inverse decays. In \sect\ref{NEQFTapproach}
we will see that the functional 
dependence on $f_{\lepton\higgs}=(1+\feq{\higgs}{\momhig}-\feq{\lepton}{\momlep})$ in the integrated 
collision terms is the same as that encountered in the {\CP}-violating parameter $\epsilon_i$ itself.

Considering the last two terms in \eqn\eqref{divlepCPscatt} we find for the scattering contributions:
\begin{align}
	\label{eqn:quantum corrected Boltzmann equation 2--2}
	&\frac{s\hubbleratem }{z}\frac{d Y_{L}}{dz}\Big|_{S,\frac{\mu_\lepton}{T}}=
	-4\frac{\mu_\lepton}{T}(1+c_{\higgs\lepton}) 
	\big[ \langle\gamma'^{\lepton\higgs}_{\bar{\lepton}\bar{\higgs}}\rangle 
	+ \langle \gamma^{\lepton\lepton}_{\bar{\lepton}\bar{\higgs}}\rangle\big] \kend
\end{align}
where we defined the `RIS subtracted reaction density' 
$\langle\gamma'^{\lepton\higgs}_{\bar{\lepton}\bar{\higgs}}\rangle$. If we adopt the amplitudes defined in \eqn\eqref{eqn:matrix element RIS subtracted} in the framework of RIS subtraction, it is given by 
\begin{align}
	& \langle\gamma'^{\lepton\higgs}_{\bar{\lepton}\bar{\higgs}}\rangle\equiv\bigg<
	\big[{1-\sum_i\frac{\pi (1+\epsilon_i^2)
	\EffAmplitude{}{\majneutrino_i}}{4\EffAmplitude{}{\lepton\higgs\leftrightarrow 
	\bar{\lepton}\bar{\higgs}}M_i\width{i}}\delta (s-M_i^2)}\big]
	\gamma^{\lepton\higgs}_{\bar{\lepton}\bar{\higgs}}\bigg> \dend\nonumber
\end{align}
Note that the contribution proportional to $\epsilon_i^2$ is of higher order in $h$.
Furthermore, we get for the extra term:
\begin{align}
	\label{eqn:quantum corrected Boltzmann equation 2--2 extra term}
	& \frac{s\hubbleratem }{z}\frac{d Y_{L}}{dz}\Big|_{S,\text{extra}}\approx {}-2\sum_i \big<\qstatfeqf{\majneutrino_i}{}\epsilon_i \gamma^D_{\majneutrino_i}\big>\\
	& +2\frac{\mu_\lepton}{T}\sum_i \big<\epsilon_i \big(\feq{\lepton}{}-c_{\higgs\lepton}
	\feq{\higgs}{}+\feq{\lepton}{}-c_{\higgs\lepton}\feq{\higgs}{}\big)\gamma^W_{\majneutrino_i}\big>\nonumber\dend
\end{align}
Here we used 
\begin{align}
	\label{eqn:reduction of RIS collision terms}
	\Big<\frac{(16\pi)^2\momlep\mommaj\EffAmplitude{}{\majneutrino_i}\delta 
	(s-M_i^2)}{g_w^2 M_i^2\EffAmplitude{}{\lepton\higgs\leftrightarrow\bar\lepton\bar\higgs}}
	\gamma^{\lepton\higgs}_{\bar{\lepton}\bar{\higgs}}\Big> = 
	\langle\gamma^{W}_{\majneutrino_i}\rangle\dend
\end{align}
We have written \eqn\eqref{eqn:quantum corrected Boltzmann equation 2--2 extra term} schematically in order to show how it compares to other washout terms.
Note that the extra terms indicate that there will be a slight difference between the equations 
obtained in the 2PI approach and those obtained with RIS subtraction at finite temperature. 
Comparing \eqns\eqref{eqn:quantum corrected Boltzmann equation 2--2 extra term} and 
\eqref{eqn:quantum corrected Boltzmann equation 2--2} we see that the first term in the former
equation will cancel the latter contribution in thermal equilibrium 
($\f{\majneutrino_i}{}=\feq{\majneutrino_i}{}$) if the decay contributions are summed up. 
The second term in \eqn\eqref{eqn:quantum corrected Boltzmann equation 2--2 extra term} is 
due to quantum statistics. Since it is 
proportional to $\epsilon_i\mu/T$ it can be large only if $\epsilon_i$ is large (as in 
the case of resonant leptogenesis). Anticipating our knowledge about the structure obtained 
within NEQFT, we will ignore the extra term in what follows.

At the time being, everything is still exact with respect to deviations of $\f{\majneutrino_i}{}$ 
from equilibrium. This distribution is necessarily distorted due to the fact that it is
subject to conflicting equilibrium conditions corresponding to the decay into particles 
and antiparticles, by the effects of the expansion and, possibly, due to non-equilibrium 
initial conditions. In order to obtain the full momentum-dependent distribution function 
we would have to solve the corresponding full kinetic equations however 
\cite{Basbol:2007,Garayoa:2009my,HahnWoernle:2009qn,Garny:2009rv,Garny:2009qn}.

To proceed we shall as usual assume that the deviation of the Majorana neutrinos from 
equilibrium is small. This allows us to neglect the 
${\Delta\f{\majneutrino_i}{}\frac{\mu_\lepton}{T}}$ contribution 
\eqref{eqn:quantum corrected Boltzmann equation mu by T times Delta f Li result} and to 
replace $\f{\majneutrino_i}{q}\to \feq{\majneutrino_i}{q}$ in 
\eqn\eqref{eqn:quantum corrected Boltzmann equation mu by T Li result}. The extra terms cancel 
at this level of approximation up to the quantum-statistical term.
In order to bring the remaining source-term 
\eqn\eqref{eqn:quantum corrected Boltzmann equation Delta f Li result} into the
conventional form, we need to assume that the non-equilibrium distribution of the Majorana 
neutrino is proportional to its equilibrium value (with momentum independent 
prefactor)\footnote{This amounts to the assumption that its shape can, in terms of its 
quantitative effect on the integrated collision terms, effectively be captured by a 
Maxwell-Boltzmann distribution with (small) `pseudo-chemical potential' \cite{Bernstein:1988}. 
Strictly speaking, it implies that we need to revert to a classical distribution function 
for the Majorana neutrinos.}
\begin{equation*}
	\f{\majneutrino_i}{\mommaj} \equiv \frac{\n{\majneutrino_i}}{\nequ{\majneutrino_i}}\feq{\majneutrino_i}{\mommaj}\dend
\end{equation*}
With this approximation we can write
\begin{equation}
	\label{eqn:deviation from equilibrium distribution function abundance relation}
	\biggl\langle X\frac{\Delta\f{\majneutrino_i}{}}{\feq{\majneutrino_i}{}}\gamma\biggr\rangle
	= \bigg(\frac{\abundance{\majneutrino_i}}{\abundanceeq{\majneutrino_i}}-1\bigg)\langle X\gamma \rangle\dend
\end{equation}
The total contribution to the evolution equations for the lepton asymmetry is then given by
\begin{align}
	\label{eqn:quantum corrected Boltzmann equation all terms}
	\frac{d Y_{L}}{dz}=\sum_i\bigg(\frac{d Y_{L_i}}{dz}\Big|_{\Delta\f{\majneutrino_i}{}} 
	+\frac{d Y_{L_i}}{dz}\Big|_{\frac{\mu_\lepton}{T}}\bigg)+\frac{d Y_{L}}{dz}\Big|_{S,\frac{\mu_\lepton}{T}}
	\kend
\end{align}
i.e.~we obtain \eqn\eqref{RateEquationsa}. We see that, at this level of approximation, 
there are no contributions due to extra terms apart from those which cancel due to the 
RIS-sub\-tra\-cti\-on. Quantitative differences can arise if the deviation of the Majorana 
neutrinos from equilibrium is large or $\epsilon_i$ is of order $1$.
For the evolution of the Majorana neutrino we obtain with \app\ref{RateEquationsApp}, 
similar to \eqn\eqref{eqn:quantum corrected Boltzmann equation Li result}, the decay contributions
\begin{subequations}
	\label{eqn:quantum corrected Boltzmann equation Ni result}
	\noeqref{eqn:quantum corrected Boltzmann equation Delta f Ni result,eqn:quantum corrected Boltzmann equation mu by T times Delta f Ni result,eqn:quantum corrected Boltzmann equation mu by T Ni result}
	\begin{align}
		\label{eqn:quantum corrected Boltzmann equation Delta f Ni result}
		&\frac{s\hubbleratem }{z}\frac{d Y_{\majneutrino_i}}{dz}\Big|_{\Delta\f{\majneutrino_i}{}} 
		= -\bigg< \frac{\Delta\f{\majneutrino_i}{\mommaj}}{\feq{\majneutrino_i}{}}
		\gamma^D_{\majneutrino_i}\bigg>\kend\\
		\label{eqn:quantum corrected Boltzmann equation mu by T times Delta f Ni result}
		&\frac{s\hubbleratem }{z}\frac{d Y_{\majneutrino_i}}{dz} \Big|_{\Delta\f{\majneutrino_i}{}{\frac{\mu_\lepton}{T}}} 
		= -\frac{\mu_\lepton}{T}\big<\epsilon_i (c_{\higgs\lepton}\feq{\higgs}{\momhig}
		-\feq{\lepton}{\momlep}) \frac{\Delta\f{\majneutrino_i}{\mommaj}}{\feq{\majneutrino_i}{}}
		\gamma^D_{\majneutrino_i}\big>\kend\hphantom{aaa} \\
		\label{eqn:quantum corrected Boltzmann equation mu by T Ni result}
		&\frac{s\hubbleratem }{z}\frac{d Y_{\majneutrino_i}}{dz}\Big|_{\frac{\mu_\lepton}{T}}
		=+\frac{\mu_\lepton}{T}(1+c_{\higgs\lepton})\bigg<\epsilon_i 
		\frac{(1-\f{\majneutrino_i}{})}{(1-\feq{\majneutrino_i}{})}\gamma^W_{\majneutrino_i}\bigg>\kend
	\end{align}
\end{subequations}
and for the extra term in \eqn\eqref{divlepCPpred}:
\begin{align}
	\label{eqn:quantum corrected Boltzmann equation decay extra term Ni result}
	 \frac{s\hubbleratem }{z} \frac{d Y_{\majneutrino_i}}{dz}& \Big|_{D,\text{extra}} = 
	-2\frac{\mu_\lepton}{T}(1+c_{\higgs\lepton})\bigg<\epsilon_i 
	\frac{(1-\f{\majneutrino_i}{})}{(1-\feq{\majneutrino_i}{})}\gamma^W_{\majneutrino_i}\bigg>\nonumber\\
	&-2\frac{\mu_\lepton}{T}\big<\epsilon_i (c_{\higgs\lepton}\feq{\higgs}{\momhig}-\feq{\lepton}{\momlep}) 
	\frac{\Delta\f{\majneutrino_i}{\mommaj}}{\feq{\majneutrino_i}{}}\gamma^D_{\majneutrino_i}\big>\dend
\end{align}
Neglecting again ${\Delta\f{\majneutrino_i}{}\frac{\mu_\lepton}{T}}$ and ${\epsilon_i\frac{\mu_\lepton}{T}}$ 
contributions we obtain
\begin{align}
	\label{eqn:quantum corrected Boltzmann equation all terms Ni}
	\frac{d Y_{\majneutrino_i}}{dz}=\frac{d Y_{\majneutrino_i}}{dz}\Big|_{\Delta\f{\majneutrino_i}{}} \kend
\end{align}
i.e.~\eqn\eqref{RateEquationsb}. If higher order contributions are taken into account, we get a 
difference between the conventional equations and those derived in the 2PI-formalism. Ignoring 
the contribution \eqref{eqn:quantum corrected Boltzmann equation mu by T times Delta f Ni result} 
and the second term in \eqn\eqref{eqn:quantum corrected Boltzmann equation decay extra term Ni result},
which are due to quantum statistics, we obtain a contribution
\begin{align}
	\mp\frac{Y_L}{2\,Y_\lepton^{eq}}\langle \epsilon_i\, \gamma^W_{\majneutrino_i}\rangle\kend
\end{align}
to ${d Y_{\majneutrino_i}}/{dz}$.
Here the upper sign applies if the extra terms are included and the lower sign if not. This can therefore 
result in the inclusion of this term with wrong sign even if quantum statistics are neglected, 
compare e.g.~\cite{Pilaftsis:2003gt}.\footnote{The origin of this difference is that no RIS subtraction 
alike is performed for the Boltzmann equation of the heavy Majorana neutrino.}

The reaction densities for decay, $\langle \epsilon_i\, \gamma^D_{\majneutrino_i}\rangle$, 
$\langle \gamma^D_{\majneutrino_i}\rangle$, $\langle \gamma^W_{\majneutrino_i}\rangle$, and 
scattering, $\langle\gamma'^{\lepton\higgs}_{\bar{\lepton}\bar{\higgs}}\rangle$, 
$\langle\gamma^{\lepton\lepton}_{\bar{\higgs}\bar{\higgs}}\rangle$, represent the hydrodynamical 
coefficients which govern the evolution of the number densities (abundances). We will compute them 
numerically once the additional medium dependence of the amplitudes (in particular the 
\CP-violating parameters) has been derived. In addition, it is useful to define a thermally 
averaged {\CP}-violating parameter as 
\begin{align*} 
\langle \epsilon_i \rangle \equiv \frac{\langle\epsilon_i \gamma^D_{\majneutrino_i}\rangle}{\langle 
\gamma^D_{\majneutrino_i}\rangle}\kend
\end{align*}
which equals $\epsilon_i$ if it is momentum independent, such as in the zero temperature case, 
but will differ once thermal effects are included. This quantity is meaningful for the comparison 
with conventional results because it takes into account that the deviation of the Majorana neutrino 
abundance from equilibrium, which appears in the source-term for the lepton abundance, is 
influenced by the ({\CP}-conserving) decay reaction density in the denominator. 

Inserting 
conventional vacuum amplitudes in \eqns\eqref{RateEquations1} with \eqns\eqref{eqn:reaction density decays} 
and \eqref{eqn:reaction density washout} and dropping quantum-statistical factors one obtains the 
conventional results for the reaction densities. 
For the readers convenience we quote them here. For the decay reaction density 
we obtain
\begin{align}
	\label{eqn:decay reaction density main text}
	\langle \gamma^W_{\majneutrino_i}\rangle = 
	\langle \gamma^D_{\majneutrino_i}\rangle \approx 
	\frac{g_\majneutrino}{2\pi^2} M_i^2\Gamma_i T K_1\left(\frac{M_i}{T}\right)
	\kend
\end{align}
and $\langle \epsilon_i\, \gamma^D_{\majneutrino_i}\rangle = 
\epsilon_i \langle \gamma^D_{\majneutrino_i}\rangle$, see \app\ref{Kinematics}.
For the two-body scattering the reaction density is given by 
\begin{align}
	\label{ReactDensCanon}
	\langle\gamma^{ab}_{ij}\rangle & \approx \frac{T}{64\pi^4}\int\limits_{s_{min}}^\infty ds 
	\sqrt{s}K_1\left(\frac{\sqrt{s}}{T}\right) \hat{\sigma}(s)\kend
\end{align}
where $\hat \sigma(s)$ is so-called reduced cross section:
\begin{align}
\label{eqn:reduced cross section definition main text}
\hat \sigma(s)\equiv \frac{1}{8\pi}\int\limits_0^{2\pi} 
	\frac{d\varphi_{ai}}{2\pi}
	\int\limits_{t^-}^{t^+} \frac{dt}{s}
	\meavsqudep{ab\leftrightarrow ij} \dend
\end{align}
For the $\lepton\lepton\leftrightarrow\bar\higgs\bar\higgs$ process it reads 
\begin{align}
\label{sigmaLLHH}
\hat\sigma&=\frac{1}{2\pi}\sum \Re(h^\dagger h)^2_{ij}\sqrt{a_i a_j}\\
& \times\biggl\{
\frac{1}{a_i-a_j}\ln\left(\frac{a_i(x+a_j)}{a_j(x+a_i)}\right)\nonumber\\
&+\frac12\frac{1}{x+a_i+a_j}\ln\left(\frac{(x+a_i)(x+a_j)}{a_ia_j}\right)\biggr\}\kend\nonumber
\end{align}
where we have replaced $s$ by $x\equiv s/M_1^2$ and introduced dimensionless 
quantities $a_i\equiv M_i^2/M_1^2$ and $c_i\equiv \Gamma_i/M_i=(h^\dagger h)_{ii}/8\pi$. 
The case $i=j$ is included in this expression in the limiting sense $a_j\rightarrow a_i$.
Note that \eqn\eqref{sigmaLLHH} only contains the real part of $(h^\dagger h)^2$. The 
contribution of the imaginary part vanishes because $\Im (h^\dagger h)^2_{ij}$ is 
antisymmetric with respect to $i\leftrightarrow j$ whereas the sum in the curly brackets 
is symmetric under this transformation. The integration of \eqn\eqref{LHLH} yields for the 
reduced `RIS subtracted cross section' of the $\lepton\higgs\leftrightarrow\bar\lepton\bar\higgs$ process: 
\begin{align}
\label{sigmaLHLH}
\hat\sigma'&=\frac{1}{4\pi x}\sum\Re (h^\dagger h)^2_{ij}\sqrt{a_i a_j}\\
&\times\left\{x^2
\frac{(x-a_i)(x-a_j)+(1-2\delta_{ij})a_i a_j c_i c_j }{\left[(x-a_i)^2+(a_i c_i)^2\right]
\left[(x-a_j)^2+(a_jc_j)^2\right]}
\right.\nonumber\\
&+2\frac{x+a_i}{a_j-a_i}\ln\left(\frac{x+a_i}{a_i}\right)+
2\frac{x+a_j}{a_i-a_j}\ln\left(\frac{x+a_j}{a_j}\right)\nonumber\\
&+\frac{x-a_i}{(x-a_i)^2+(a_ic_i)^2}\left[x-(x+a_j)\ln\left(\frac{x+a_j}{a_j} 
\right)\right]\nonumber\\
&+\left.\frac{x-a_j}{(x-a_j)^2+(a_jc_j)^2}\left[x-(x+a_i)\ln\left(\frac{x+a_i}{a_i} 
\right)\right]\right\}\nonumber\dend
\end{align}
The reduced `cross section' \eqref{sigmaLHLH} is \textit{negative}\footnote{Note that $\hat\sigma'$ is not a physical cross section but denotes the contribution to the reaction density arising
from the difference of the full and the RIS term. We stress that all physical rates are manifestly positive, e.g. the washout term, to which $\hat\sigma'$ yields a sub-leading correction that is relatively suppressed by
Yukawa couplings. See also \cite{Pilaftsis:2003gt}.}
in the vicinity of the mass shells, $x\approx a_i$. This is due to the 
$-(M_i\Gamma_i)^2$ term in the numerator of the RIS subtracted propagator 
\eqref{RISpropagator}. Note that because we have not approximated this term
by the Dirac-delta the structure of \eqn\eqref{sigmaLHLH} is slightly different 
from the one usually used in the literature \cite{Pilaftsis:2003gt}.

\section{\label{NEQFTapproach}Non-equilibrium QFT approach}
In this section we briefly review the description of leptogenesis within non-equilibrium quantum
field theory \cite{Schwinger:1960qe,Keldysh:1964ud,Calzetta:1986cq,Berges:2004yj}. This framework has been shown recently to be suitable for the derivation of quantum dynamic equations for the 
lepton asymmetry within a first-principle approach, and to incorporate medium, off-shell, coherence and 
possibly further quantum effects in a self-consistent 
way~\cite{Buchmuller:2000nd,Lindner:2005kv,DeSimone:2007rw,Lindner:2007am,Anisimov:2008dz,Garny:2009rv,Garny:2009qn,Anisimov:2010aq,Garny:2010nj,Beneke:2010wd,Garny:2010nz,Beneke:2010dz,Garbrecht:2010sz,Gagnon:2010kt,Anisimov:2010dk,Drewes:2010pf}.
We continue these efforts by deriving consistent quantum corrected Boltzmann 
equations that describe the generation and washout of the lepton asymmetry and include the (inverse) decay 
as well as scattering processes mediated by Majorana neutrinos. 

\subsection{\label{CTPpropagators}CTP and propagators}

\begin{figure}[h!]
	\centering
	\includegraphics{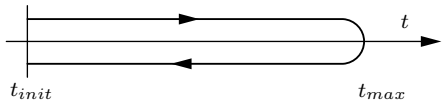}
	\caption{\label{fig:CTP}Closed time path.}
\end{figure}

The lepton asymmetry is given by the
$\mu=0$-com\-po\-ne\-nt of the expectation value of the lepton-current operator:
\begin{equation*}
 j^\mu_L(x) = \bigl\langle { \textstyle \sum\limits_{\alpha,a} } \, 
 \bar\lepton^a_\alpha(x) \gamma^\mu \lepton^a_\alpha(x) \bigr\rangle \dend
\end{equation*}
It can be expressed in terms of the leptonic two-point function. We define the 
two-point functions for the Higgs, lepton and Majorana fields with time arguments 
attached to the closed time path (CTP) shown in \fig\ref{fig:CTP} by
\begin{subequations}
	\label{TwoPointFunctions}
	\noeqref{TwoPointFunctionHiggs,TwoPointFunctionLept,TwoPointFunctionMaj}
	\begin{align}
		\label{TwoPointFunctionHiggs}
		\pHiggs{}{ab}(x,y) & = \langle T_\CTP \higgs^a(x) \higgs^{*b}(y) \rangle \kend \\
		\label{TwoPointFunctionLept}
		\pLept{\alpha\beta}{ab}(x,y) & = \langle T_\CTP \lepton_{\alpha}^a(x) \bar \lepton_{\beta}^b(y) \rangle \kend \\
		\label{TwoPointFunctionMaj}
		\pMaj{ij}{}(x,y) & = \langle T_\CTP \majneutrino_{i}(x) \bar\majneutrino_{j}(y) \rangle \kend
	\end{align}
\end{subequations}
where the sub- and superscripts refer to $SU(2)_L$ and flavor indices and $T_\CTP$ denotes 
time-ordering with respect to the CTP. 
We will frequently use matrix notation for the flavor indices,
where e.g. $\pMajmat{}{}$ denotes the flavor-matrix $\pMaj{ij}{}$, etc.
Using the definition \eqref{TwoPointFunctionLept} we find for the lepton-current:
\begin{equation*}
	j_L^\mu(x) = - \sum_{\alpha,a} \tr \left[\gamma_\mu \pLept{\alpha\alpha}{a a}(x,x) \right]\dend
\end{equation*}
Two-point functions $G(x,y)$ (where $G$ stands for $\pHiggs{}{}$, $\pLept{}{}$ or $\pMaj{}{}$) defined on the CTP can be decomposed into a \emph{spectral function} $G_\rho$
and \emph{statistical propagator} $G_F$:
\begin{equation}
	\label{FRhoDecomposition}
	G(x,y) = G_F(x,y) - \frac{i}{2} \sgn_\CTP(x^0-y^0) G_\rho(x,y)\dend
\end{equation}
The signum function is either $+1$ or $-1$ depending on whether $x^0$ or $y^0$ occur `later' on the contour $\CTP$. $G_F$ and $G_\rho$ encode information on the state and the spectrum of the system, respectively. For example, for the leptons they are given by
\begin{align*}
		\pLept{\alpha\beta}{ab\,F}(x,y) & = \frac12 \langle \left[ \lepton_{\alpha}^a(x), \bar \lepton_{\beta}^b(y) \right]_- \rangle \kend \\
		\pLept{\alpha\beta}{ab\,\rho}(x,y) & = i \langle \left[ \lepton_{\alpha}^a(x), \bar \lepton_{\beta}^b(y) \right]_+ \rangle \kend
\end{align*}
where $[.,.]_\pm$ denote (anti-)commutators. Statistical and spectral functions of Majorana neutrino and Higgs can be expressed similarly, with $+$ and $-$ exchanged for bosons.
Although there are only two independent two-point functions for each species, it is convenient to introduce additional combinations of them, namely the \emph{Wightman} functions
\begin{align}
\label{Wightman}
G_\gtrless(x,y) & = G_F(x,y) \mp \frac{i}{2}G_\rho(x,y) \kend 
\end{align}
as well as \emph{retarded} and \emph{advanced} functions,
\begin{subequations}
	\label{RetAdvDef}
	\noeqref{GR,GA}
	\begin{align}
		\label{GR}
		G_R(x,y) & = \Theta(x^0-y^0)G_\rho(x,y) \kend \\
		\label{GA}
		G_A(x,y) & = - \Theta(y^0-x^0)G_\rho(x,y) \dend
	\end{align}
\end{subequations}
From the above definitions one can see that the difference of the retarded and advanced propagators gives the spectral one, whereas the sum yields the \textit{hermitian} propagator $G_h(x,y)$:
\begin{subequations}
\label{hermDef}
\noeqref{GrhoDef,GhDef}
\begin{align}
 \label{GrhoDef}
 G_R(x,y)-G_A(x,y)& = G_\rho(x,y) \kend \\
 \label{GhDef}
 G_R(x,y)+G_A(x,y)& = 2G_h(x,y) \dend
\end{align}
\end{subequations}
Finally, we will also need the {\CP} conjugated propagators on the CTP:
\begin{subequations}
\label{CPconjugates}
\noeqref{HiggsCPconj,LeptCPconj,MajCPconj}
\begin{align}
 \label{HiggsCPconj}
 \pHiggscp{}{ab}(x,y) & \equiv \pHiggs{}{ba}(\bar y,\bar x) \kend \\
 \label{LeptCPconj}
 \pLeptcp{\alpha\beta}{ab}(x,y) \equiv & (CP) \pLept{\beta\alpha}{ba}(\bar y,\bar x)^T (CP)^{-1} \kend \\
 \label{MajCPconj}
 \pMajcp{ij}{}(x,y) & \equiv (CP) \pMaj{ji}{}(\bar y,\bar x)^T (CP)^{-1} \dend
\end{align}
\end{subequations}
Here $\bar x = (x^0,-\vec{x})$, $C=i\gamma^2 \gamma^0$ and $P=\gamma^0$ are the charge conjugation
and parity matrices, respectively, and the transposition refers to spinor indices. 
{\CP} conjugated statistical and spectral functions immediately follow from the above definition
by inserting the decomposition \eqref{FRhoDecomposition}.

\subsection{Kadanoff-Baym equations for leptons}

\begin{figure}[h!]
	\includegraphics{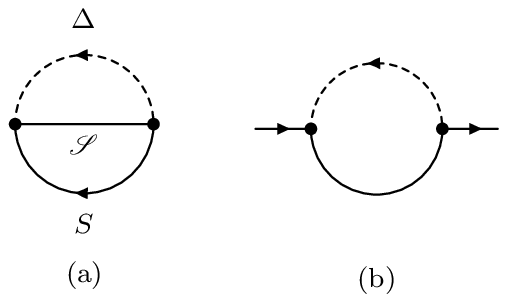}
	\\[5mm]
	\includegraphics{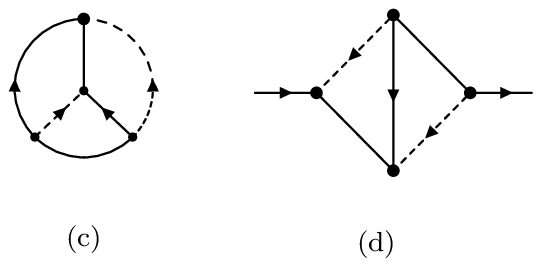}
	\caption{\label{fig:2PI contributions}Two- and three-loop contributions to the 2PI 
	effective action and the corresponding contributions to the lepton self-energy.
	Note that the propagator lines used here denote full resummed propagators in contrast to
	those employed in the previous Feynman graphs. The contributions (a) and (c) to the 2PI
	effective action are known as `setting-sun' and `mercedes' diagrams respectively.}
\end{figure}

The time-evolution of the two-point functions is described self-consistently by the Kadanoff-Baym 
(KB) equations. These equations can be obtained from a variational principle using the so-called 2PI 
effective action~\cite{PhysRevD.10.2428}. The resulting equations of motion have the form of 
Schwinger-Dyson equations for the non-equilibrium propagators formulated on the CTP:
\begin{align}
	\label{SDCTP}
	\pLeptmat{-1}{}(x,y) & = \pLeptmat{-1}{0}(x,y) - \sLeptmat{}{}(x,y)\dend
\end{align}
Here $\pLeptmat{-1}{}(x,y)$ is the inverse of the \emph{full} lepton propagator in coordinate 
space, and $\pLeptmat{-1}{0}(x,y)$ is the inverse of the \emph{free} lepton propagator,
\begin{equation*}
	\pLeptmat{-1}{0}(x,y) = \delta_{ab}\delta^{\alpha\beta} \slashed{\partial}_x \delta_\CTP(x-y) P_L \dend
\end{equation*}
The information about the interaction processes is encoded in the self-energies $\sLept{}{}$. 
They can be obtained by cutting one line of the 2PI contributions to the effective actions. 
The two- and three-loop contributions are presented in \fig\ref{fig:2PI contributions}.

The KB equations can be obtained by convoluting the Schwinger-Dyson equation with the full propagator, which yields:
\begin{align*}
	 \slashed{\partial}_x \pLept{\alpha\beta}{}(x,y) & = \\
	 \delta^{\alpha\beta}&\delta_\CTP(x-y) + \int_\CTP d^4z 
	 \, \sLept{\alpha\gamma}{}(x,z) \pLept{\gamma\beta}{}(z,y)\dend\nonumber
\end{align*}
Here $\int_\CTP d^4z=\int_\CTP dz^0 \int d^3z$.
After decomposing the resulting equation into statistical and spectral components, one obtains:
\begin{subequations}
	\label{KBlepton}
	\noeqref{KBSFlept,KBSrholept}
	\begin{align}
	 \label{KBSFlept}
		i \slashed{\partial}_x \pLept{\alpha\beta}{F}(x,y) & = \int_0^{x^0} \!\! d^4z \, \sLept{\alpha\gamma}{\rho}(x,z) \pLept{\gamma\beta}{F}(z,y) \nonumber\\
& - \int_0^{y^0} \!\! d^4z \, \sLept{\alpha\gamma}{F}(x,z) \pLept{\gamma\beta}{\rho}(z,y) \kend \\
	 \label{KBSrholept}
		i \slashed{\partial}_x \pLept{\alpha\beta}{\rho}(x,y) & = \int_{y^0}^{x^0} \!\! d^4z \, \sLept{\alpha\gamma}{\rho}(x,z) \pLept{\gamma\beta}{\rho}(z,y) \dend
	\end{align}
\end{subequations}
The equations for Majorana and Higgs propagators have a similar structure, with the Klein-Gordon 
instead of the Dirac operator for the latter. The Schwinger-Dyson equations \eqref{SDCTP} and 
the corresponding Kadanoff-Baym equations \eqref{KBlepton} are formally very similar to the 
Schwinger-Dyson equation in vacuum. However, out of equilibrium the propagators depend not only 
on the relative coordinate $s=x-y$, but also on the central coordinate $X=(x+y)/2$, which makes 
their solution much more involved. In contrast to the Schwinger-Dyson equation in vacuum, 
the KB equations determine the spectral properties of the system including medium corrections, 
as well as the non-equilibrium dynamics of the statistical propagator self-consistently. Since 
the latter represents the quantum field theoretical generalization of the classical particle 
distribution functions, KB equations can be seen as the quantum field theoretical generalizations 
of Boltzmann equations.

As pointed out above, an equation of motion for the lepton asymmetry can be derived by considering 
the divergence of the lepton-current ${\cal D}_\mu j_L^\mu(x)$. Using the KB equations 
\eqref{KBlepton} one obtains\footnote{We assume here that in FRW space-time the effects of the Universe
expansion can be captured, to the required accuracy, by introducing the invariant
integration measure $\dg^4z\equiv\sqrt{-g}d^4z$ and using
the covariant derivative ${\cal D}_\mu$. As has been demonstrated in \cite{Hohenegger:2008zk}, this is the case
for scalar fields. A manifestly covariant generalization of center and relative coordinates $X$
and $s$ to curved space-time can be found in \cite{PhysRevD.35.3796}.}:
\begin{align}
	\label{MasterEquation}
	{\cal D}_\mu j_L^\mu(x) & = -g_w
	\lim_{y\rightarrow x}({\cal D}_x^\mu + {\cal D}_y^\mu ) 
	\tr\left[\gamma_\mu \pLept{\alpha\alpha}{}(x,y) \right] \nonumber\\
	& = g_w\, i\int_0^{x^0} \dg^4z \, \tr\bigl[ \sLept{\alpha\beta}{\rho}(x,z) \pLept{\beta\alpha}{F}(z,x) \nonumber\\
	& - \sLept{\alpha\beta}{F}(x,z) \pLept{\beta\alpha}{\rho}(z,x) - \pLept{\alpha\beta}{\rho}(x,z) \sLept{\beta\alpha}{F}(z,x) \nonumber\\
	& + \pLept{\alpha\beta}{F}(x,z) \sLept{\beta\alpha}{\rho}(z,x) \bigr] \dend
\end{align}
Here summation over repeated indices is implicitly assumed. The two equations above represent 
the quantum generalization of the Boltzmann equation for the lepton asymmetry. Thus, they may 
be considered as the master equations for a quantum field theoretical treatment of leptogenesis 
\cite{Buchmuller:2000nd,DeSimone:2007rw}.

The dependence of the two-point functions on the relative coordinate $s$ is characterized 
by the hard scales like the Majorana neutrino mass $M_1$ or the temperature $T$ of the 
surrounding plasma. In contrast to that, the variation with the central coordinate $X$ is 
given by the macroscopic time-evolution of the system, e.g. the Hubble rate $H$ or the 
Majorana decay rate $\Gamma$. Therefore, it is possible to perform an expansion in slow 
relative to fast time-scales, i.e. in powers of e.g. $\Gamma/M_1$ or $H/T$. Technically, 
this can be realized by a so-called gradient or derivative expansion with respect to $X$, 
and a Fourier transformation with respect to $s$, known as Wigner transformation, see 
\app\ref{SelfEnergy} for more details. Then, to leading order in the gradients, 
the evolution equation \eqref{MasterEquation} for the lepton asymmetry becomes Markovian, 
and after some straightforward algebra, can be written as
\begin{align}
	\label{MasterEquationWigner}
	{\cal D}_\mu j^\mu_\lepton(t) & = g_w \int\limits_0^\infty\frac{dp^0}{2\pi} \int \frac{d^3\momlep}{(2\pi)^3} \\
	& \times \tr\bigl\{ \bigl[ \sLept{\alpha\beta}{<}(t,\momlep) \pLept{\beta\alpha}{>}(t,\momlep) 
	- \sLept{\alpha\beta}{>}(t,\momlep) \pLept{\beta\alpha}{<}(t,\momlep)\bigr]\nonumber\\
	& - \bigl[\sLeptcp{\beta\alpha}{<}(t,\momlep) \pLeptcp{\alpha\beta}{>}(t,\momlep) -
	\sLeptcp{\beta\alpha}{>}(t,\momlep) \pLeptcp{\alpha\beta}{<}(t,\momlep) \bigr]\bigr\}\dend\nonumber
\end{align}
Note that it is possible to investigate higher orders in the derivative expansion 
systematically \cite{Garny:2010nz}. In \eqn\eqref{MasterEquationWigner}
all two-point functions are evaluated in Wigner space, where $\momlep$
is the \textit{physical} momentum \cite{Hohenegger:2008zk} that corresponds to $s$. 
For a spatially homogeneous system (like FRW) the two-point functions depend only on 
the time coordinate $t=X^0$, and on the momentum $\momlep$, because of spatial 
translational invariance. Strictly speaking, this is true only in the rest frame 
of the medium (comoving frame). In a general frame the two-point functions depend 
on $X\cdot u$, where $u_\mu$ is the four-velocity of the medium. The latter satisfies 
the normalization condition $u^\mu u_\mu=1$, and is given by $u=(1,0,0,0)$ in the 
medium rest frame. 

In order to allow for a physical interpretation of \eqn\eqref{MasterEquationWigner} 
we have written it such that the integration is over positive frequencies only, and 
expressed the lepton propagator and self-energy in terms of the Wigner transformed 
Wightman functions \eqn\eqref{Wightman}. In thermal equilibrium, the Wightman functions 
depend only on the momentum $\momlep$ and satisfy the Kubo-Martin-Schwinger (KMS) 
relation $G_>^{th}(\momlep) = \pm e^{\momlep\cdot u/T} G_<^{th}(\momlep)$ for 
fermions/bosons, respectively. When inserting the KMS relations for propagators and 
self-energies into \eqn\eqref{MasterEquationWigner}, one immediately finds that the 
divergence of the lepton-current vanishes in thermal equilibrium as it should 
(see also \cite{Beneke:2010wd}). In other words, the quantum equation for the lepton 
asymmetry is in accordance with the third Sakharov condition. We emphasize that it is 
\emph{not} necessary to apply RIS subtraction to obtain this 
result within the CTP approach \cite{Garny:2009rv,Garny:2009qn,Beneke:2010wd}.

The four terms on the right-hand side of \eqn\eqref{MasterEquationWigner} may be 
interpreted as gain and loss terms of leptons and anti-leptons respectively 
\cite{Garny:2009rv,Garny:2009qn}. In particular, one may define generalized 
lepton distribution functions $f_\lepton^{\alpha\beta}(t,\momlep)$ via the 
so-called Kadanoff-Baym ansatz
\begin{equation}
	\label{KBLepton}
	\pLept{\alpha\beta}{>} = (1-f_\lepton^{\alpha\beta})\pLept{\alpha\beta}{\rho}\kend\quad
	\pLept{\alpha\beta}{<} = -f_\lepton^{\alpha\beta}\pLept{\alpha\beta}{\rho} \dend
\end{equation}
Thus the contribution on the right-hand side of \eqn\eqref{MasterEquationWigner} that 
contains $\pLept{\beta\alpha}{<}$ corresponds to the lepton loss term, while the contribution 
proportional to $\pLept{\beta\alpha}{>}$ represents the lepton gain term. Analogous 
definitions relate the {\CP} conjugate propagators with the anti-lepton distribution.
Note that the KMS relations ensure that in equilibrium $\f{}{}$ approaches the Fermi-Dirac 
distribution $f_\lepton^{\alpha\beta}\to \delta^{\alpha\beta}\feq{FD}{}$. The flavor 
off-diagonal components encode coherent flavor correlations \cite{Beneke:2010dz}. In the 
unflavored regime considered here $f_\lepton^{\alpha\beta}=
\delta^{\alpha\beta}f_\lepton$ and $\pLept{\alpha\beta}{}=\delta^{\alpha\beta}\pLept{}{}$. 
In the quasiparticle (QP) approximation, the spectral function is given by
\begin{align}
	\label{leptonQPspectralfunc}
	\pLept{\alpha\beta}{\rho}(t,\momlep) & = (2\pi)\,\sign(p^0)\, \delta(p^2-m_\lepton^2)\,
	\delta^{\alpha\beta}\,P_L\slashed{\momlep}P_R \nonumber\\
	&\equiv \pLeptsc{}{\rho}\,\delta^{\alpha \beta}\,P_L \slashed{\momlep}P_R \kend
\end{align}
where we assume that leptons obey conventional dispersion relation and $m_\lepton$ 
is the effective thermal mass. These assumptions might be modified in the presence of 
a medium \cite{Weldon:1989pa158,Giudice:2003jh,Kiessig:2010jopv259}.

Due to the presence of the Dirac-delta-function in \eqn\eqref{leptonQPspectralfunc} 
the integration over $\momp^0$ in \eqn\eqref{MasterEquationWigner} is trivial and leaves 
only the integration over spatial momenta of on-shell leptons.
Therefore the right-hand side of \eqn\eqref{MasterEquationWigner} can be interpreted as 
a difference of two (integrated) Boltzmann-like equations -- one for the particles and 
one for the antiparticles \cite{Garny:2009qn}. According to the physical interpretation 
of \eqn\eqref{MasterEquationWigner} in terms of gain and loss terms, the Wightman components 
of the lepton self-energy and of its {\CP} conjugate are the analogs of the collision integrals. 
Since we limit our analysis to the unflavored regime, it is convenient to perform the 
 summation over the flavor indices: $\delta^{\alpha\beta}\sLept{\alpha\beta}{}
=\sLept{\alpha\alpha}{}\equiv \sLept{}{}$. Then the one-loop contribution, see 
\fig\ref{fig:2PI contributions} (b), takes the form:
\begin{align}
	\label{Sigma1WT}
	\sLept{(1)}{\gtrless}(t,\momlep) 
	= & -\int \lorentzdd{\momhig}\lorentzdd{\mommaj}(2\pi)^4 
	\deltafour{\momlep+\momhig-\mommaj}\nonumber\\
	& \times 
	(\yu^\dagger\yu)_{ji} \, P_R \pMaj{ij}{\gtrless}(t,\mommaj) P_L \pHiggs{}{\lessgtr}(t,\momhig)\kend
\end{align}
where $\lorentzdd{\mommaj}\equiv {d^4\mommaj}/{(2\pi)^4}$. The explicit expression for 
the two-loop contribution is rather lengthy and it is convenient to split it into three
distinct terms:
\begin{align}
	\label{SE3pMT}
	\sLept{(2)}{\gtrless} = \sLept{(2.1)}{\gtrless} + 
	\sLept{(2.2)}{\gtrless} + \sLept{(2.3)}{\gtrless} \dend
\end{align}
The first term on the right-hand side reads
\begin{align}
	\label{SEDMT}
	&\sLept{(2.1)}{\gtrless}(t,\momlep) = \int \lorentzdd{\mommaj} \lorentzdd{\momhig}\
	(2\pi)^4\delta(\momlep+\momhig-\mommaj) \\
	&\times\big[ (\yu^\dagger \yu)_{in}(\yu^\dagger \yu)_{jm}
	\varLambda_{mn}(t,q,k)P_L C \pMaj{ij}{\gtrless}(t,\mommaj) 
	P_L\pHiggs{}{\lessgtr}(t,\momhig)\nonumber\\
	&+ (\yu^\dagger \yu)_{ni} (\yu^\dagger \yu)_{mj}
	P_R \pMaj{ji}{\gtrless}(t,\mommaj) C P_R V_{nm}(t,\mommaj,\momlep)
	\pHiggs{}{\lessgtr}(t,\momhig)\big]\nonumber \kend
\end{align}
where we have introduced two functions containing loop corrections:
\begin{align}
	\label{LC1MT}
	&\varLambda_{mn}(t,\mommaj,\momhig) \equiv \int \lorentzdd{k_1} \lorentzdd{k_2} \lorentzdd{k_3}\nonumber\\
	& \times (2\pi)^4\delta(\mommaj+k_1+k_2)\ (2\pi)^4 \delta(\momhig+k_2-k_3)\nonumber\\
	& \times \bigl[ P_R \pMaj{mn}{R}(t,-k_3) C P_R \pLept{T}{F}(t,k_2)\pHiggs{}{A}(t,k_1)\nonumber\\
	& +\ P_R \pMaj{mn}{F}(t,-k_3) C P_R \pLept{T}{R}(t,k_2) \pHiggs{}{A}(t,k_1)\nonumber\\
	& +\ P_R \pMaj{mn}{R}(t,-k_3) C P_R \pLept{T}{A}(t,k_2) \pHiggs{}{F}(t,k_1) \bigr]\kend
\end{align}
and $V_{nm}(t,q,k)\equiv P\, \varLambda^\dagger_{nm}(t,q,k)\, P$ to shorten the notation. 
Comparing \eqns\eqref{Sigma1WT} and \eqref{SEDMT} we see that they have a very similar structure. 
First, the integration is over momenta of the Higgs and Majorana neutrino and the delta-function 
contains the same combination of the momenta. Second, both self-energies include one Wightman 
propagator of the Higgs field and one Wightman propagator of the Majorana field. Upon the use 
of the Kadanoff-Baym ansatz the Wightman propagators can be interpreted as cut-propagators 
which describe on-shell particles created from or absorbed by the plasma \cite{Carrington:2004tm}.
On the other hand, the retarded and advanced propagators can be associated with the off-shell 
intermediate states. We therefore conclude that \eqns\eqref{Sigma1WT} and \eqref{SEDMT} describe 
(inverse) decays of the heavy neutrino into a lepton-Higgs pair. 

The second term on the right-hand side of \eqn\eqref{SE3pMT} contains two Wightman propagators 
of the Higgs field and one Wightman propagator of the lepton field. The Majorana propagator
appears only in the intermediate state: 
\begin{align}
	\label{SeScaMT}
	&\sLept{(2.2)}{\gtrless}(t,\momlep) = \int \lorentzdd{\momlep_2} 
	\lorentzdd{\momhig_1} \lorentzdd{\momhig_2}(2\pi)^4\delta(\momlep+\momhig_1-\momlep_2-\momhig_2)\nonumber\\
	&\times (\yu^\dagger \yu)_{ni}(\yu^\dagger \yu)_{mj}
	\bigl[P_R\pMaj{ij}{R}(t,\momlep_2+\momhig_2) CP_R\pLept{T}{\lessgtr}(t,-\momlep_2)P_L\nonumber\\
	& \times C\pMaj{mn}{A}(t,\momlep_2-\momhig_1)P_L \pHiggs{}{\lessgtr}(t,\momhig_1)
	\pHiggs{}{\lessgtr}(t,-\momhig_2)\bigr]\dend
\end{align}
We therefore conclude that this term describes lepton number violating scattering processes 
mediated by the heavy neutrino.
Finally the last term in \eqn\eqref{SE3pMT} contains two Wightman propagators of 
the Majorana field and one of the lepton field, whereas the Higgs field is in 
the intermediate state: 
\begin{align}
	\label{SeLCMT}
	&\sLept{(2.3)}{\gtrless}(t,\momlep) = \int d\Pi_{\momlep_2} d\Pi_{\momhig_1} 
	d\Pi_{\momhig_2}(2\pi)^4\delta(\momlep+\mommaj_1-\momlep_2-\mommaj_2)\nonumber\\
	&\times (\yu^\dagger \yu)_{ij} (\yu^\dagger \yu)_{lk}
	\bigl[P_R\pMaj{jk}{\gtrless}(t,-\mommaj_1) CP_R
	\pLept{T}{\gtrless}(t,\momlep_2)P_LC \nonumber\\
	&\times \pMaj{li}{\gtrless}(t,\mommaj_2)P_L 
	\pHiggs{}{A}(t,-\mommaj_2-\momlep_2)\pHiggs{}{R}(t,\mommaj_1-\momlep_2)\bigr]\dend
\end{align}
Therefore it can be identified with the Higgs mediated scattering processes. These 
conserve lepton number and do not contribute to generation of the lepton asymmetry.

The {\CP} conjugate of the Wigner transforms can be obtained using \eqn\eqref{CPconjugates}. 
In practice this amounts to replacing the propagators by their {\CP} conjugate and the couplings 
by their complex conjugate in the above expressions. For instance for the {\CP} conjugate of the
one-loop self-energy we find:
\begin{align}
\label{Sigma1WTCP}
	\sLeptcp{(1)}{\gtrless}(t,\momlep) = & - \int 
	\lorentzdd{\momhig}\lorentzdd{\mommaj}(2\pi)^4 \deltafour{\momhig+\momlep-\mommaj}\nonumber\\
	& \times (\yu^\dagger \yu)^*_{ji}\, P_R \pMajcp{ij}{\gtrless}(t,\mommaj) 
	P_L \pHiggscp{}{\lessgtr}(t,\momhig)\dend
\end{align}
Expression for the {\CP} conjugate of the two-loop lepton self-energy can be obtained in a 
similar way.

For the Higgs propagators in the above self-energies we can also use the Kadanoff-Baym ansatz, 
\begin{align}
	\label{KBHiggs}
	\pHiggs{}{>}=(1+f_\higgs)\pHiggs{}{\rho}\kend\quad 
	\pHiggs{}{<}=f_\higgs\pHiggs{}{\rho}\kend
\end{align}
and the simple quasiparticle approximation for the spectral function,
\begin{align}
	\label{QPforHiggs}
	\pHiggs{}{\rho}(t,\momhig)=\,(2\pi)\,\sign(\momhig^0)\,\delta(\momhig^2-m_\higgs^2)\kend
\end{align}
where $m_\higgs$ is the effective thermal mass. Effects of the finite thermal Higgs 
mass will be studied in \sect\ref{HiggsDecay}.

\section{\label{Majorana}Majorana contribution}

In this section we will analyze the lepton number and \CP-violating (inverse) decay 
of the Majorana neutrino as well as the two-body scattering processes mediated by 
the heavy neutrino. In particular, we will derive expressions for the in-medium 
\CP-violating parameters, decay widths and scattering amplitudes. We will also 
explicitly demonstrate that the obtained equation for the lepton asymmetry is free 
of the double-counting problem.

\subsection{\label{TreeLevelDecay}Decay at tree-level approximation}

In the previous section we have used the Kadanoff-Baym ansatz and quasiparticle
approximation for the Higgs and lepton fields. Let us now assume that similar 
approximations also hold for Majorana neutrinos. That is, we assume that in 
\eqns\eqref{Sigma1WT} and \eqref{Sigma1WTCP} the spectral function $\pMaj{ij}{\rho}$ 
is diagonal in flavor space and can be approximated by
\begin{align}
\label{QPMaj}
\pMaj{ij}{\rho}&=(2\pi)\,\sign(\mommaj^0)\delta(\mommaj^2-M_i^2)\,\delta_{ij}\,(\slashed{\mommaj}+M_i) \kend
\end{align}
and that it is related to the Wightmann components via the Kadanoff-Baym ansatz:
\begin{equation}
\label{KBMaj}
\pMaj{ii}{>} = (1-f_{\majneutrino_i})\pMaj{ii}{\rho}, \quad
\pMaj{ii}{<} = -f_{\majneutrino_i}\pMaj{ii}{\rho} \dend
\end{equation}
Substituting \eqns\eqref{Sigma1WT} and \eqref{Sigma1WTCP} in \eqn\eqref{MasterEquationWigner} and 
making the above approximations we find after some algebra that the lepton-current can be 
represented in the form:
\begin{align}
	\label{asymLtree}
	{\cal D}_\mu j_L^{\mu}(t) & = \sum_i \int 
	\dpi{\majneutrino_i}{\lepton\higgs}{q}{pk}\nonumber \\ 
 	& \times \bigl[\EffAmpl{}{\lepton\higgs}{\majneutrino_i} 
	\F{\lepton\higgs}{\majneutrino_i}{\momlep \momhig}{\mommaj} 
	- \EffAmpl{}{\bar\lepton\bar\higgs}{\majneutrino_i} 
	\F{\bar \lepton \bar \higgs}{\majneutrino_i }{\momlep \momhig}{\mommaj } \bigr]\kend
\end{align}
where $\cal F$ have been introduced in \eqn\eqref{eqn:definition of F} and we have defined:
\begin{subequations}
	\label{EffectiveDecayAmplitudesTree}
	\noeqref{EffAmplSELH,EffAmplSEbarLbarH}
	\begin{align}
		\label{EffAmplSELH}
		\EffAmpl{T}{\lepton\higgs}{\majneutrino_i}&\equiv 
		g_w (h^\dagger h)_{ii} \tr[(\slashed{\mommaj}+M_i)P_L\slashed{\momlep}\,]\kend\\
		\label{EffAmplSEbarLbarH}
		\EffAmpl{T}{\bar \lepton \bar \higgs}{\majneutrino_i}&\equiv 
		g_w (h^\dagger h)_{ii} \tr[(\slashed{\mommaj}+M_i)P_L\slashed{\momlep}\,]\dend
	\end{align}
\end{subequations}
The superscript `T' stands for `tree-level'. 
The expression \eqref{asymLtree} strongly resembles the Boltzmann equation. 
Therefore the functions $\EffAmpl{T}{\lepton\higgs}{\majneutrino_i}$ 
and $\EffAmpl{T}{\bar \lepton \bar \higgs}{\majneutrino_i}$ can be interpreted as 
effective in-medium amplitudes squared, summed over internal degrees of freedom, 
for the decays into leptons and antileptons respectively. 
\begin{figure}[h!]
	\centering
	\includegraphics{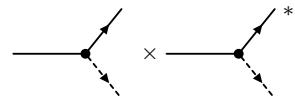}
	\caption{\label{TtimesT}Tree-level contribution.}
\end{figure} 
The two effective amplitudes \eqref{EffectiveDecayAmplitudesTree} can be replaced
by the total decay amplitude and the \CP-violating parameter.
Using \eqn\eqref{EffectiveDecayAmplitudesTree} we find that within the used approximations the 
resulting decay amplitude coincides with the outcome of the vacuum calculation, 
$\EffAmplitude{}{\majneutrino_i}=4g_w (h^\dagger h)_{ii}(\momlep\mommaj)$, and that $\epsilon_i=0$. 

In the presence of a nonzero lepton asymmetry $f_\lepton\neq f_{\bar \lepton}$ 
and $f_\higgs\neq f_{\bar \higgs}$. Therefore $\F{\lepton\higgs}{\majneutrino_i}{\momlep \momhig}{\mommaj}
\neq \F{\bar \lepton \bar \higgs}{\majneutrino_i }{\momlep \momhig}{\mommaj }$ and this leads to 
a washout of the asymmetry. 
Despite the fact that \eqn\eqref{asymLtree} correctly describes the (leading-order)
washout processes, it fails to describe processes which generate lepton asymmetry: 
in the considered approximation $\epsilon_i=0$ because the \CP-violating effects, which are 
required to produce the asymmetry, are of fourth order in the Yukawa couplings of the 
Majorana neutrino. In \eqn\eqref{EffectiveDecayAmplitudesTree} we have taken into account 
only terms quadratic in the coupling. In other words, this approximation corresponds to 
the tree-level approximation in the canonical approach. 

Terms of higher order in the couplings emerge from three- and higher-loop contributions 
to the lepton self-energy, see \eqns\eqref{SE3pMT}-\eqref{SeScaMT}, as well as from 
expansion of the full Majorana propagators entering the self-energies. 

\subsection{\label{EqSolKB}Equilibrium solution for Majorana propagator}

In order to define an effective CP-violating parameter and decay width that 
incorporate medium corrections we have to identify the quasiparticle 
excitations in the system. To perform this analysis we follow the discussion of the 
self-energy contribution within a toy-model as presented in \cite{Garny:2009qn}.
As has been demonstrated there, it is important to take the matrix structure 
of the Majorana propagator in flavor space into account. Our starting point is the 
Schwinger-Dyson equation for the Majorana two-point function:
\begin{align}
\pMajmat{-1}{}(x,y) & = \pMajmat{-1}{0}(x,y) - \sMajmat{}{}(x,y)\label{SDCTPb}\dend 
\end{align} 
Let us split the self-energy into diagonal and off-diagonal components in 
flavor space and introduce a diagonal propagator $\pMajdiag{}{}$ defined by 
the equation:
\begin{equation}
	\label{eqn: SD for diagonal propagators}
	\pMajdiagmat{-1}{}(x,y)=\pMajdiagmat{-1}{0}(x,y)-\sMajmat{d}{}(x,y)\kend
\end{equation}
where $\pMajdiagmat{}{0}$ is the free propagator and $\sMajmat{d}{}$
denotes the diagonal components of the self-energy. 
The poles of the diagonal propagator define the quasiparticle excitations. 
It can be shown that the dynamics of these is described by a Boltzmann-like 
quantum kinetic equation.

Inserting this decomposition into the Schwinger-Dyson equation we find,
using matrix notation:
\begin{equation}
	\label{SDequ}
	\pMajmat{-1}{}(x,y)=\pMajdiagmat{-1}{}(x,y)-\sMajmat{'}{}(x,y)\kend
\end{equation}
where $\sMajmat{'}{}$ denotes the off-diagonal components of the 
self-energy and $\pMajmat{}{}$ the full neutrino propagator including flavor-diagonal
and flavor off-diagonal contributions.	
Multiplying \eqn\eqref{SDequ} by $\pMajmat{}{}$ from the left, by 
$\pMajdiagmat{}{}$ from the right and integrating over the contour 
$\cal C$ we obtain a formal solution for the full non-equilibrium 
propagator:
\begin{align}
	\label{SDint}
	\pMajmat{}{}(x,y)&= \pMajdiagmat{}{}(x,y) \notag \\ &+\int_\mathcal{C} 
	\dg^4u \dg^4v \pMajmat{}{}(x,u)\sMajmat{'}{}(u,v)\pMajdiagmat{}{}(v,y)\dend
\end{align}
After decomposing the propagators and self-energies into the spectral and statistical 
components, we can rewrite \eqn\eqref{SDint} in the form:
\begin{align}
	\label{propFRho}
	\pMajmat{}{F(\rho)}(x,y)&=\pMajdiagmat{}{F(\rho)}(x,y)-\int \dg^4u \dg^4v \theta(u^0) \theta(v^0) \notag \\
	&\times \bigl[ \pMajmat{}{R}(x,u) \sMajmat{'}{R} (u,v) \pMajdiagmat{}{F(\rho)}(v,y) \notag\\ 
	& +\pMajmat{}{R}(x,u) \sMajmat{'}{F(\rho)} (u,v) \pMajdiagmat{}{A}(v,y) \notag\\ 
	&+ \pMajmat{}{F(\rho)}(x,u) \sMajmat{'}{A} (u,v) \pMajdiagmat{}{A}(v,y) \bigr]\dend
\end{align}
Here we are using the retarded and advanced propagators defined by \eqn\eqref{RetAdvDef}, 
so that the integration can be extended to the whole $uv$-plane. Using their definitions 
and \eqn\eqref{propFRho}, we can also derive formal solutions for the retarded and advanced 
propagators:
\begin{align}
	\label{propRA}
	\pMajmat{}{R(A)}(x,y)&=\pMajdiagmat{}{R(A)}(x,y)-\int \dg^4u \dg^4v \theta(u^0) \theta(v^0) \notag \\
	\times & \pMajmat{}{R(A)}(x,u) \sMajmat{'}{R(A)} (u,v) \pMajdiagmat{}{R(A)}(v,y)\dend
\end{align}
Next we Wigner transform \eqns\eqref{propFRho} and \eqref{propRA} and perform the leading 
order gradient expansion as has been outlined in \sect\ref{NEQFTapproach}. Combining both 
results, we find for the full statistical and spectral propagators and the corresponding 
causal two-point functions of the system in to equilibrium:
\begin{subequations}
\label{FullPropagator}
\noeqref{FullPropagatorFrho,FullPropagatorRA}
\begin{align}
	\label{FullPropagatorFrho} 
	\pMajmat{}{F(\rho)}&= \pMajdiagmat{}{F(\rho)} -\pMajmat{}{R} \sMajmat{'}{R}\pMajdiagmat{}{F(\rho)}\notag \\ 
	&-\pMajmat{}{R} \sMajmat{'}{F(\rho)} \pMajdiagmat{}{A}-\pMajmat{}{F(\rho)}\sMajmat{'}{A} \pMajdiagmat{}{A}, \\
	\label{FullPropagatorRA} 
	\pMajmat{}{R(A)}& =\pMajdiagmat{}{R(A)}-\pMajmat{}{R(A)}\sMajmat{'}{R(A)}\pMajdiagmat{}{R(A)}\kend
\end{align}
\end{subequations}
where all propagators and self-energies are evaluated at the same point $(X,\mommaj)$ in configuration space. 
We can express the full statistical and spectral propagators in terms of the diagonal ones and the off-diagonal self-energies,
\begin{align}
	\label{ODprop}
	\pMajmat{}{F(\rho)}= \HatMatTheta{}{R} \bigl[\pMajdiagmat{}{F(\rho)}-\pMajdiagmat{}{R} \sMajmat{'}{F(\rho)} \pMajdiagmat{}{A} \bigr] \HatMatTheta{}{A}\kend
\end{align}
where $\HatMatTheta{}{R}$ and $\HatMatTheta{}{A}$ are defined by 
$\HatMatTheta{}{R}\equiv \bigl( \textbf{I}+\pMajdiagmat{}{R}\sMajmat{'}{R} \bigr)^{-1}$
and $\HatMatTheta{}{A}\equiv \bigl(\textbf{I}+\sMajmat{'}{A} \pMajdiagmat{}{A} \bigr)^{-1}$ 
respectively, with $\textbf{I}$ being the $4n\times4n$ unit matrix in the Dirac and flavor 
space of the $n$ generations. Solution \eqn\eqref{ODprop} reduces the dynamics of the full
statistical and spectral propagators to the dynamics of two quasiparticle excitations. 
Their masses, decay widths and \CP-violating parameters are determined by the medium and 
the abundances are described by the corresponding one-particle distribution functions. 
Strictly speaking, the solution \eqref{ODprop} is valid only in thermal equilibrium. 
However, we assume that it also holds for small deviations from equilibrium. 

To consistently analyze processes of the fourth order in the coupling one has to use 
so-called extended quasiparticle approximation (eQP) for the statistical propagator and 
spectral function \cite{Spicka:1994zz,Spicka:1995zz,CondMatPhys2006_9_473,PhysRevC.48.1034,JPhys2006_35_110}. 
The eQP approximation represents the diagonal propagator as a sum of two terms:
\begin{align}
\label{eQP}
\pMajdiagmat{}{\gtrless}&=\pMajEQPmat{\gtrless} -{\textstyle\frac{1}{2}} 
\bigl(\pMajdiagmat{}{R} \sMajmat{d}{\gtrless}\pMajdiagmat{}{R} + \pMajdiagmat{}{A} 
\sMajmat{d}{\gtrless}\pMajdiagmat{}{A} \bigr) \dend
\end{align}
The first describes decay processes, whereas the second can be associated with scattering 
processes. Inserting \eqn\eqref{eQP} into \eqn\eqref{ODprop} we get a solution for the resummed 
Majorana propagator consistent up to the fourth order in the couplings:
\begin{align}
	\label{eQPequ}
	\pMajmat{}{\gtrless} & =\HatMatTheta{}{R} \bigl[\,\pMajEQPmat{\gtrless}\nonumber\\
	&-\pMajdiagmat{}{R} \sMajmat{'}{\gtrless}\pMajdiagmat{}{A} 
	- {\textstyle\frac{1}{2}}\bigl(\pMajdiagmat{}{R} \sMajmat{d}{\gtrless}\pMajdiagmat{}{R}+
	\pMajdiagmat{}{A} \sMajmat{d}{\gtrless}\pMajdiagmat{}{A} \bigr) 
	\bigr]\HatMatTheta{}{A}\dend 
\end{align}
The first term in the above formula describes Majorana decay, see \sect\ref{CPViolationInDecay},
whereas the remaining three terms describe the two-body scattering processes mediated by the 
Majorana neutrino. These are discussed in \sect\ref{MajoranaMediatedScattering}.

Using definition of the retarded and advanced two-point functions, \eqn\eqref{RetAdvDef},
and the Schwinger-Dyson equation for the diagonal propagators, 
\eqn\eqref{eqn: SD for diagonal propagators}, we find that the causal propagators 
in \eqn\eqref{eQPequ} are given by
\begin{align}
	\label{diagRA}	
	\pMajdiagmat{}{R(A)}=-\bigl( \slashed{\mommaj} -\hat{M} - \sMajmat{}{R(A)}\bigr)^{-1}\dend
\end{align}
Splitting the retarded and advanced self-energies into the vector and scalar
components we can write the solution of \eqn\eqref{diagRA} in the form:
\begin{align}
\label{SRAsol}
\pMajdiag{}{R(A)}&=-\frac{\bigl(\slashed{q}-\sMajsl{v}{R(A)}\bigr)+
\bigl(M+\sMaj{s}{R(A)}\bigr)}{\bigl(q-\sMaj{v}{R(A)}\bigr)^2-
\bigl(M+\sMaj{s}{R(A)}\bigr)^2}\nonumber\\
&\equiv -\frac{\OmegaNum{}{h}\mp \frac{i}2\sMajsl{v}{\rho}}{\OmegaDen{}{h}\pm i\PiDen{}{\rho}}
\kend
\end{align}
where we have omitted flavor indices to shorten the notation and introduced 
\begin{align*}
\OmegaNum{}{h}&\equiv \bigl(\slashed{q}-\sMajsl{v}{h}\bigr)+
\bigl(M+\sMaj{s}{h}\bigr)\kend\\
\OmegaDen{}{h}&\equiv \bigl(q-\sMaj{v}{h}\bigr)^2-
\bigl(M+\sMaj{s}{h}\bigr)^2-\bigl({\textstyle\frac12}\sMaj{v}{\rho}\bigr)^2\kend\\
\PiDen{}{\rho}&\equiv \sMaj{v}{h}\sMaj{v}{\rho}-q\sMaj{v}{\rho}\dend
\end{align*}
From \eqn\eqref{SRAsol} we can extract the spectral and hermitian propagators. To leading order 
in the Yukawas they read
\begin{subequations}
\label{ShrhoApprox}
\noeqref{SrhoApprox,ShApprox}
\begin{align}
\label{SrhoApprox}
\pMajdiag{}{\rho}\approx+\OmegaNum{}{h}\,\frac{2\,\PiDen{}{\rho}}{
\OmegaDen{2}{h}+\PiDen{2}{\rho}}\equiv \Omega_h \pMajdiagsc{}{\rho}\kend\\
\label{ShApprox}
\pMajdiag{}{h}\approx-\OmegaNum{}{h}\,\frac{\OmegaDen{}{h}}{
\OmegaDen{2}{h}+\PiDen{2}{\rho}}\equiv \Omega_h \pMajdiagsc{}{h}\dend
\end{align}
\end{subequations}
The on-shell condition is defined by $\OmegaDen{}{h}=0$. Expanding 
$\OmegaDen{}{h}$ to linear order in the Yukawas we find:
\begin{align*}
\OmegaDen{}{h}&\approx \bigl(q^2-M^2\bigr)\left(1-2\Gamma/\pi M\right)
\nonumber\\
&+(\Gamma/\pi M)\,q^2\ln\left(|q^2|/M^2\right)-2 q\sMaj{v,med}{h}\kend
\end{align*}
where $\sMajsl{v,med}{h}$ is the medium-induced component of the hermitian self-energy
 in the on-shell renormalization scheme.
In vacuum the on-shell condition is fulfilled for $q^2=M^2$, i.e. $M$ is the physical vacuum mass. 
At non-zero temperatures the mass receives medium-induced corrections. To linear order in the Yukawas
the effective mass is given by $\Meff{2}{}\approx M^2+2 q\sMaj{v,med}{h}$. 
For a hierarchical mass spectrum, which we consider here, the contributions of the 
hermitian self-energy are always negligible and we will use $\OmegaDen{}{h}\approx \mommaj^2-M^2$
and $\OmegaNum{}{h}\approx \slashed{\mommaj}-M$ in the following. From \eqn\eqref{ShrhoApprox} we can also deduce 
the effective width. To leading order in the Yukawas it is given by 
$\Meff{}{}\Gammaeff{}{}\approx-q\sMaj{v}{\rho}$. The minus sign in this definition 
ensures that the effective decay width is positive. One-loop contribution 
to the Majorana self-energy is derived in \app\ref{SelfEnergy}. In a 
\CP-symmetric medium it is given by
\begin{align}
	\label{eqn: Prho}
	\sMaj{ij}{\rho}&=-{\frac{g_w}{16\pi}}
	\bigl[(h^\dagger h)_{ij}P_L+(h^\dagger h)^*_{ij} P_R\,\bigr] L_{\rho}\dend
\end{align}
Therefore we can write the effective decay width in the form 
$\Gammaeff{}{i}=\Gamma_i\cdot (q L_\rho/M^2_i)$, where $\Gamma_i$ is the total vacuum 
decay width. For positive $q^2$ and $q^0$ the loop integral $L_\rho$ takes the form:
\begin{align}
	\label{eqn:Lrho}
	L^\mu_{\rho}=16\pi \int \lorentzd{\higgs}{\momhig} 
	\lorentzd{\lepton}{\momlep} & (2\pi)^4 \deltafour{\mommaj-\momhig-\momlep}\,
	\momlep^\mu\nonumber\\
	&\times \bigl[1+\f{\higgs}{}(E_\momhig)-\f{\lepton}{}(E_\momlep)\bigr]\dend
\end{align}
For massless final states $2(qp)=M_i^2$. Therefore the definition of the effective decay 
width inferred in \sect\ref{RISQuantStat} from the requirement of successful RIS subtraction
is consistent with that implied by \eqn\eqref{ShrhoApprox}.
 
For the eQP Wightman propagators we can use the Ka\-da\-noff-Baym ansatz. As can be 
inferred from \eqn\eqref{eQP}, the corresponding spectral function reads
\begin{align}
\label{eQPSpectrFunc}
\pMajEQPmat{\rho}=-{\textstyle\frac12}\pMajdiagmat{}{R}\sMajmat{d}{\rho}
\pMajdiagmat{}{R}\sMajmat{d}{\rho}\pMajdiagmat{}{A}\sMajmat{d}{\rho}
\pMajdiagmat{}{A}\dend
\end{align}
Substituting \eqn\eqref{SRAsol} into \eqn\eqref{eQPSpectrFunc} we obtain
\begin{align}
\label{eQPSpectrFull}
\pMajEQP{}{\rho}=&-{\textstyle\frac12}\bigl(
\OmegaNum{}{h}\sMajsl{v}{\rho}\OmegaNum{}{h}
\PiDen{2}{\rho}
+4\,\OmegaNum{}{h}\sMajsl{v}{\rho}\sMajsl{v}{\rho}
\OmegaDen{}{\rho}\OmegaDen{}{h}\nonumber\\
&+\sMajsl{v}{\rho}\sMajsl{v}{\rho}\sMajsl{v}{\rho}
\OmegaDen{2}{h}\bigr)/\bigl(\OmegaDen{2}{h}+\PiDen{2}{\rho}\bigr)^2
\kend
\end{align}
where we have again omitted the flavor indices. 
The second and the third terms in \eqn\eqref{eQPSpectrFull} vanish 
on the mass shell and can be neglected. Commuting $\OmegaNum{}{h}$ 
and $\sMajsl{v}{\rho}$ in the first term and again neglecting contributions
which are tiny on the mass shell we finally obtain for the eQP spectral function:
\begin{align}
\label{eQPSpectrFirstTerm}
\pMajEQP{}{\rho} & \approx \OmegaNum{}{h}
\frac{4 \PiDen{3}{\rho}}{\left[\OmegaDen{2}{h}+\PiDen{2}{\rho}\right]^2}
\dend
\end{align}
Note that structures of \eqns\eqref{eQPSpectrFirstTerm} and \eqref{SrhoApprox} are very 
similar. Furthermore, as follows from \eqn\eqref{deltalimit}, in the limit of vanishing 
decay width both of them approach the delta-function. However, for a small but \textit{finite} 
decay width the eQP spectral function is a better approximation to the delta-function 
than \eqn\eqref{SrhoApprox}. Therefore, we can approximate it by the usual expression,
\begin{align}
\label{eQPSpectralFunction}
\pMajEQP{}{\rho} & \approx (2\pi)\,\sign(q^0) \delta (q^2-M^2)(\slashed{q}+M) \kend
\end{align}
and at the same time keep finite-width terms in the diagonal propagators.

\subsection{\label{CPViolationInDecay} CP-violation in Majorana decay}

To go beyond the tree-level approximation and take into account \CP-violating effects 
we need to consider contributions to the lepton self-energy that are of the fourth 
order in the Yukawa couplings. 

One of them comes from expansion of the Majorana propagator in the one-loop self-energy. 
Substituting the decay term of \eqn\eqref{eQPequ} into \eqn\eqref{Sigma1WT} we can write it in 
the form: 
\begin{align}
	\label{Sigma1WTExpansion}
	&\sLept{(1)}{\gtrless} (t,\momlep) = 
	-\int \lorentzdd{\mommaj}\lorentzdd{\momhig}(2\pi)^4 \deltafour{\mommaj-\momhig-\momlep} \\
	& \times (\yu^\dagger \yu)_{mn} P_R \MatTheta{ni}{R}(t,\mommaj) 
	\pMajEQP{ii}{\gtrless}(t,\mommaj) \pHiggs{}{\lessgtr}(t,\momhig)
	\MatTheta{im}{A}(t,\mommaj) P_L \nonumber \dend
\end{align}
Substituting \eqn\eqref{Sigma1WTExpansion} and its {\CP} conjugate into \eqn\eqref{MasterEquation} 
we find that the resulting contribution to the divergence of the lepton-current has precisely 
the form \eqref{asymLtree}. However, the corresponding effective amplitudes are no longer equal:
\begin{subequations}
	\label{EffectiveDecayAmplitudesSE}
	\noeqref{EffAmplSECPLH,EffAmplSECPbarLbarH}
	\begin{align}
		\label{EffAmplSECPLH}
		\EffAmpl{T}{\lepton\higgs}{\majneutrino_i}&+\EffAmpl{S}{\lepton\higgs}{\majneutrino_i}
		\equiv g_w{\textstyle\sum}_{mn}(h^\dagger h)_{mn}\nonumber\\
		\times&\tr[\MatTheta{ni}{R}(t,\mommaj)(\slashed{\mommaj}+M_i)
		\MatTheta{im}{A}(t,\mommaj)P_L\slashed{\momlep}P_R\,]\kend\\
		\label{EffAmplSECPbarLbarH}
		\EffAmpl{T}{\bar\lepton\bar\higgs}{\majneutrino_i}&
		+\EffAmpl{S}{\bar\lepton\bar\higgs}{\majneutrino_i}\equiv 
		g_w {\textstyle\sum}_{mn}(h^\dagger h)^*_{mn}\nonumber\\
		\times&\tr[\MatThetacp{ni}{R}(t,\mommaj)(\slashed{\mommaj}+M_i)
		\MatThetacp{im}{A}(t,\mommaj)P_L\slashed{\momlep}P_R\,]\dend
	\end{align}
\end{subequations}
The matrices $\HatMatTheta{}{R}$ and $\HatMatTheta{}{A}$ are evaluated on the mass shell 
of the $i$'th Majorana neutrino. The bar denotes \CP-con\-ju\-gation and the trace is over 
Dirac indices. 
\begin{figure}[h!]
	\centering
	\includegraphics{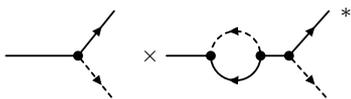}
	\caption{\label{TplusS}Interference of tree-level and one-loop self-energy corrections.}
\end{figure} 
As compared to tree-level result \eqref{EffectiveDecayAmplitudesTree} it additionally 
contains interference of the tree-level and one-loop self-energy contributions to the 
Majorana decay amplitude, see \fig\ref{TplusS}. For a hierarchical mass spectrum we can 
use the approximation 
\begin{align}
	\HatMatTheta{}{R}\approx \textbf{I}-\pMajdiagmat{}{R} \sMajmat{'}{R}
	\approx \textbf{I}-{\textstyle\frac{i}{2}}\pMajdiagmat{}{h} \sMajmat{'}{\rho}\kend
\end{align}
and a similar approximation for $\HatMatTheta{}{A}$. Using furthermore \eqns\eqref{ShrhoApprox} 
and \eqref{eqn: Prho} we find for the \CP-violating parameter:
\begin{align}
	\label{SEepsEpsilon}
	\epsilon_i^{S}=\sum&
	\frac{\Im\, (h^\dagger h)_{ij}^2}{(h^\dagger h)_{ii}(h^\dagger h)_{jj}} 
	\frac{\Delta M_{ij}^2 M_i \Gamma_j}{\big(\Delta M_{ij}^2\big)^2+ 
	\bigl(\Gamma_j/M_j\, \mommaj L_\rho\bigr)^2} \notag \\
	& \times \frac{\momlep L_\rho}{\mommaj\momlep}\kend
\end{align}
where $\momlep$ and $\mommaj$ are on-shell momenta of the outgoing lepton and decaying 
Majorana neutrino respectively. In vacuum $L_\rho^\mu = \theta(q^2)\,\sign(\mommaj^0) q^\mu$ 
and the \CP-violating parameter takes the form:
\begin{align}
	\label{eqn: self-energycp-violation vacuum}
	\hskip -3mm
	\epsilon_i^{S}=\sum&
	\frac{\Im\, (h^\dagger h)_{ij}^2}{(h^\dagger h)_{ii}(h^\dagger h)_{jj}} 
	\frac{\Delta M_{ij}^2 M_i \Gamma_j}{\big(\Delta M_{ij}^2\big)^2+ 
	\bigl(\Gamma_j M^2_i/M_j\bigr)^2} \dend
\end{align}
The `regulator' in the denominator of \eqn\eqref{eqn: self-energycp-violation vacuum}
differs from the result $M_i \Gamma_j$ found in \cite{Plumacher:1997ru,Pilaftsis:2003gt} 
by the ratio of the masses. For a hierarchical neutrino mass spectrum the `regulator' term 
is sub-dominant and this difference is numerically small. Note also that although 
\eqn\eqref{SEepsEpsilon} does not diverge 
in the limit of vanishing mass difference the approximations made in the course of its 
derivation are not applicable for a quasidegenerate mass spectrum \cite{Garny:2009qn}.
For a consistent treatment of resonant enhancement within NEQFT we refer to~\cite{Garny:2011hg}.

\begin{figure}[h!]
	\centering
	\includegraphics{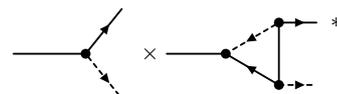}
	\caption{\label{InterferenceV}Interference of tree-level and one-loop vertex corrections.}
\end{figure}

The two-loop lepton self-energy is of the fourth order in the couplings to begin with. 
Therefore, for a hierarchical mass spectrum one can safely neglect the off-diagonal 
components of the Majorana propagators and replace $\pMaj{}{}$ by the eQP one:
\begin{align}
	\label{Sigma2WTExpansion}
	&\sLept{(2.1)}{\gtrless}(t,\momlep) = \int \lorentzdd{\mommaj} \lorentzdd{\momhig}\
	(2\pi)^4\delta(\momlep+\momhig-\mommaj)\nonumber \\
	&\times\big[ (\yu^\dagger \yu)^2_{ij}
	\varLambda_{jj}(t,q,k)P_L C \pMajEQP{ii}{\gtrless}(t,\mommaj) 
	P_L\pHiggs{}{\lessgtr}(t,\momhig)\nonumber\\
	&+ (\yu^\dagger \yu)^2_{ji}
	P_R \pMajEQP{ii}{\gtrless}(t,\mommaj) C P_R V_{jj}(t,\mommaj,\momlep)
	\pHiggs{}{\lessgtr}(t,\momhig)\big] \dend
\end{align}
Substituting \eqn\eqref{Sigma2WTExpansion} and its {\CP} conjugate into \eqn\eqref{MasterEquation} 
we again find that the resulting contribution to the divergence of the lepton-current has 
the form \eqref{asymLtree}. The corresponding effective amplitudes read
\begin{subequations}
\label{MajoranaVertexAmplitudes}
\noeqref{EffAmplVertLH,EffAmplVertbarLbarH}
\begin{align}
	\label{EffAmplVertLH}
	\EffAmpl{V}{N_i}{\lepton \higgs} \equiv & -g_w(h^{\dagger}h)_{ij}^2\ 
	M_i\,\tr\bigl[\varLambda_{jj}(q,k) C P_L\slashed pP_R\bigr]\\
	&-g_w(h^{\dagger}h)_{ji}^{2}\,M_i\, \tr\bigr[ C V_{jj}(q,k)P_L\slashed p P_R\bigr]\kend\nonumber\\
	\label{EffAmplVertbarLbarH}
	\EffAmpl{V}{N_i}{\bar \lepton \bar \higgs} 
	\equiv & -g_w(h^{\dagger}h)_{ij}^2\ M_i\,
	\tr\bigr[ C V_{jj}(q,k)P_L\slashed p P_R\bigr]\\
	&-g_w(h^{\dagger}h)_{ji}^{2}\,M_i\, \tr\bigl[\varLambda_{jj}(q,k)
	C P_L\slashed pP_R\bigr]\nonumber\dend\hphantom{aa}
\end{align}
\end{subequations}
They describe 
interference of the tree-level and one-loop vertex contributions to the Majorana decay 
amplitude, see \fig\ref{InterferenceV}.
To account for the contribution of the vertex correction to the decay width and 
the \CP-violating parameter we have to substitute the sum of $ \EffAmpl{V}{N_i}{\lepton \higgs}$
and $\EffAmpl{T}{N_i}{\lepton \higgs}+\EffAmpl{S}{N_i}{\lepton \higgs}$ 
and a similar sum for the antiparticles into \eqn\eqref{AmplsqAndEpsDef}. The vertex contribution to the decay 
amplitude is of fourth order in the coupling and is negligible compared to the tree-level term.
Since we assume the medium to be almost \CP-symmetric we can use, at leading order, \CP-symmetric 
two-point functions in the loop integrals $\varLambda_{jj}$ and $V_{jj}$. Then, at leading order in 
the Yukawa couplings, we find for the vertex contribution to the \CP-violating parameter:
\begin{align}
	\label{MajoranaCPVertex}
	\epsilon^V_i &= -\sum \frac{\Im\,(h^\dagger h)_{ij}^2}{(h^\dagger h)_{ii}}
	\frac{M_iM_j}{\mommaj\momlep}
	\ \int \lorentzdd{k_1} \lorentzdd{k_2} \lorentzdd{k_3} \\
	&\times (2\pi)^4 \delta(q+k_1+k_2) (2\pi)^4 \delta(k+k_2-k_3) (\momlep k_2)\nonumber\\ 
	&\times \bigl[ \pHiggs{}{\rho}(k_1) \pLeptsc{}{F}(k_2) \pMajdiagsc{jj}{h}(k_3) 
	+ \pHiggs{}{F}(k_1) \pLeptsc{}{\rho}(k_2) \pMajdiagsc{jj}{h}(k_3)\nonumber\\
	& - \pHiggs{}{h}(k_1) \pLeptsc{}{\rho}(k_2) \pMajdiagsc{jj}{F}(k_3) 
	+ \pHiggs{}{h}(k_1) \pLeptsc{}{F}(k_2) \pMajdiagsc{jj}{\rho}(k_3)
	\nonumber\\
	& + \pHiggs{}{\rho}(k_1) \pLeptsc{}{h}(k_2) \pMajdiagsc{jj}{F}(k_3) 
	+ \pHiggs{}{F}(k_1) \pLeptsc{}{h}(k_2) \pMajdiagsc{jj}{\rho}(k_3)\bigr]\nonumber\dend
\end{align}
The quasiparticle approximation and the KB-ansatz enforce two of the intermediate lines of the vertex 
loop to be on-shell whereas the remaining line described by the hermitian part of the retarded 
and advanced propagators remains off-shell. The three lines in square brackets in \eqn\eqref{MajoranaCPVertex} 
therefore correspond to different cuts through two of the three internal lines of the loop 
diagram \fig\ref{treevertexself}.(c). Note also that only for one of the three internal lines 
the corresponding distribution function enters the result. 

The first possible cut described 
by the first line in square brackets corresponds to cutting the propagators of Higgs and 
lepton. One can interpret this cut as decay of the Majorana neutrino into a lepton-Higgs pair 
which is followed by a subsequent $t$-channel scattering mediated by a virtual Majorana neutrino. 
Introducing
\begin{align}
	\label{Majorana vertex loop integral}
	K^\mu_{i}&(\mommaj,\momhig)=16\pi \int \lorentzd{\higgs}{\momhig_2} 
	\lorentzd{\lepton}{\momlep_2} (2\pi)^4 \deltafour{\mommaj-\momhig_2-\momlep_2}\,
	\momlep^\mu_2\nonumber\\
	&\times \bigl[{1+\f{\higgs}{}(E_{\momhig_2})-\f{\lepton}{}(E_{\momlep_2})}\bigr]
	M_i^2 \pMajdiagsc{ii}{h}(\momhig-\momlep_2) \kend
\end{align}
we can rewrite the first term in \eqn\eqref{MajoranaCPVertex} in a form which stro\-ngly resembles the form 
of the self-energy \CP-vi\-o\-la\-ting parameter: 
\begin{align}
	\label{eqn: vertex cp-violation}
	\epsilon_i^V=-\frac12\sum\frac{\Im\, (h^\dagger h)_{ij}^2}{(h^\dagger h)_{ii}(h^\dagger h)_{jj}}
	&\frac{M_i \Gamma_j}{M_j^2} \frac{p K_j}{qp}\dend
\end{align}
In vacuum $K_j$ can be computed explicitly and we recover the well-known result 
\cite{Fukugita:1986hr}: 
\begin{align}
	\label{eqn: vertex cp-violation vacuum}
	\epsilon_i^V& = \sum\frac{\Im\,(h^\dagger h)_{ij}^2 }{(h^\dagger h)_{ii}(h^\dagger h)_{jj}}
	\frac{\Gamma_j}{M_i}\nonumber\\
	& \times \bigl[1-\left(1+M_j^2/M_i^2\right)\ln\left(1+M_i^2/M_j^2\right)\bigr]\dend
\end{align}
Adding up \eqns\eqref{eqn: self-energycp-violation vacuum} and 
\eqref{eqn: vertex cp-violation vacuum} we obtain the canonical 
expression for the vacuum \CP-violating parameter, \eqn\eqref{epsilon vacuum}.

If the intermediate Majorana neutrino is much heavier than the decaying one 
then $M_j^2 \pMajdiagsc{jj}{h}\approx 1$ 
and therefore $K_j(\mommaj,\momhig)\approx L_\rho(\mommaj)$. In this case 
we can also neglect the `regulator' term in the denominator of 
\eqn\eqref{SEepsEpsilon}. In this approximation the two \CP-violating 
parameters have the same structure and their sum can be written in the 
form: 
\begin{align}
\epsilon_i=\epsilon_i^{vac}\,\frac{\momlep L_\rho}{\mommaj \momlep}\dend
\end{align}
Note that the combination of the distribution functions that enters the 
self-energy and vertex \CP-violating parameters, see \eqns\eqref{eqn:Lrho} and 
\eqref{Majorana vertex loop integral}, is the same as that of
$f_{\lepton\higgs}=1+\feq{\higgs}{}-\feq{\lepton}{}$ encountered in the derivation of the rate 
equations, see \eqn\eqref{eqn:linear expansion of product of quantum-statistical terms}. 
This result is in agreement with the findings of~\cite{Garny:2009rv,Garny:2009qn,Beneke:2010wd,Anisimov:2010dk}
using NEQFT and of \cite{Kiessig:2011fw} based on imaginary-time thermal QFT. Note that older results featured a different dependence on the distribution functions, 
with an additional term quadratic in the one-particle distribution functions which is 
absent in \eqn\eqref{eqn: vertex cp-violation} as well as in \eqn\eqref{SEepsEpsilon}:
\begin{align}
	\label{finiteTCP}
	 1+\bar{f}_\higgs-\bar{f}_\lepton+2\bar{f}_\higgs \bar{f}_\lepton\
 \rightarrow\
	1+\bar{f}_\higgs-\bar{f}_\lepton\
	\dend
\end{align} In \cite{Garny:2010nj} it was demonstrated that the result obtained 
using thermal field theory can be reconciled with the result of NEQFT calculation once causal Green's functions are used in the former. 
The two other cuts in \eqn\eqref{MajoranaCPVertex} are proportional to $\f{\majneutrino}{}-\f{\lepton}{}$
and to $\f{\majneutrino}{}+\f{\higgs}{}$ respectively. They vanish in the zero temperature limit and are 
usually Boltzmann-suppressed at finite temperatures, but can be relevant in specific cases 
\cite{Garbrecht:2010sz}.
 
\begin{figure}[h!]
	\begin{center}
	\includegraphics[width=\columnwidth]{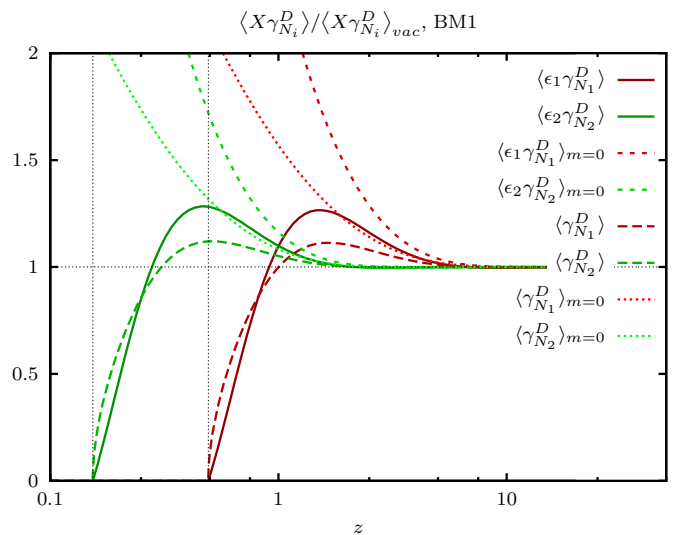}
	\end{center}
	\caption{\label{GammaDRatio} Decay and \CP-violating reaction densities 
	with thermal lepton and Higgs masses, ${\langle{X\gamma^D_{\majneutrino_i}}\rangle}$, 
	and with zero masses, ${\langle{X\gamma^D_{\majneutrino_i}}\rangle_{m=0}}$, 
	for the two Majorana neutrinos $\majneutrino_1$ and $\majneutrino_2$. The 
	values are normalized to the corresponding reaction density in the conventional 
	approximation ${\big<{X\gamma^{D}_{\majneutrino_i}}\big>_{vac}}$. The thermal 
	enhancement due to quantum-statistical factors is overcompensated by the phase space 
	suppression due to thermal masses at high temperatures. Note that we show only the self-energy contribution to the \CP-violating reaction densities.}
\end{figure}

The quantities that enter the rate equations are the decay, washout and 
\CP-violating decay reaction densities. In the canonical approximation, i.e. 
when the quantum-statistical effects and effective masses of the Higgs and leptons 
are neglected, they are given by \eqn\eqref{eqn:decay reaction density main text}.
If the thermal masses are neglected but the quantum-statistical effects are taken 
into account, there is an enhancement of the decay and washout reaction densities
at high temperature, see \fig\ref{GammaDRatio}. 
However, the inclusion of the thermal masses turns this enhancement into 
a suppression at high temperatures. It is explained by the decrease of the decay phase space. 
At intermediate temperatures the thermal 
masses become small relative to the Majorana mass and we observe a minor 
enhancement. For the \CP-violating reaction density we observe a very 
similar behavior. Given that for a hierarchical mass spectrum most 
of the asymmetry is typically generated by the lightest Majorana neutrino at 
$z_f \sim \ln K_1\sim {\cal O}(1)$, where $K_1$ is the washout parameter 
(see \app\ref{NumericalParameters}), 
we expect the medium effects to induce a moderate enhancement of the total 
generated asymmetry.

\subsection{\label{MajoranaMediatedScattering} Majorana-mediated scattering}

Two-body scattering processes mediated by Majorana neutrinos violate 
lepton number by two units and play an important role in the washout of the generated 
asymmetry. In this section we derive the effective scattering amplitudes using NEQFT.
This is an important part of our results.

The last three terms in \eqn\eqref{eQPequ} contain the Wigner-tran\-s\-for\-med one-loop 
Majorana self-energy:
\begin{align}
	\label{SEmaj}
	\sMaj{ij}{\gtrless}(t,\mommaj) = & -g_w \int \lorentzdd{\momhig} 
	\lorentzdd{\momlep}(2\pi)^4\delta(\mommaj-\momlep-\momhig) \nonumber\\ 
	& \times \bigl[ \, (\yu^\dagger \yu)_{ij} P_L 
	\pLept{}{\gtrless} (t,\momlep)P_R 
	\pHiggs{}{\gtrless}(t,\momhig) \nonumber \\ 
	& + (\yu^\dagger \yu)_{ji}P_R P 
	\pLeptcp{}{\gtrless}(t,\bar{\momlep}) P P_L 
	\pHiggscp{}{\gtrless} (t,\bar{\momhig}) \bigr] \kend \hphantom{a}
\end{align}
see \app\ref{MajoranaSelfEn} for more details. Combining them with \eqn\eqref{Sigma1WT} we 
find that their contribution to the divergence of the lepton-current 
\eqref{MasterEquationWigner} contains two Wightman propagators of leptons and two 
of the Higgs field. As we have argued above, these correspond to initial and final 
states in the kinetic equations. Therefore, we conclude that these terms describe 
scattering processes depicted in \fig\ref{LHbarLbarH} and \fig\ref{LLbarHbarH}.
As Higgs and leptons are maintained close to equilibrium we can safely use the 
Kadanoff-Baym ansatz for their propagators in the Majorana self-energy \eqref{SEmaj}.
Inserting \eqn\eqref{SEmaj} into the scattering terms of \eqn\eqref{eQPequ} we can then split the Majorana 
propagator into a lepton number conserving and lepton number violating part:
\begin{subequations}
	\label{propMajapSp}
	\noeqref{propMajapSp>,propMajapSp<}
	\begin{align}
		\label{propMajapSp>}
		\pMaj{ij}{>}(\mommaj)&= g_w\int \lorentzdd{\momhig} \lorentzdd{\momlep}(2\pi)^4 
		\delta(q-p-k)\pHiggs{}{\rho}(\momhig) \pLeptsc{}{\rho}(\momlep) \nonumber \\ 
		&\times \big[ (1-f_{\bar \lepton}^\momlep)(1+f_{\bar \higgs}^\momhig) 
		\pMaj{ij}{LC}(\mommaj,\momlep)\nonumber \\
		&+(1-f_{\bar \lepton}^\momlep)(1+f_{\bar \higgs}^\momhig)\pMaj{ij}{LV}(\mommaj,\momlep)\big] \kend \\
		\label{propMajapSp<}
		\pMaj{ij}{<}(\mommaj)&=- g_w\int \lorentzdd{\momhig} 
		\lorentzdd{\momlep}(2\pi)^4 \delta(q-p-k)\pHiggs{}{\rho}(\momhig) 
		\pLeptsc{}{\rho}(\momlep) \nonumber \\ 
		&\times \bigl[f_\lepton^\momlep f_\higgs^\momhig 
		\pMaj{ij}{LC}(\mommaj,\momlep) +f_{\bar \lepton}^\momlep f_{\bar \higgs}^\momhig 
		\pMaj{ij}{LV}(\mommaj,\momlep) \bigr] \kend
	\end{align}
\end{subequations}
where we have defined:
\begin{subequations}
	\label{LVLC}
	\begin{align}
		\label{LC}
		&\pMaj{ij}{LC}=(h^\dagger h)_{ij} 
		\bigl[ (1-\delta^{ij})\pMajdiag{ii}{R}(q)P_L\slashed{\momlep} 
		P_R \pMajdiag{jj}{A}(\mommaj) \\ 
		& + \frac{\delta^{ij}}{2}
		\big( \pMajdiag{ii}{R} (q)P_L\slashed{\momlep} P_R 
		\pMajdiag{jj}{R}(\mommaj) +\pMajdiag{ii}{A}(q)P_L\slashed{\momlep} 
		P_R \pMajdiag{jj}{A}(\mommaj)\big) \bigr]\kend \nonumber \\
		\label{LV}
		&\pMaj{ij}{LV}=(h^\dagger h)_{ji} 
		\bigl[ (1-\delta^{ij})\pMajdiag{ii}{R}(q)P_R\slashed{\momlep} 
		P_L \pMajdiag{jj}{A}(\mommaj) \\ 
		&+ \frac{\delta^{ij}}{2}
		\big(\pMajdiag{ii}{R} (q)P_R\slashed{\momlep} P_L 
		\pMajdiag{jj}{R}(\mommaj) +\pMajdiag{ii}{A}(q)P_R
		\slashed{\momlep} P_L \pMajdiag{jj}{A}(\mommaj)\big) \bigr]\dend \nonumber
	\end{align}
\end{subequations}
Here we neglected higher order terms coming from the matrices $\hat{\Theta}_R$ and $\hat{\Theta}_A$. 
The first terms in \eqns\eqref{LC} and \eqref{LV} corresponds to the second term 
in \eqn\eqref{eQPequ}, whereas the remaining terms correspond to the last two terms 
in \eqn\eqref{eQPequ}. Substituting the Majorana propagators \eqref{propMajapSp} into 
the lepton-current \eqref{MasterEquationWigner} together with the one-loop lepton 
self-energy \eqref{Sigma1WT} we finally obtain
\begin{align}
	\label{lepdivAP}
	{\cal D}_\mu j^\mu_L(t) & = g_w^2 \int \!\lorentzdd{\momlep_1} 
	\lorentzdd{\momhig_1} \lorentzdd{\momlep_2} 
	\lorentzdd{\momhig_2} \nonumber \\ & 
	\times 	\pHiggs{}{\rho}(\momhig_1)\pHiggs{}{\rho}(\momhig_2)\pLeptsc{}{\rho}(\momlep_1)
	\pLeptsc{}{\rho}(\momlep_2) \nonumber \\
	&\times \bigl[A^{LV}(\momlep_1+\momhig_1,\momlep_1,\momlep_2) 
	\F{\lepton\higgs}{\bar\lepton\bar\higgs}{\momlep_1 \momhig_1}{\momlep_2\momhig_2}\nonumber\\
	&+A^{LC}(\momlep_1+\momhig_1,\momlep_1,\momlep_2)
	\F{\lepton\higgs}{\lepton\higgs}{\momlep_1 \momhig_1}{\momlep_2\momhig_2}\bigr]\dend
\end{align}
Note that the zeroth component of the momenta in the above equation can have both signs. 
The effective amplitudes of the lepton number conserving and lepton number violating 
processes read
\begin{subequations}
 \label{traceAP}
 \noeqref{traceAPLV,traceAPLC}
\begin{align}
	\label{traceAPLV}
	A^{LV}(\mommaj,\momlep_1,\momlep_2)\equiv& (h^\dagger h)_{ji}\,
	\tr\bigl[\pMaj{ij}{LV}(q,\momlep_2) P_L \slashed{\momlep}_1 P_R\bigr]\kend \hphantom{aaaa}\\
	\label{traceAPLC}
	A^{LC}(\mommaj,\momlep_1,\momlep_2)\equiv& (h^\dagger h)_{ji}\,
	\tr\bigl[\pMaj{ij}{LC}(q,\momlep_2) P_L \slashed{\momlep}_1P_R\bigr]\dend \hphantom{aaaa}
\end{align}
\end{subequations}
The functions $A^{LC}(\mommaj,\momlep_1,\momlep_2)$ and $A^{LV}(\mommaj,\momlep_1,\momlep_2)$ 
are symmetric under the exchange of the momenta $\momlep_1$ and $\momlep_2$. This implies that 
the contribution of $A^{LC}$ in \eqn\eqref{lepdivAP} vanishes. The terms of $\pMaj{}{LV}$ 
diagonal in flavor space correspond to the RIS-propagator. Substituting \eqn\eqref{SRAsol} 
into \eqn\eqref{traceAPLV} and taking the trace we find that it contains only the scalar 
components of the retarded and advanced propagators and is proportional to: 
\begin{align}
	\label{eQPRIS}
	\pMajdiagsc{2}{R}+\pMajdiagsc{2}{A}&\approx 2 M^2
	\frac{\OmegaDen{2}{h}-\PiDen{2}{\rho}}{\left[\OmegaDen{2}{h}+\PiDen{2}{\rho}\right]^2}\nonumber\\
	&\approx 2M^2\frac{(q^2-M^2)^2-(M\Gammaeff{}{})^2}{\left[(q^2-M^2)^2
	+(M\Gammaeff{}{})^2\right]^2}\dend
\end{align}
Equation~\eqref{eQPRIS} differs from the canonical result \eqref{RISpropagator} only in that 
the vacuum masses and decay widths are replaced by thermal ones. For a hierarchical mass spectrum 
this difference can be safely neglected. Introducing an analogue of the RIS subtracted propagator,
\begin{align}
	\label{KB_RISAP}
	\mathscr{P}_{ij}(q)\equiv 
	\pMajdiagsc{ii}{A}(q)\pMajdiagsc{jj}{R}(q) -{\textstyle\frac12} \delta^{ij} \pMajdiagsc{ii}{\rho}(q)\pMajdiagsc{ii}{\rho}(q)\kend
\end{align}
we can rewrite the lepton number violating effective amplitude in a compact form:
\begin{align*}
	A^{LV}(\mommaj,\momlep_1,\momlep_2)&=2(h^\dagger h)_{ij}^2(\momlep_1 \momlep_2)M_i M_j \mathscr{P}_{ij}(q)\dend 
\end{align*}

Next we perform the trivial integrations over the 
frequencies using the Dirac-deltas in the quasiparticle spectral functions \eqref{leptonQPspectralfunc} and \eqref{QPforHiggs}. 
Each Dirac-delta can be decomposed into two terms, one with positive and one with negative 
frequency. Therefore, the integration over the four frequencies gives rise to $2^4$ terms, but only 
six of them satisfy energy conservation ensured by the remaining delta-function. In a homogeneous 
and isotropic medium the one-particle distribution functions satisfy 
\begin{align}
	\label{oddp0}
	1-f_\lepton(-\momlep)=f_{\bar \lepton}(\momlep) \kend \quad 1+f_\higgs(-\momhig)=-f_{\bar \higgs}(\momhig)\kend
\end{align}
and the diagonal Majorana propagators have the properties
\begin{align}
	\label{symAR}
	\pMajdiagsc{ii}{\rho}(-q)=-\pMajdiagsc{ii}{\rho}(q)\kend\quad
	\pMajdiagsc{ii}{R}(-q)=\pMajdiagsc{ii}{A}(q)\dend
\end{align}
Upon substitution of the resulting self-energy into \eqn\eqref{MasterEquationWigner}
and the use of \eqns\eqref{oddp0} and \eqref{symAR} the remaining six contributions in 
the lepton-current can be conveniently written as
\begin{align}
	\label{SEScattering}
	{\cal D}_\mu j_L^{\mu}(t)\Big|_{S}^{S} =&- \int
	\dpi{\lepton\lepton}{\higgs\higgs}{\momlep_1\momlep_2}{\momhig_1 \momhig_2}\nonumber \\ 
	&\times\Big[ 
	\EffAmpl{(t\times t)}{\bar \higgs \bar \higgs}{\lepton \lepton} 
	\F{\bar \higgs \bar \higgs}{\lepton \lepton}{\momhig_1 \momhig_2}{\momlep_1 \momlep_2}
	+ 
	\EffAmpl{(t\times t)}{\bar \lepton \bar \lepton}{\higgs \higgs}
	\F{\bar \lepton \bar \lepton}{\higgs \higgs}{\momlep_1 \momlep_2}{\momhig_1 \momhig_2} \nonumber \\ 
	&+ 2 \big( 
	\EffAmpl{(s\times s)}{\bar \lepton \bar \higgs}{\lepton \higgs} 
	+\EffAmpl{(t\times t)}{\bar \lepton \bar \higgs}{\lepton \higgs} \big)
	\F{\bar \lepton \bar \higgs}{\lepton \higgs}{\momlep_1 \momhig_1 }{\momlep_2 \momhig_2} \Big]\kend
\end{align}
where we have defined the effective scattering amplitudes:
\begin{subequations}
 \label{M1}
 \noeqref{LHLHss,LHLHtt,LLHHtt,HHLLtt}
\begin{align}
	\label{LHLHss}
	\EffAmpl{(s\times s)}{\bar \lepton \bar \higgs}{\lepton \higgs}&= 
	2 g_w^2 (\momlep_1 \momlep_2) \sum \Re (h^\dagger h)_{ij}^2 M_i M_j \mathscr{P}_{ij}(q_s)\kend
	\hphantom{aaaa}\\
	\label{LHLHtt}
	\EffAmpl{(t\times t)}{\bar \lepton \bar \higgs}{\lepton \higgs} &= 
	2 g_w^2 (\momlep_1 \momlep_2) \sum \Re (h^\dagger h)_{ij}^2 M_i M_j \mathscr{P}_{ij}(q_t)\kend \\
	\label{LLHHtt}
	\EffAmpl{(t\times t)}{\bar \lepton \bar \lepton}{\higgs \higgs}&= 
	 2 g_w^2 (\momlep_1 \momlep_2) \sum(h^\dagger h)_{ij}^2 M_i M_j \mathscr{P}_{ij}(q_t)\kend \\
	\label{HHLLtt}
	\EffAmpl{(t\times t)}{\bar \higgs \bar \higgs}{\lepton \lepton}&= 
	 2 g_w^2 (\momlep_1 \momlep_2) \sum (h^\dagger h)_{ji}^2 M_i M_j \mathscr{P}_{ij}(q_t)\dend 
\end{align}
\end{subequations}
The momenta of the Majorana neutrinos are related to the momenta of the initial and final states 
by $q_s=\momlep_1+\momhig_1$ and $q_t=\momlep_1-\momhig_2$. From \eqn\eqref{M1} we see that the 
obtained amplitudes contain only $s\times s$ and $t\times t$ interference terms. Indeed, in the 
products of the Majorana propagators in \eqn\eqref{KB_RISAP} both of them depend on the same momentum. 

The missing cross terms emerge from the two-loop (vertex) contribution to the lepton self-energy.
As we have mentioned in \sect\ref{NEQFTapproach}, within the discussed assumptions and 
approximations the second and third terms on the right-hand side of \eqn\eqref{SE3pMT} 
describe scattering processes. Since they are of the fourth order in the 
Yukawas, we can replace the full Majorana propagators by the diagonal propagators.
The third term, \eqn\eqref{SeLCMT}, corresponds to lepton number conserving processes and 
does not need to be discussed further. The second one, \eqn\eqref{SeScaMT}, is given by
\begin{align}
	\label{SeSca2}
	\sLept{(2.2)}{\gtrless}&(t,\momlep) = \int \lorentzdd{\momlep_2} 
	\lorentzdd{\momhig_1} \lorentzdd{\momhig_2}(2\pi)^4\delta(\momlep+\momhig_1-\momlep_2-\momhig_2)\nonumber\\
	\times & (\yu^\dagger \yu)^2_{ij} 
	\bigl[P_R\pMajdiag{jj}{R}(t,\momlep_2+\momhig_2) CP_R\pLept{T}{\lessgtr}(t,-\momlep_2)P_L\nonumber\\
	 \times & C\pMajdiag{ii}{A}(t,\momlep_2-\momhig_1)P_L \pHiggs{}{\lessgtr}(t,\momhig_1)
	\pHiggs{}{\lessgtr}(t,-\momhig_2)\bigr]\dend
\end{align}
We substitute \eqn\eqref{SeSca2} into the equation for the lepton-current \eqref{MasterEquationWigner} 
and perform the steps preceding \eqn\eqref{SEScattering}. Using furthermore relations \eqref{oddp0}
we find
\begin{align}
 \label{VertScattering}
	{\cal D}_\mu j^{\mu}_L(t)\Big|_{S}^{V}= &- \int 
	\dpi{\lepton\higgs}{\lepton\higgs}{\momlep_1 \momhig_1}{\momlep_2 \momhig_2} \nonumber\\
	& \times \Bigl[ 2 \EffAmpl{(s\times t)}{\bar \lepton \bar \higgs}{\lepton \higgs} 
	\F{\bar \lepton \bar \higgs}{\lepton \higgs}{\momlep_1 \momhig_1 }{\momlep_2 \momhig_2}
	+ \EffAmpl{(u\times t)}{\bar \higgs \bar \higgs}{\lepton \lepton} 
	\F{\bar \higgs \bar \higgs }{\lepton \lepton}{\momhig_1 \momhig_2 }{\momlep_1 \momlep_2}\nonumber\\ 
	& +\EffAmpl{(u\times t)}{\bar \lepton \bar \lepton}{\higgs \higgs} 
	\F{\bar \lepton \bar \lepton }{ \higgs \higgs}{\momlep_1 \momlep_2 }{\momhig_1 \momhig_2} \Bigr]\kend
\end{align}
where we have introduced 
\begin{subequations}
\label{M2}
\noeqref{LHLHst,LLHHut,HHLLut}
\begin{align}
\label{LHLHst}
	 \EffAmpl{(s\times t)}{\bar \lepton \bar \higgs}{\lepton \higgs} &\equiv 
	 4g_w (\momlep_1\momlep_2) \sum M_iM_j \Re\, (h^{\dagger}h)_{ij}^2 \nonumber\\
	 & \times \Re\,\bigl\{\pMajdiagsc{ii}{A}(q_t)\pMajdiagsc{jj}{R}(q_s)\bigr\}\kend\\
\label{LLHHut}
	 \EffAmpl{(u\times t)}{\bar \lepton \bar \lepton}{\higgs \higgs} &
	 \equiv 2 g_w (\momlep_1\momlep_2) \sum M_iM_j \nonumber\\
	 &\times \Re \bigl\{(h^{\dagger}h)_{ij}^2\, \pMajdiagsc{ii}{A}(q_t)\pMajdiagsc{jj}{R}(q_u)\bigr\}\kend\\
\label{HHLLut}
	\EffAmpl{(u\times t)}{\bar \higgs \bar \higgs}{\lepton \lepton} & 
	\equiv 2 g_w (\momlep_1\momlep_2) \sum M_iM_j \nonumber \\
	 &\times \Re \bigl\{(h^{\dagger}h)_{ji}^2\, \pMajdiagsc{ii}{A}(q_u)\pMajdiagsc{jj}{R}(q_t)\bigr\}\kend
\end{align}
\end{subequations}
and $q_u=\momlep_2-\momhig_2$.
The combinations of the momenta appearing in the products of the Majorana propagators 
clearly indicate that the above amplitudes correspond to the interference terms of the 
$s$-, $t$-, and $u$-channel contributions. 
 
Combining \eqns\eqref{M1} and \eqref{M2} we obtain for the effective amplitude 
of $\lepton \lepton \leftrightarrow \bar\higgs \bar\higgs$ scattering:
\begin{align}
	\label{LLHHKB}
	\EffAmpl{}{\lepton\lepton}{\bar\higgs\bar\higgs}&= 2 (\momlep_1 \momlep_2)\sum 
	(h^\dagger h)^2_{ij} M_i M_j \nonumber\\
	&\times \bigl[ 4\mathscr{P}_{ij}(q_t)\nonumber\\
	&+ \pMajdiagsc{ii}{A}(q_t)\pMajdiagsc{jj}{R}(q_u)+
	\pMajdiagsc{ii}{A}(q_u)\pMajdiagsc{jj}{R}(q_t)\bigr]\kend
\end{align}
whereas the effective amplitude of $\lepton\higgs\leftrightarrow \bar \lepton \bar \higgs$ 
scattering reads
\begin{align}
	\label{LHLHKB}
	\EffAmpl{}{\lepton\higgs}{\bar\lepton\bar\higgs} 
	&=4(\momlep_1 \momlep_2) \sum \Re (h^\dagger h)^2_{ij} M_i M_j \nonumber\\
	&\times\bigl[ 2\mathscr{P}_{ij}(q_s)+2\mathscr{P}_{ij}(q_t)\nonumber\\
	&+ \pMajdiagsc{ii}{A}(q_t)\pMajdiagsc{jj}{R}(q_s) 
	+\pMajdiagsc{ii}{A}(q_s)\pMajdiagsc{jj}{R}(q_t)\bigr]\dend
\end{align}
To compare the obtained expressions with the vacuum results of the 
canonical computation \eqns\eqref{LHLH} and \eqref{LLHH}, we evaluate the retarded and advanced propagators
at zero temperature. In vacuum $L_\rho(t,\mommaj)=\theta(q^2)\,
\sign(\mommaj^0)\,\slashed{\mommaj}$. Therefore for positive $q_0$ it follows 
from \eqn\eqref{diagRA} that
\begin{align}
	\label{SRAvac}
	\pMajdiagsc{ii}{R/A}(q)\approx-\left[q^2-M_i^2\pm i\theta(q^2) \Gamma_i/M_i\,q^2\right]^{-1}\dend
\end{align}
In the vicinity of the mass shell of the respective Majorana neutrino $q^2=M_i^2$ and we find
\begin{align}
	\label{SRAPrel}
	\pMajdiagsc{ii}{R}=-P_i\kend\quad \pMajdiagsc{ii}{A}=-P^*_i\dend
\end{align}
For $t$- and $u$-channels the imaginary parts of \eqns\eqref{InversePropagator} 
and \eqref{SRAvac} vanish, so that in the vacuum limit
\begin{align}
	\mathscr{P}_{ij}(q)=P^*_i(q^2)P_j(q^2)\kend 
\end{align}
and $\mathscr{P}_{ij}$ is symmetric with respect to $i\leftrightarrow j$. The squares of the 
$t$- and $u$-channel propagators in \eqn\eqref{LLHH} give identical contributions 
to the reduced cross section of $\lepton\lepton\leftrightarrow\bar\higgs\bar\higgs$ 
process. Therefore, upon substitution of \eqn\eqref{LLHHKB} into 
\eqn\eqref{eqn:reduced cross section definition main text}
and the use of the $i\leftrightarrow j$ symmetry, we recover for the reduced cross section 
the canonical result \eqref{sigmaLLHH}.
Relations \eqref{SRAPrel} imply that the off-diagonal components of $\mathcal{P}_{ij}$ and 
$\mathscr{P}_{ij}$ coincide. For the $s$-channel the diagonal components of $\mathscr{P}_{ii}$ 
are given in the vacuum limit by
\begin{align}
	\mathscr{P}_{ii}(q_{s})=\frac{(s-M_i^2)^2-(\Gamma_i/M_i\,s)^2}{[(s-M_i^2)^2
	+(\Gamma_i/M_i\,s)^2]^2}\dend
\end{align}
Thus, in the vicinity of the mass shell, $s\approx M_i^2$, the canonical expression for the RIS-sub\-tracted 
propagator, \eqn\eqref{RISpropagator}, coincides with the expression obtained from first 
principles. This justifies results of the earlier calculations. For the reduced cross section 
of $\lepton\higgs\leftrightarrow\bar\lepton\bar\higgs$ scattering we recover \eqn\eqref{sigmaLHLH}.

At finite temperatures $L_\rho(t,\slashed{\mommaj})$ is not zero even for $t$- and $u$-channels. 
In other words, the medium effects induce additional contributions to the effective decay amplitudes. 
However, these contributions are proportional to the coefficients $c_i$. Numerical analysis shows 
that for the two chosen sets of parameters, see \app\ref{NumericalParameters}, the additional correction 
typically do not have any sizable impact on the reaction densities.

\begin{figure}[h!]
	\includegraphics[width=\columnwidth]{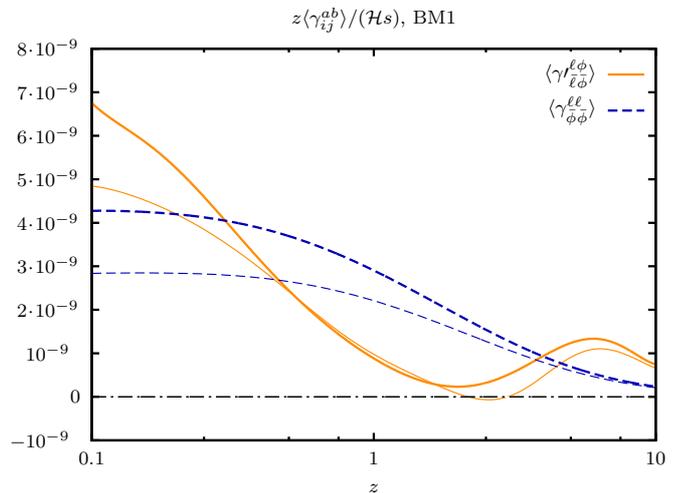}
	\caption{\label{BlzmnVsFDBE1}
	Washout reaction densities due to $\lepton\lepton\leftrightarrow\bar\higgs\bar\higgs$ 
	and $\lepton\higgs\leftrightarrow\bar\lepton\bar\higgs$ scattering processes for 
	benchmark point~1 in the approximation of massless leptons and Higgs. Shown are the 
	reaction densities computed using the Boltzmann approximation (\ref{ReactDensCanon})
	(thin lines), and taking into account the quantum-statistical effects, 
	(\ref{ReactDensExact}) (thick lines). The (RIS subtracted) reaction densities 
	$\langle\gamma'^{\lepton\higgs}_{\bar{\lepton}\bar{\higgs}}\rangle$ may be negative 
	as they are not the physical rates for the $2\leftrightarrow 2$ scattering process.}
\end{figure}

\begin{figure}[h!]
	\includegraphics[width=\columnwidth]{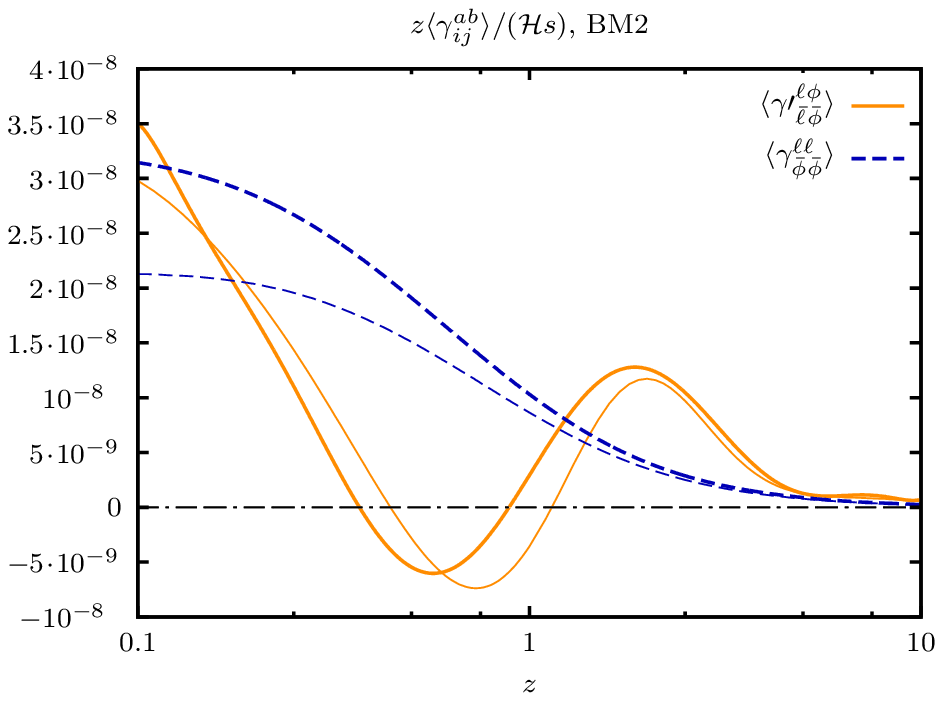} 
	\caption{\label{BlzmnVsFDBE2}
	Washout reaction densities due to $\lepton\lepton\leftrightarrow\bar\higgs\bar\higgs$ 
	and $\lepton\higgs\leftrightarrow\bar\lepton\bar\higgs$ scattering processes for benchmark 
	point~2. Compare \fig\ref{BlzmnVsFDBE1}.}
\end{figure}

A quantity relevant for the numerical analysis is the ratio
$\x\langle\gamma_{ij}^{ab}\rangle/\hubbleratem s$. The dependence of this ratio
on the dimensionless inverse temperature is presented in 
\fig\ref{BlzmnVsFDBE1} and \fig\ref{BlzmnVsFDBE2}. If the approximate expression 
\eqref{ReactDensCanon} is used, then the reaction density of 
$\lepton\higgs\leftrightarrow\bar\lepton\bar\higgs$ scattering becomes
negative for $2\lesssim z \lesssim 3$ for the first set of the parameters 
whereas for the second set of the parameters it turns negative for 
$0.5\lesssim z \lesssim 1$. A qualitatively similar behavior has also been 
observed in \cite{Pilaftsis:2003gt}. 
This rather counter-intuitive result can be traced back
to the behavior of the RIS part of the effective amplitude \eqref{LHLH}
which is negative in the vicinity of the mass shell. Its sign is not fixed by physical
requirements, since it constitutes a sub-leading contribution to the washout rate.
As can be inferred from \fig\ref{BlzmnVsFDBE1}, for the first set of model 
parameters the quantum-statistical corrections render the RIS subtracted reaction 
density of $\lepton\higgs\leftrightarrow\bar\lepton\bar\higgs$ scattering positive 
in the whole range of temperatures. However, this is merely a numerical coincidence. 
For the second set of parameters, see \fig\ref{BlzmnVsFDBE2}, the 
reaction density of $\lepton\higgs\leftrightarrow\bar\lepton\bar\higgs$
scattering remains negative for $0.4\lesssim z \lesssim 0.9$. It is nevertheless 
important to note that even in the region where the reaction density is negative 
the quantum-statistical corrections shift it upwards as compared to the result of 
the canonical computation.

As far as $\bar \lepton \bar \lepton\leftrightarrow \higgs \higgs$ scattering 
is concerned, the quantum-statistical corrections enhance the reduced reaction
density at high temperatures by about 50\%. As the temperature decreases, the 
reaction density computed using \eqn\eqref{ReactDensExact} approaches the 
one computed using \eqn\eqref{ReactDensCanon} as one would expect. A similar behavior 
is also observed for the reaction density of $\lepton\higgs\leftrightarrow\bar\lepton\bar\higgs$
scattering. 

In \figs\ref{BlzmnVsFDBE1} and \ref{BlzmnVsFDBE2} we neglected the thermal 
masses of initial and final states. 
\begin{figure}[h!]
	\includegraphics[width=\columnwidth]{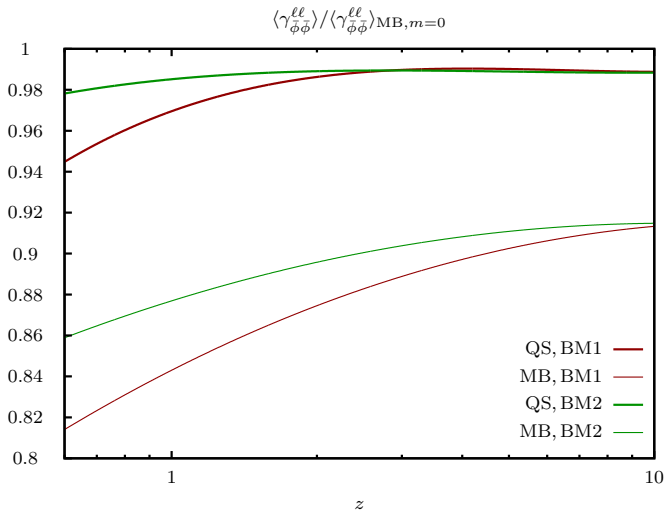} 
	\caption{\label{ScatteringWithMasses}
	Ratio of the reaction density computed using thermal masses with (thick lines) and 
	without (thin lines) the quantum-statistical terms to the canonical one for the two 
	sets of parameters.}
\end{figure} 
To estimate the size of the mass corrections, in \fig\ref{ScatteringWithMasses} we plot ratio 
of the reaction density of the $\lepton\lepton\leftrightarrow\bar\higgs\bar\higgs$ scattering 
computed using thermal masses of the Higgs and leptons with and without the quantum-statistical 
terms to the canonical one for the two sets of parameters. If the quantum-statistical effects 
are neglected, the thermal masses lead to a $\sim 15\%$ suppression of the reaction 
densities. On the other hand, an enhancement induced by the quantum-statistical terms to a 
large extent compensates the mass-induced suppression. As a result, the deviation from the 
canonical reaction density does not exceed $\sim 5\%$ in the whole range of temperatures.

\begin{figure}[h!]
	\includegraphics[width=\columnwidth]{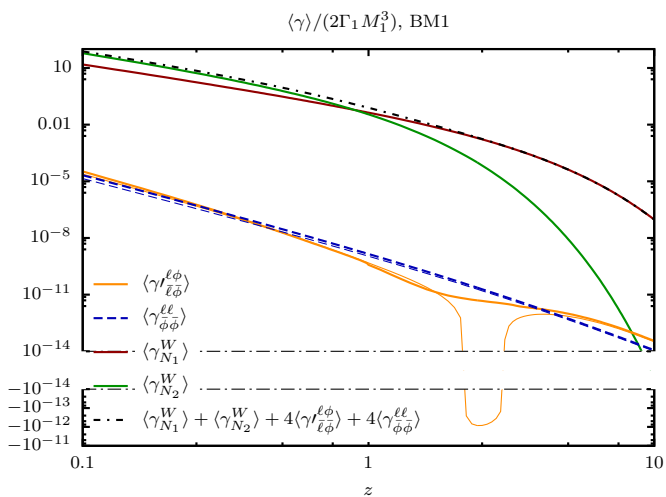} 
	\caption{\label{WashoutRDBM1}
	Washout reaction densities due to $\lepton\lepton\leftrightarrow\bar\higgs\bar\higgs$ 
	and $\lepton\higgs\leftrightarrow\bar\lepton\bar\higgs$ scattering processes for benchmark 
	point~1 in the approximation of massless leptons and Higgs. For comparison the washout 
	reaction densities for $\majneutrino_1$ and $\majneutrino_2$ (inverse) decays are shown
	as well. Note that the normalization differs from the one used in \fig\ref{BlzmnVsFDBE1} 
	and \fig\ref{BlzmnVsFDBE2} For the present choice of parameters contributions by 
	$2\leftrightarrow 2$ scatterings are strongly suppressed by the smallness of the couplings.
	}
\end{figure}

\begin{figure}[h!]
	\includegraphics[width=\columnwidth]{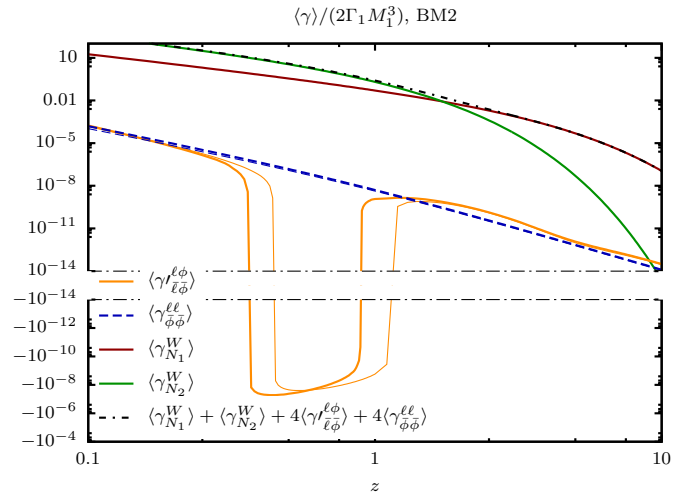} 
	\caption{\label{WashoutRDBM2}
	Washout reaction densities due to 
	$\lepton\lepton\leftrightarrow\bar\higgs\bar\higgs$ and 
	$\lepton\higgs\leftrightarrow\bar\lepton\bar\higgs$ scattering processes as well 
	as (inverse) decays for benchmark point~2. Compare \fig\ref{WashoutRDBM1}.
	}
\end{figure}

To compare the relative importance of the (inverse) decay and scattering processes in 
\fig\ref{WashoutRDBM1} and \fig\ref{WashoutRDBM2} we also present the uniformly 
normalized reaction densities.
In both cases we observe a qualitatively similar picture: for the chosen sets of parameters 
reaction densities of the scattering processes are strongly suppressed by the smallness of 
the Yukawa couplings as compared to those of the decay processes.

In the rate equation for the lepton asymmetry, see \eqn\eqref{RateEquationsa}, the total washout 
rate is given by a sum of reaction densities for decays and scattering. Whereas 
$\langle \gamma^W_{\majneutrino_i}\rangle$ and $\langle \gamma^{\lepton\lepton}_{\bar\higgs\bar\higgs}\rangle$
are positive, the RIS subtracted reaction density $\langle \gamma'^{\lepton\higgs}_{\bar\lepton\bar\higgs}\rangle$ 
can be negative at some temperatures. If the total washout rate would turn negative, it would 
lead to a spurious self-enhanced generation of the asymmetry. In \figs\ref{WashoutRDBM1} and 
\ref{WashoutRDBM2} we show the sum of the reactions densities. For both parameter sets it is positive. 
This sum should always be positive. If the quantum-statistical terms 
are neglected this can be demonstrated explicitly. In the absence of the quantum-statistical
corrections, the results of this section revert to the ones discussed in \sect\ref{ConventionalApproach}.
Using the expression for the RIS-propagator, \eqn\eqref{RIS_zero_width}, we can rewrite the RIS subtracted 
scattering amplitude \eqref{LHLH} as a difference of the unsubtracted one and the RIS term. 
Since the latter is proportional to $\delta(s-M_i^2)$ the integration in 
\eqns\eqref{eqn:reduced cross section definition main text}
and \eqref{ReactDensCanon} is trivial and we obtain after some algebra: 
\begin{align}
\label{UnsubtractedReactDens}
4\langle \gamma'^{\lepton\higgs}_{\bar\lepton\bar\higgs}\rangle\approx
4\langle \gamma^{\lepton\higgs}_{\bar\lepton\bar\higgs}\rangle
-{\textstyle \sum_i} \langle\gamma^W_{\majneutrino_i}\rangle \dend
\end{align}
The total washout rate is then given by $4\langle \gamma^{\lepton\higgs}_{\bar\lepton\bar\higgs}\rangle+
4\langle \gamma^{\lepton\lepton}_{\bar\higgs\bar\higgs}\rangle$ and 
is positive as a sum of two positive functions. In other words, even though
$\langle \gamma'^{\lepton\higgs}_{\bar\lepton\bar\higgs}\rangle$ can be negative 
at some temperatures, the total washout rate is always positive. The form of the 
unsubtracted scattering reaction density that can be inferred from 
\eqn\eqref{UnsubtractedReactDens} is another manifestation of the double-counting. 
If we had not subtracted the RIS contribution, we would have 
counted contributions of the inverse decay processes twice and ended up with an 
incorrect prediction for the generated asymmetry.

\section{\label{HiggsDecay} Higgs contribution}

In the preceding section we have approximately taken the gauge interactions into 
account in the form of effective masses of the Higgs and leptons.\footnote{We will not attempt a fully consistent inclusion of
gauge interactions here. Their effect has been addressed systematically for the production rate of Majorana neutrinos
in various temperature regimes in~\cite{Anisimov:2010gy,Laine:2011pq,Besak:2012qm,Laine:2012nq}, 
but not for \CP-violating rates up to now. In this work we model
gauge interactions in a simplified way by including thermal masses, similar as has been done previously
in the context of thermal field theory computations~\cite{Giudice:2003jh}. In this way the results obtained within
NEQFT can be compared to previous computations, and may ultimately be compared to a full treatment of
gauge interactions within NEQFT.}
The thermal masses are of order of $g T$ and large enough to influence the values of the reaction densities quantitatively. 
In particular, the Majorana neutrino decay can become kinematically forbidden when the sum of the masses of lepton 
and Higgs exceeds the heavy Majorana neutrino mass. For even higher temperatures (with 
$\m{\higgs}(T)>\m{\lepton}(T)+{M_i}(T)$) the Higgs decay channel into a lepton-Majorana pair becomes 
kinematically allowed instead and can contribute to the asymmetry since it violates {\CP}. 
For simplicity we do not take modified dispersion relations into account here, see~\cite{Weldon:1989pa158,Giudice:2003jh,Kiessig:2010pr,Kiessig:2011fw,Kiessig:2011ga}, but use the simple picture of temperature 
dependent thermal masses as an estimate:
\begin{align}
\label{thermalmasses}
	\m{\higgs}^2 &=\bigg(\frac{3}{16} \gtwo^2 + \frac{1}{16} \gprime^2 + \frac{1}{4} \gyt^2 
	+ \frac12 \lambda\bigg){T^2}, \nonumber\\
	\m{\lepton}^2 &=\bigg(\frac{3}{32} \gtwo^2 + \frac{1}{32} \gprime^2\bigg){T^2}\kend
\end{align}
where we use the temperature dependent values of the $SU(2)_L$, $U(1)$, top Yukawa and Higgs 
self-couplings $\gtwo$, $\gprime$, $\gyt$ and $\lambda$ assuming a Higgs mass of $115\GeV$ \cite{Kajantie:1995dw}. We also ignore that the thermal mass of leptons might be better
approximated by the `asymptotic thermal mass' $\sqrt{2}\m{\lepton}$ in a kinematic regime in 
which their momentum is such that $\momlep^2\sim (gT)^2$ \cite{Weldon:1982bn}. We do also not take into account, 
in our quantitative analysis, the thermal corrections to the Majorana neutrino masses as 
they are negligible compared to their vacuum masses.\footnote{Note that they may be 
relevant in the different context of resonant leptogenesis if the size of the thermal correction 
is comparable to the mass splittings $\abs{M_i -M_j}
$ as they can influence the resonance in this case.}

\begin{figure}[!ht]
	\centering
	\includegraphics[width=\columnwidth]{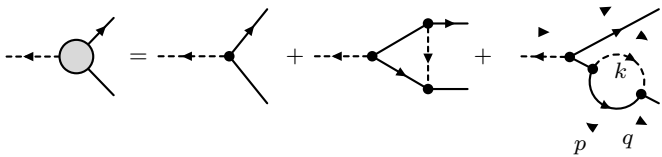}
	\caption{\label{HiggsDecayPlot} Tree level contribution and one-loop corrections 
	to the (anti-)Higgs decay amplitude. The additional arrows illustrate the 
	 direction of momentum flow.}
\end{figure}
The {\CP}-violating decay of the Higgs arises from the interference of the
tree-level, self-energy and vertex graphs depicted in \fig\ref{HiggsDecayPlot}.
Although in this case the decaying particle -- the Higgs doublet --
is very close to thermal equilibrium due to the Yukawa and gauge interactions
of the Standard Model, the Majorana neutrino in the final state
may deviate from equilibrium, so that the third Sakharov condition is fulfilled.

Using the expression for the divergence of the lepton-current, \eqn\eqref{MasterEquationWigner}, 
we can extract the corresponding {\CP}-violating parameter. To calculate 
the self-energy contribution it is convenient to 
rewrite the one-loop lepton self-energy \eqref{Sigma1WT} in the form: 
	\begin{align}
	\label{LSelfEnergyForHiggs}	
		\sLept{}{\gtrless}(t,\momlep)&=-(\yu^\dagger\yu)_{ji}
		{\textstyle\int} \lorentzdd{\momhig} \lorentzdd{\mommaj} 
		(2\pi)^4\deltafour{\mommaj+\momlep-\momhig}\nonumber\\
		&\times \pHiggscp{}{\gtrless}(t,\bar \momhig)P_R C 
		(\pMaj{ji}{\lessgtr})^T(t,\mommaj)C^{-1}P_L\kend
	\end{align} 
where the transposition is only in Dirac space and we have used 
one of the properties of Majorana propagator: 
\begin{align}
	\label{MajoranaTransposition}
	(\pMaj{ij}{\gtrless})(X,\mommaj)=C(\pMaj{ji}{\lessgtr})^T(X,-\mommaj)C^{-1}\dend
\end{align}
In its {\CP} conjugate the Yukawa couplings are replaced by their complex
conjugates and the propagators by the {\CP} conjugate ones. 

Similar to the case of the Majorana decay, substituting \eqn\eqref{LSelfEnergyForHiggs} 
into \eqn\eqref{MasterEquationWigner} we can define effective Higgs decay amplitudes: 
\begin{subequations}
	\label{HiggsDecayAmplitides}
	\noeqref{HiggsDecayAmplitidesLN,HiggsDecayAmplitidesbarLN}
	\begin{align}
		\label{HiggsDecayAmplitidesLN}
		\EffAmpl{}{\bar \higgs}{\lepton\majneutrino_i}&\equiv 
		g_w\sum_{mn} (h^\dagger h)_{mn}\nonumber\\
		&\times \tr\left[\MatTheta{mi}{R}(\mommaj)
		(\slashed{\mommaj}+M_i)\MatTheta{in}{A}(\mommaj)
		P_R\slashed{\momlep}P_L\right]\kend\hphantom{aa}\\
		\label{HiggsDecayAmplitidesbarLN}
		\EffAmpl{}{\higgs}{\bar \lepton\majneutrino_i}&\equiv 
		g_w\sum_{mn} (h^\dagger h)^*_{mn}\nonumber\\
		& \times \tr\left[\MatThetacp{mi}{R}(\mommaj)
		(\slashed{\mommaj}+M_i)\MatThetacp{in}{A}(\mommaj)
		P_R\slashed{\momlep}P_L\right]\dend
	\end{align}
\end{subequations}
The overall factor $g_w$ in \eqns\eqref{HiggsDecayAmplitides} comes from summation 
over the doublet components of the (decaying) Higgs particle. To leading order 
in the couplings: 
\begin{subequations}
	\label{HiggsAmplitudes}
	\noeqref{HiggsAmplitude,AntihiggsAmplitude}
	\begin{align}
		\label{HiggsAmplitude}
		\EffAmpl{}{\bar \higgs}{\lepton\majneutrino_i}& \approx
		2g_w\bigl[(h^\dagger h)_{ii}(\momlep\cdot \mommaj)\nonumber\\
		& + {\textstyle\frac{g_w}{16\pi}}
		\Im(h^\dagger h)_{ij}^2 M_i M_j \pMajdiagsc{jj}{h} (\momlep\, L_\rho)\bigr]\kend\\
		\label{AntihiggsAmplitude}
		\EffAmpl{}{\higgs}{\bar\lepton\majneutrino_i} & \approx
		2g_w\bigl[(h^\dagger h)_{ii}(\momlep\cdot \mommaj)\nonumber\\
		& - {\textstyle\frac{g_w}{16\pi}}
		\Im (h^\dagger h)_{ij}^2 M_i M_j \pMajdiagsc{jj}{h} (\momlep\, L_\rho)\bigr]\kend
	\end{align}
\end{subequations}
which, up to the relative sign in square brackets, coincides with the amplitudes 
\eqref{EffectiveDecayAmplitudesSE}. The corresponding \CP-violating parameter reads 
\begin{align}
	\label{epsilonhiggs}
	\epsilon_{\higgs,i}^S=-\frac{\Im(h^\dagger h)^2_{ij}}{(h^\dagger h)_{ii}(h^\dagger h)_{jj}}
	\frac{M_i\Gamma_j}{M_i^2-M_j^2}\frac{\momlep L_\rho}{\momlep \mommaj}\kend
\end{align}
where $\momlep$ and $\mommaj$ are the momenta of the on-shell final lepton and Majorana
neutrino with positive zeroth components respectively. The direction of momentum flow is 
as defined in \fig\ref{HiggsDecayPlot}. Although $L_\rho$ in 
\eqns\eqref{SEepsEpsilon} and \eqref{epsilonhiggs} is one and the same function, because of the 
different kinematic regimes the explicit result in terms of the distribution functions differs 
for the Higgs decay: 
\begin{align}
	\label{LRhoHiggs}
	L_\rho(t,\mommaj)= 16\pi\int & \lorentzd{\higgs}{\momhig} \lorentzd{\lepton}{\momleploop} 
	(2\pi)^4 \delta(\mommaj+\momlep-\momhig)\,\slashed{\momlep}\notag \\
	\times & [\,f_\higgs(\e{\higgs}{\momhig})+f_\lepton(\e{\lepton}{\momleploop})]\kend
\end{align}
see \app\ref{MajoranaSelfEn}.
Note that our result is different from the one presented in \cite{Covi:1997dr,Giudice:2003jh,Davidson:pr2008}.
Instead of the $\f{\higgs}{}-\f{\lepton}{}-2\f{\higgs}{} \f{\lepton}{}$ dependence, it 
is proportional to a sum, $f_\higgs+f_\lepton$, of the two distribution 
functions. This dependence can also be obtained in the framework of
real time thermal field theory using causal $n$-point functions, compare \cite{Garny:2010nj}. 
The derivation within the Kadanoff-Baym formalism gives certainty concerning the sign of the 
contribution by Higgs decay. The \CP-violating parameter \eqref{epsilonhiggs} has an opposite 
sign relative to that for Majorana neutrino decay. However, it is canceled by the relative 
sign in \eqn\eqref{EpsilonThermalHiggs}.

To calculate the vertex contribution we use \eqn\eqref{MajoranaTransposition}
and represent the two-loop self-energy \eqref{Sigma2WTExpansion} in the form
\begin{align}
	\label{VertexHiggsDecay}
	&\sLept{(2.1)}{\gtrless}(t,\momlep) = \int \lorentzdd{\mommaj} \lorentzdd{\momhig}\
	(2\pi)^4\delta(\momlep+\mommaj-\momhig) \\
	&\times\big[ (\yu^\dagger \yu)^2_{ij}
	\varLambda_{jj}(t,-q,-k)P_L (\pMajEQP{ii}{\lessgtr})^T(t,\mommaj) 
	C P_L\pHiggscp{}{\gtrless}(t,\bar\momhig)\nonumber\\
	&+ (\yu^\dagger \yu)^2_{ji}
	P_R C(\pMajEQP{ii}{\lessgtr})^T(t,\mommaj) P_R V_{jj}(t,-\mommaj,-\momlep)
	\pHiggscp{}{\gtrless}(t,\momhig)\big] \nonumber\dend
\end{align}
Its {\CP} conjugate again differs by the conjugation of the Yukawas and 
propagators. Substituting \eqn\eqref{VertexHiggsDecay} and its {\CP} conjugate 
into \eqn\eqref{MasterEquationWigner} we obtain for the corresponding 
effective amplitudes:
\begin{subequations}
\label{HiggsVertexAmplitudes}
\noeqref{EffAmplHLN,EffAmplbarHbarLN}
\begin{align}
	\label{EffAmplHLN}
	&\EffAmpl{V}{\bar\higgs}{\lepton N_i} \equiv g_w(h^{\dagger}h)_{ij}^2\ M_i\,
	\tr\bigl[\varLambda_{jj}(-q,-k) C P_L\slashed{\momlep} P_R\bigr]\nonumber\\
	&\hspace{3mm}+g_w(h^{\dagger}h)_{ji}^{2}\,M_i\, 
	\tr\bigr[C V_{jj}(-q,-k)P_L\slashed{\momlep} P_R\bigr]\kend\\
	\label{EffAmplbarHbarLN}
	&\EffAmpl{V}{\higgs}{\bar \lepton N_i} 
	\equiv g_w(h^{\dagger}h)_{ij}^2\ M_i
	\tr\bigr[C V_{jj}(-q,-k)P_L\slashed{\momlep} P_R\bigr]\nonumber\\
	&\hspace{3mm}+g_w(h^{\dagger}h)_{ji}^{2}\, M_i \tr\bigl[\varLambda_{jj}(-q,-k)
	C P_L\slashed{\momlep} P_R\bigr]\dend
\end{align}
\end{subequations}
Just like for the self-energy contribution we observe that the overall sign of the vertex 
contribution to the Higgs decay amplitude is opposite to that in the Majorana decay, compare 
\eqns\eqref{MajoranaVertexAmplitudes} and \eqref{HiggsVertexAmplitudes}. The 
corresponding \CP-violating parameter reads
\begin{align}
	\label{HiggsCPVertex}
	\epsilon_{\higgs,i}^V &= \sum \frac{\Im\,(h^\dagger h)_{ij}^2}{(h^\dagger h)_{ii}}
	\frac{M_iM_j}{\mommaj\momlep}
	\ \int \lorentzdd{k_1} \lorentzdd{k_2} \lorentzdd{k_3} \\
	&\times (2\pi)^4 \delta(q-k_1-k_2) (2\pi)^4 \delta(k-k_2-k_3) (\momlep k_2)\nonumber\\ 
	&\times \bigl[ \pHiggs{}{\rho}(k_1) \pLeptsc{}{F}(k_2) \pMajdiagsc{jj}{h}(k_3) 
	+ \pHiggs{}{F}(k_1) \pLeptsc{}{\rho}(k_2) \pMajdiagsc{jj}{h}(k_3)\nonumber\\
	& - \pHiggs{}{h}(k_1) \pLeptsc{}{\rho}(k_2) \pMajdiagsc{jj}{F}(k_3) 
	- \pHiggs{}{h}(k_1) \pLeptsc{}{F}(k_2) \pMajdiagsc{jj}{\rho}(k_3)
	\nonumber\\
	& + \pHiggs{}{\rho}(k_1) \pLeptsc{}{h}(k_2) \pMajdiagsc{jj}{F}(k_3) 
	- \pHiggs{}{F}(k_1) \pLeptsc{}{h}(k_2) \pMajdiagsc{jj}{\rho}(k_3)\bigr]\nonumber\dend
\end{align}
The last three lines of \eqn\eqref{HiggsCPVertex} correspond to the three possible cuts of 
the vertex graph. Similarly to the Majorana decay only two of the intermediate states 
can be on-shell and for only one of them the corresponding distribution function 
enters the result. The value of the vertex \CP-violating parameter depends on the 
temperature as well as on masses of the Majorana neutrinos. For definiteness, let us 
assume a strongly hierarchical mass spectrum, $M_j\gg m_\higgs > M_i$. In this case 
contribution of the last two lines in \eqn\eqref{HiggsCPVertex} is strongly suppressed. 
Integrating out the delta-functions we find for the contribution of the first cut: 
\begin{align}
	\label{eqn: higgs vertex cp-violation}
	\epsilon_{\higgs,i}^V=\frac12 \sum \frac{\Im\, (h^\dagger h)_{ij}^2}{(h^\dagger h)_{ii}(h^\dagger h)_{jj}}
	&\frac{M_i \Gamma_j}{M^2_j} \frac{p K_j}{qp}\kend
\end{align}
where the loop function $ K_j$ is now defined as:
\begin{align}
	\label{Higgs vertex loop integral}
	K^\mu_{i}(\mommaj,\momhig)&=16\pi \int \lorentzd{\higgs}{\momhig_2} 
	\lorentzd{\lepton}{\momlep_2} (2\pi)^4 \deltafour{\mommaj+\momlep_2-\momhig_2}\,
	\momlep^\mu_2\nonumber\\
	&\times \bigl[\f{\higgs}{}(E_{\momhig_2})+\f{\lepton}{}(E_{\momlep_2})\bigr]
	M_i^2 \pMajdiagsc{ii}{h}(\momhig-\momlep_2) \dend
\end{align}
Note that the vertex \CP-violating parameter \eqref{eqn: higgs vertex cp-violation} has an opposite 
sign relative to that for Majorana neutrino decay. Similarly to the self-energy contribution 
we observe that the $1-\f{\lepton}{}+\f{\higgs}{}$ combination is replaced in 
\eqref{Higgs vertex loop integral} by $\f{\lepton}{}+\f{\higgs}{}$. For a milder mass hierarchy 
the two other cuts can become important. Their contributions are proportional
to $1-\f{\lepton}{}-\f{\majneutrino}{}$ and $\f{\higgs}{}+\f{\majneutrino}{}$ respectively.

\begin{figure}
	\includegraphics[width=\columnwidth]{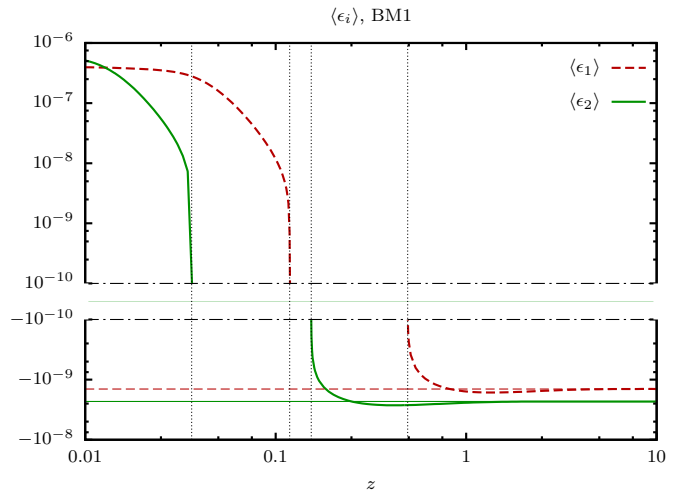}
	\caption{\label{EpsilonGammaHiggsBM1} Averaged self-energy \CP-violating parameters for Majorana neutrino 
	and Higgs decay for benchmark point~1 as a function of the inverse temperature. Thin lines 
	represent the value in the zero temperature limit. With conventional dispersion relations, 
	the decay $\majneutrino_1\rightarrow\lepton\higgs$ is active at $\x >4.93\cdot 10^{-1}$ but 
	replaced at high temperatures ($\x <1.18\cdot 10^{-1}$) by $\bar\higgs\rightarrow\majneutrino_1\lepton$. 
	The decay $\majneutrino_2\rightarrow\lepton\higgs$ is active at $\x >1.54\cdot 10^{-1}$ but 
	replaced at high temperatures ($\x <3.62\cdot 10^{-2}$) by $\bar\higgs\rightarrow\majneutrino_2\lepton$. 
	}
\end{figure}

\begin{figure}
	\includegraphics[width=\columnwidth]{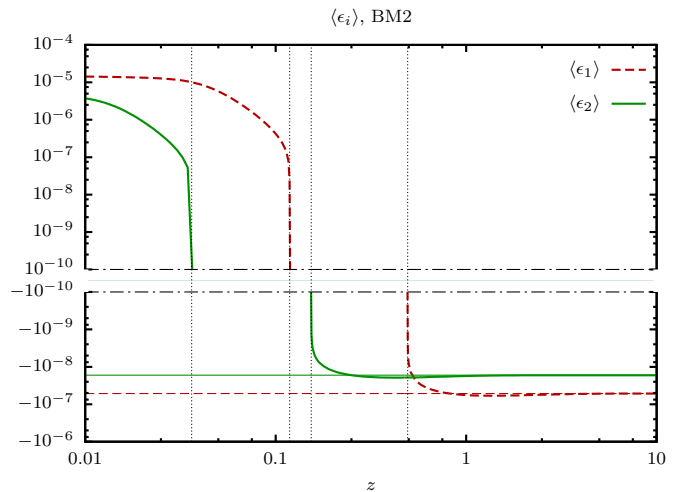}
	\caption{\label{EpsilonGammaHiggsBM2} Averaged self-energy \CP-violating parameters for Majorana 
	neutrino and Higgs decay for benchmark point~2. See caption of \fig\ref{EpsilonGammaHiggsBM1}.}
\end{figure}

The first-principle computation 
gives for the Higgs decay contribution to the evolution of the lepton 
current an expression similar to \eqn\eqref{divlepCPpred}:
\begin{align}
	\label{divlepCPHiggspred}
	\frac{d Y_{Li\atop\majneutrino_i}}{dz}&=\frac{z}{s\hubbleratem }
	\int\dpi{\majneutrino_i\higgs\lepton}{}{\mommaj\momhig\momlep}{}\nonumber\\
	&\times\bigl[\EffAmplitude{}{\bar{\higgs}\rightarrow\lepton\majneutrino_i }
	\F{\majneutrino_i\lepton}{\bar{\higgs}}{\mommaj\momlep}{\momhig} \mp
	\EffAmplitude{}{\higgs\rightarrow\bar\lepton\majneutrino_i }
	\F{\majneutrino_i\bar{\lepton}}{{\higgs}}{\mommaj\momlep}{\momhig}\bigr]\dend
\end{align}
We do not discuss `extra' terms here which would arise from `naive' Boltzmann equations.
To write this as a rate equation we need to repeat the steps in \sect\ref{RateEquations} which lead to 
\eqn\eqref{eqn:quantum corrected Boltzmann equation Li result}. For a general process $aN\leftrightarrow b$ 
(where we allow for deviations from equilibrium in $\f{N}{}$) we have 
\eqn\eqref{eqn:linear order expansion Higgs decay term}. Therefore we obtain, for Higgs decay, 
the following contributions to the rate equations for lepton number and Majorana neutrino 
abundance (see \app\ref{RateEquationsApp}):
\begin{subequations}
\noeqref{eqn:rate equation Higgs decay contributions Li,eqn:rate equation Higgs decay term 2,eqn:rate equation Higgs decay term 3}
\begin{align}
	\label{eqn:rate equation Higgs decay contributions Li}
	&{\frac{s\hubbleratem }{z}}\frac{d Y_{Li}}{dz}\Big|_{\Delta\f{\majneutrino_i}{}} =
	-\bigg< \epsilon_{\higgs,i}\frac{\Delta\f{\majneutrino_i}{\mommaj}}{\feq{\majneutrino_i}{}}
	\gamma^D_{\higgs,i}\bigg>\kend\\
	\label{eqn:rate equation Higgs decay term 2}
	&{\frac{s\hubbleratem }{z}}\frac{d Y_{Li}}{dz} \Big|_{\Delta\f{\majneutrino_i}{}{\frac{\mu_\lepton}{T}}} = 
	\frac{\mu_\lepton}{T}\bigg<\epsilon_{\higgs,i}(1-\feq{\lepton}{\momlep} -c_{\higgs\lepton}\feq{\higgs}{\momhig}) \frac{\Delta\f{\majneutrino_i}{\mommaj}}{\feq{\majneutrino_i}{}}
	\gamma^D_{\higgs,i}\bigg>\kend\\
	\label{eqn:rate equation Higgs decay term 3}
	&{\frac{s\hubbleratem }{z}}\frac{d Y_{Li\atop\majneutrino_i}}{dz}\Big|_{\frac{\mu_\lepton}{T}} =
	-\frac{\mu_\lepton}{T}(1+c_{\higgs\lepton})\bigg<\frac{(1-\f{\majneutrino_i}{})}{(1-
	\feq{\majneutrino_i}{})}\gamma^W_{\higgs,i}\bigg>\dend
	\hphantom{aaaa}
\end{align}
\end{subequations}
We introduced the decay and washout reaction densities, $\langle X\gamma^D_{\higgs,i}\rangle$ 
and $\langle X\gamma^W_{\higgs,i}\rangle$, for Higgs decay:
\begin{align}
	\label{EpsilonThermalHiggs}
	\langle X\gamma^D_{\higgs,i}\rangle \equiv &\int \dpi{{\lepton}{\majneutrino_i}{\higgs}}{}{{\momlep}{\mommaj}{\momhig}}{} 
	\, (2\pi)^4 \deltafour{\mommaj+\momlep-\momk} X\EffAmplitude{}{\higgs,i} \feq{\majneutrino_i}{q} f_{\lepton\higgs}\kend\nonumber\\
	\langle X\gamma^W_{\higgs,i}\rangle \equiv &\langle X(1-\feq{\majneutrino_i}{})\gamma^D_{\higgs,i}\rangle\kend
\end{align}
where now $f_{\lepton\higgs}=(\feq{\higgs}{\momhig}+\feq{\lepton}{\momlep})$.
In complete analogy to \eqn\eqref{AmplsqAndEpsDef} the total amplitude and 
the {\CP}-violating parameter for (anti-)Higgs decay are defined as
\begin{subequations}
\label{AmplsqAndEpsDefHiggs}
	\begin{align}
		\EffAmplitude{}{\higgs,i} & \equiv 
		\EffAmplitude{}{\bar\higgs \rightarrow \lepton\majneutrino_i}+
		\EffAmplitude{}{\higgs \rightarrow \bar\lepton\majneutrino_i}\approx 2\EffAmplitude{}{\bar\higgs \leftrightarrow \lepton\majneutrino_i}\kend\\
		\label{HiggsEpsilonUnintegrated}
		\epsilon_{\higgs,i} & \equiv
		\frac{\EffAmplitude{}{\bar\higgs \rightarrow \lepton\majneutrino_i}-
		\EffAmplitude{}{\higgs \rightarrow \bar\lepton\majneutrino_i}}{\EffAmplitude{}{\higgs,i}}\dend
	\end{align}
\end{subequations}
Similarly to the Majorana neutrino decays we also define an averaged \CP-violating parameter as 
\begin{align*} 
\langle \epsilon_{\higgs,i} \rangle \equiv \frac{\langle\epsilon_{\higgs,i} \gamma^D_{\higgs,i}\rangle}{\langle \gamma^D_{\higgs,i}\rangle}\dend
\end{align*}
By comparing \eqns\eqref{eqn:rate equation Higgs decay contributions Li} and 
\eqref{eqn:quantum corrected Boltzmann equation Li result} we observe that, ignoring 
${\Delta\f{\majneutrino_i}{}{{\mu_\lepton}/{T}}}$ contributions, the difference 
to the Majorana neutrino decay contributions amounts to the replacements 
$\langle X\gamma^D_{\majneutrino_i}\rangle\rightarrow -\langle X\gamma^D_{\higgs,i}\rangle$ 
and $\langle X\gamma^W_{\majneutrino_i}\rangle\rightarrow\langle X\gamma^W_{\higgs,i}\rangle$. 
We therefore obtain \eqn\eqref{RateEquationsa} with an opposite sign for the \CP-violating source 
term. This sign cancels the relative sign of the \CP-violating parameter such that Majorana 
neutrino decay and Higgs decay contribute effectively with same sign.
Similarly, for the contribution to the Majorana neutrino rate equation: 
\begin{align}
	\label{eqn:rate equation Higgs decay contributions N}
	&{\frac{s\hubbleratem }{z}}\frac{d Y_{Li}}{dz}\Big|_{\Delta\f{\majneutrino_i}{}} = -\bigg< \frac{\Delta\f{\majneutrino_i}{\mommaj}}{\feq{\majneutrino_i}{}}\gamma^D_{\higgs,i}\bigg>\kend\nonumber\\
	&{\frac{s\hubbleratem }{z}}\frac{d Y_{Li}}{dz} \Big|_{\Delta\f{\majneutrino_i}{}{\frac{\mu_\lepton}{T}}} = \frac{\mu_\lepton}{T}\bigg<(1-\feq{\lepton}{\momlep} - c_{\higgs\lepton}\feq{\higgs}{\momhig}) \frac{\Delta\f{\majneutrino_i}{\mommaj}}{\feq{\majneutrino_i}{}}\gamma^D_{\higgs,i}\bigg>\kend\nonumber\\
	&{\frac{s\hubbleratem }{z}}\frac{d Y_{Li\atop\majneutrino_i}}{dz}\Big|_{\frac{\mu_\lepton}{T}} = -\frac{\mu_\lepton}{T}(1+c_{\higgs\lepton})\bigg<\epsilon_{\higgs,i}\frac{(1-\f{\majneutrino_i}{})}{(1-\feq{\majneutrino_i}{})}\gamma^W_{\higgs,i}\bigg>\dend\nonumber
\end{align}
To compute the thermally averaged \CP-violating parameter we take into account 
the temperature dependent evolution of the lepton and Higgs masses
 \eqref{thermalmasses} \cite{Giudice:2003jh}. 
The averaged \CP-violating parameter in the Higgs decay and in the Majorana decay 
as functions of the inverse temperature are presented in \fig\ref{EpsilonGammaHiggsBM1} 
and \fig\ref{EpsilonGammaHiggsBM2}.
At very high temperatures the magnitude of 
$\langle\epsilon_i\rangle$ can be much larger for Higgs decay. 
As the temperature decreases the 
\CP-violating parameter for the Higgs decay approaches zero. This is explained by the shrinking of 
the available phase space in the loop integrals \eqref{LRhoHiggs} and \eqref{Higgs vertex loop integral}. 

\section{\label{Summary}Conclusions}

In this paper we have studied leptogenesis in the type-I seesaw extension of the 
Standard Model using the 2PI-formalism of non-equilibrium quantum field theory. 

The asymmetry generation 
can, in the case of thermal leptogenesis, be approximately described by rate equations. 
Usually these statistical equations 
are treated as a `black-box' in 
the sense that their form is assumed given and model specific amplitudes are inserted 
by hand. Indeed, this approach is supported by the observation that it describes the 
free decay in the zero temperature limit correctly and inherent inconsistencies (namely 
the `double-counting problem') can be resolved in exact equilibrium. However,
out of equilibrium it is not obvious whether the subtraction of real intermediate states works to all orders and how amplitudes
computed in thermal field theory enter kinetic equations. These issues can be completely
avoided in a systematic treatment within non-equilibrium quantum field theory.

Only in recent years fundamental questions related to the 
non-equilibrium statistical description received more attention. The progress here is 
mainly based on the 2PI-formalism which is known to yield consistent quantum kinetic 
equations without double-counting. 
These equations can be reduced to a system of Boltzmann-like kinetic 
equations for quasiparticles which can easily be compared to the conventional results. 
In the course of the derivation necessary approximations and the related physical 
assumptions have to be specified explicitly. Therefore, this 
approach enables a deeper insight into the dynamics of the asymmetry generation. 

In the conventional analysis the minimal set of interactions 
is obtained at order $\order{h^4}$ of the perturbative expansion. We have complemented 
existing analyses based on the 2PI-formalism 
by the computation of further processes which appear at this order.
Starting from a system of Kadanoff-Baym and (equivalent) Schwinger-Dyson equations 
for leptons and heavy Majorana neutrinos we have derived 
Boltzmann-like quantum-kinetic equations for the lepton asymmetry. They include 
(inverse) decays of the heavy Majorana neutrinos as well as two-body scattering 
processes mediated by the heavy neutrinos:
\begin{align*}
	\frac{s\hubbleratem }{z}\frac{d Y_L}{dz} &=
	\sum_i \int \dpi{\lepton\higgs}{N_i}{\momlep\momhig}{\mommaj}
	\F{\lepton\higgs}{N_i}{\momlep\momhig}{\mommaj}\EffAmpl{}{\lepton\higgs}{\majneutrino_i}\nonumber\\
	&-\sum_i \int \dpi{\bar\lepton\bar\higgs}{N_i}{\momlep\momhig}{\mommaj}
	\F{\bar\lepton\bar\higgs}{N_i}{\momlep\momhig}{\mommaj}\EffAmpl{}{\bar\lepton\bar\higgs}{\majneutrino_i}\nonumber\\
	&-2\int \dpi{\lepton\higgs}{\bar\lepton\bar\higgs}{\momlep_1\momhig_1}{\momlep_2\momhig_2}
	\F{\bar\lepton\bar\higgs}{\lepton\higgs}{\momlep_2\momhig_2}{\momlep_1\momhig_1}
	\EffAmpl{}{\bar\lepton\bar\higgs}{\lepton\higgs}\nonumber\\
	&-\int \dpi{\lepton\lepton}{\bar\higgs\bar\higgs}{\momlep_1\momlep_2}{\momhig_1\momhig_2}
	\F{\bar\higgs\bar\higgs}{\lepton\lepton}{\momhig_1\momhig_2}{\momlep_1\momlep_2}
	\EffAmpl{}{\bar\higgs\bar\higgs}{\lepton\lepton}\nonumber\\
	&-\int \dpi{\bar\lepton\bar\lepton}{\higgs\higgs}{\momlep_1\momlep_2}{\momhig_1\momhig_2}
	\F{\bar\lepton\bar\lepton}{\higgs\higgs}{\momlep_1\momlep_2}{\momhig_1\momhig_2}
	\EffAmpl{}{\bar\lepton\bar\lepton}{\higgs\higgs}\dend
\end{align*}
Because all terms in this equation are proportional to $\cal F$, a combination of the 
distribution functions which vanishes in equilibrium, the obtained equations are free of 
the double-counting problem and no need for the real intermediate state subtraction arises. 
Together with the
systematic derivation of the effective decay and scattering amplitudes $\EffAmplitude{}{}$ 
this is the main result of the present work. The individual amplitudes arise as combinations 
of different 2PI contributions. The vertex contribution to the {\CP}-violating 
decay amplitude 
is obtained as cut of the 2PI `mercedes' diagram. The same graph yields also the $s\times t$ 
contributions to $\lepton\higgs\leftrightarrow\bar\lepton\bar\higgs$ and the $u\times t$ 
contribution to $\lepton\lepton\leftrightarrow\bar\higgs\bar\higgs$ process. To extract the 
self-energy contribution, the off-diagonal elements of the Majorana neutrino propagator 
have to be taken into account. In addition, an extended quasiparticle approximation 
needs to be employed in order to obtain the $s\times s$ and $t\times t$ contributions 
to $\lepton\higgs\leftrightarrow\bar\lepton\bar\higgs$ as well as $u\times u$ and 
$t\times t$ contributions to $\lepton\lepton\leftrightarrow\bar\higgs\bar\higgs$ 
scattering from the `setting-sun' diagram.

In the zero temperature limit the effective amplitudes reduce to the canonical ones. In particular, the form of 
the resulting amplitudes for $\lepton\higgs\leftrightarrow\bar\lepton\bar\higgs$ scattering coincides with the RIS subtracted amplitudes encountered in existing 
calculations. At finite temperatures the effective amplitudes receive thermal 
corrections. Medium corrections to the Majorana decay amplitudes into leptons and 
antileptons are $\order{h^4}$. They are small 
compared to the tree-level vacuum contribution and are therefore negligible for the 
total decay width. On the other hand, they play an important role for the \CP-violating 
source-terms, which are proportional to the difference of the two amplitudes. We 
find that medium corrections to the \CP-violating parameter are linear in 
the particle number densities. Although there is a partial cancellation of the bosonic 
and fermionic contribution, the \CP-violating parameter is enhanced. In the effective
scattering amplitudes the medium corrections affect only the `regulator' term in
the denominator of the Breit-Wigner propagators. Due to the smallness of the 
Majorana decay width, which is constrained by the light neutrino masses, numerically
these corrections are very small in the case of non-degenerate Majorana neutrinos.

Taking SM interactions into account in the form of thermal lepton and Higgs 
masses results in a suppression of the phase space for the Majorana neutrino 
decay and the enhancement of the \CP-violating parameters is overcompensated. At even higher
temperatures, when the effective Higgs mass exceeds the Majorana masses, the \CP-violating 
decay of the Higgs into a lepton-Majorana pair can become kinematically allowed instead. 
At these temperatures the averaged \CP-violating parameters for Higgs decay exceeds that 
obtained for Majorana decay in vacuum by orders of magnitude. The signs 
of the corresponding \CP-violating parameters are opposite but their contribution 
to the lepton asymmetry has the same sign (at least in the limit of hierarchical 
Majorana masses). These results are in qualitative agreement with earlier studies
based on thermal field theory and may ultimately be compared to a full treatment of
\CP-violating decays out of equilibrium, including gauge interactions.

We have also derived the corresponding rate equations for abundances of the participating 
species. They are obtained as expansion in small deviations from equilibrium ($\mu/T$ and 
$\Delta\f{\majneutrino_i}{}$) and represent the hydrodynamical approximation of the Boltzmann 
kinetic equations. As compared to the standard (zero temperature) 
result they are improved in that the obtained coefficients include medium corrections to 
the quasiparticle properties and take into account quantum-statistical effects. 
We compare with the result obtained if the amplitudes are computed in thermal quantum field 
theory and the RIS subtraction is performed manually. We find that there are differences at 
higher order in the expansion parameters. The coefficients -- reaction densities -- reflect 
the interplay between the medium enhancement of the effective amplitudes and the phase space 
suppression induced by the thermal masses of Higgs and leptons. At very low temperatures the
reaction densities approach their canonical limit. 

Since for a hierarchical mass spectrum most 
of the asymmetry is typically generated by the lightest Majorana neutrino at temperatures
of the order and smaller than its mass, we expect a moderate enhancement
of the total generated asymmetry is possible for a typical average to strong washout 
scenario. For a detailed phenomenological analysis it is necessary to include further 
phenomena such as flavour effects and $\Delta L=1$ scattering processes which contribute 
to the washout at $\order{g^2 h^2}$. Additional quantum effects beyond the present 
analysis are relevant for non-standard scenarios in which the Majorana neutrinos have 
degenerate masses or if they are not as close to thermal equilibrium.

\subsection*{Acknowledgements}
\noindent
This work has been supported by the German Science Foundation (DFG) under Grant 
KA-3274/1-1 ``Systematic analysis of baryogenesis in non-equilibrium quantum field theory''
and the Collaborative Research Center 676 ``Particles, Strings and the Early Universe''.
T.F. acknowledges support by the IMPRS-PTFS. We thank S.~Blanchet and M.~Shaposhnikov for useful discussions.

\begin{appendix}

\section{\label{Kinematics}Kinematics}

The reaction densities contain distribution functions of the 
initial and final states, which depend on the corresponding energies. Therefore to compute 
the reaction densities we need to analyze the kinematics of the decay
and scattering processes. 

\subsection{Decay}

To compute the \CP-violating reaction density and the decay reaction density 
we need to evaluate the integral: 
\begin{align}
	\langle X \gamma^D_{\majneutrino_i}\rangle & = \int \lorentzd{\majneutrino}{\mommaj}
	\,\feq{\majneutrino}{\mommaj} \int
	\lorentzd{\higgs}{\momhig} \lorentzd{\lepton}{\momlep} (2\pi)^4\delta(\momhig+\momlep-\mommaj) \nonumber\\
	&\times X \EffAmplitude{}{\majneutrino}\bigl[1-\feq{\lepton}{\momlep}+\feq{\higgs}{\momhig}\bigr]\dend
\end{align}
For the washout reaction densities or reaction densities for Higgs decay we have similar expressions.
The integration over $d^3 k$ can be performed trivially and yields 
$\vec k= \vec q - \vec p$. Using $|\vec k|=(\vec q\,^2 +\vec p\,^2 - 
|\vec q||\vec p|\cos\theta)^\frac12$ and integrating over $\theta$ we 
remove the remaining Dirac-delta and obtain
\begin{align}
	\label{DecIntMeasure}
	\int \lorentzd{\higgs}{\momlep} \lorentzd{\lepton}{\momhig} (2\pi)^4& \delta(\momhig+\momlep-\mommaj)
	& \rightarrow \frac{1}{8\pi\,|\vec \mommaj|} \int\limits_{E^{-}}^{E^{+}}dE_\momlep
	\int\limits_0^{2\pi}\frac{d\varphi}{2\pi}\dend
\end{align}
The integration limits are given by 
\begin{align}
\label{eqn: decay integration limits}
E^\pm =
{\textstyle\frac12}\bigl[E_\mommaj\bigl(1+x_\lepton-x_\higgs)
\pm |\vec \mommaj| \lambda^\frac12(1,x_\lepton,x_\higgs)\bigr]\kend
\end{align}
where $x_\lepton\equiv m_\lepton^2/q^2$, $x_\higgs\equiv m_\higgs^2/q^2$ and 
$\lambda(x,y,z)\equiv x^2+y^2+z^2-2xy-2xz-2yz$ is the usual kinematical function. 
For an on-shell heavy neutrino $q^2=M^2$. 
If $m_\lepton=m_\higgs=0$ then $\lambda(1,x_\lepton,x_\higgs)=1$ and the 
above expression simplifies to $E_\pm=\frac12(E_\mommaj\pm |\vec \mommaj|)$.
On the other hand, if $M=m_\lepton+m_\higgs$ then $\lambda(1,x_\lepton,x_\higgs)=0$
and therefore $E^{+}=E^{-}=E_\mommaj(m_\lepton/M)$. Since the integration limits coincide
in this case, the integral vanishes. 

Combining it with the integration over $d^3 q$ and using the isotropy of the medium we find 
\begin{align}
 \langle X \gamma^D_{\majneutrino_i}\rangle = \frac{1}{32\pi^3}
	&\int\limits_M^\infty d E_\mommaj \,\feq{\majneutrino}{E_\mommaj} 
	\int\limits_{E_{-}}^{E_{+}} d E_\momlep
	X \EffAmplitude{}{\majneutrino}\bigl[1-\feq{\lepton}{E_\momlep}+\feq{\higgs}{E_\momhig}\bigr]
	\dend
\end{align}
If quantum-statistical effects are neglected then both the \CP-violating parameter and the 
total tree-level decay amplitude are momentum independent and the integration can be 
performed analytically. We get:
\begin{align}
	\label{eqn: canonical decay reaction density}
	\langle X\,\gamma^D_{\majneutrino_i}\rangle 
	\approx \frac{g_\majneutrino}{2\pi^2}X M_i^2\Gamma_i T K_1\left(\frac{M_i}{T}\right)
	\kend
\end{align}
where $g_\majneutrino=2$ is the number of the Majorana spin degrees of freedom.
Similar results are obtained for Higgs decay and washout reaction densities.

\subsection{Two-body scattering}

For $2\leftrightarrow 2$ scattering processes the reaction density 
is defined by 
\begin{align}
	\label{ReactDensExactApp}
	\langle\gamma^{ab}_{ij}\rangle&\equiv\int \dpi{a b}{i j}{p_a p_b}{p_i p_j}
	(2\pi)^4 \delta(p_a+p_b-p_i-p_j)\nonumber\\
	&\times \feq{a}{}\feq{b}{}(1\pm \feq{i}{})(1\pm \feq{j}{})
	\EffAmplitude{}{ab\leftrightarrow ij}\dend
\end{align}
To reduce it to a form suitable for the numerical analysis we 
insert an identity:
\begin{align}
	1=\int ds \int d^4 q \delta(p_a+p_b-\mommaj)\delta_+(\mommaj^2-s)\kend
\end{align}
into \eqn\eqref{ReactDensExactApp}. The resulting expression can 
be interpreted as a product of the inverse decay and decay 
amplitudes integrated over the `mass' and energy of the intermediate
state:
\begin{align}
	\label{InsertedIdentity}
	\int \dpi{a b}{i j}{p_a p_b}{p_i p_j}&
	(2\pi)^4 \delta(p_a+p_b-p_i-p_j)\nonumber\\
	\rightarrow \int ds \int &\frac{d^4q}{(2\pi)^4} \delta_+(\mommaj^2-s) \nonumber\\
	\times & \int \lorentzd{a}{p_a} \lorentzd{b}{p_b} (2\pi)^4 \delta(p_a+p_b-q)\nonumber\\
	\times & \int \lorentzd{i}{p_i} \lorentzd{j}{p_j} (2\pi)^4 \delta(q-p_i-p_j)\dend
\end{align}
For the third line we will use \eqn\eqref{DecIntMeasure}. 
For the last line it is more convenient to us a different representation. Integrating out 
the delta-function we obtain for the last line in \eqn\eqref{InsertedIdentity}:
\begin{align}
\label{eqn: second integration measure}
\frac{1}{4\pi}\int\frac{d\Omega_i}{4\pi}
\frac{\vec\momlep^2_i}{q_0 |\vec p_i| - E_i|\vec q|\cos\Theta}\dend
\end{align}
Note that not all angles are kinematically allowed, see \eqn\eqref{eqn: Ei and Theta_qi}
below. As a product of Lorentz-invariant quantities the integral is also Lorentz-invariant.
We can therefore boost to the center-of-mass frame where $\vec q=0$ and $q_0=\sqrt{s}$. By energy-momentum 
conservation $|\vec\momlep_i|/q_0=\frac12 \lambda^{\frac12}(1,x_i,x_j)$, where $q^2=s$ 
now. The angle integration can be partially reduced to integration over the Mandelstam 
variable $t=(p_a-p_i)^2$ using the relation $dt=2|\vec\momlep_a||\vec\momlep_i|d\cos\theta_{ai}$. 
Note that the azimuthal angle $\varphi_{ai}$ is Lorentz-invariant by itself.
Therefore, boosting back to the rest frame of the medium we can write the left-hand side of 
\eqn\eqref{eqn: second integration measure} in the form
\begin{align}
\lambda^{-\frac12}(1,x_a,x_b)\cdot \frac1{8\pi}\int \frac{d\varphi_{ai}}{2\pi}\int \frac{dt}{s}\dend
\end{align}
Integrating furthermore over $d q^0$ and using the fact the the integrand 
is independent of the orientation of $\vec{q}$, we finally 
obtain
\begin{align}
	\label{eqn: scattering reaction density simplified}
	\langle\gamma^{ab}_{ij}\rangle &=\frac{1}{64\pi^4}\int\limits_{s_{min}}^\infty ds 
	\int\limits_{\sqrt{s}}^\infty dE_q \\
	&\times \lambda^{-\frac12}(1,x_a,x_b)\int\limits_{E^{-}}^{E^{+}} dE_a \, 
	f^{eq}_a f^{eq}_b \nonumber\\
	&\times\frac{1}{8\pi}\int\limits_0^{2\pi} 
	\frac{d\varphi_{ai}}{2\pi}
	\int\limits_{t^-}^{t^+} \frac{dt}{s}
	\EffAmplitude{}{ab\leftrightarrow ij} 
	(1\pm \feq{i}{})(1\pm \feq{j}{})\nonumber\kend
\end{align}
where $\sqrt{s}_{min}=\max(\sum m_{init}, \sum m_{fin})$. Just like for particle decay, the 
integration limits $E^\pm$ are given by \eqn\eqref{eqn: decay integration limits} but with 
$x_\lepton$ and $x_\higgs$ replaced by $x_a$ and $x_b$ respectively and the three-momentum 
given by $|\vec q|=(E_q^2-s)^\frac12$. The range of integration over $t$ is given by 
\begin{align}
\label{eqn: t range}
t^\pm=m_a^2+m_i^2&-\frac{s}{2}\bigl[(1+x_a-x_b)(1+x_i-x_j)\nonumber\\
&\mp \lambda^\frac12(1,x_a,x_b)\lambda^\frac12(1,x_i,x_j)\bigr]\dend
\end{align}	
In particular, for massless initial and finial states it reduces 
to $t^+=0$ and $t^-=-s$.

If the quantum-statistical effects are neglected 
then $f_a^{eq}(E_a)f_b^{eq}(E_b)=f_\majneutrino^{eq}(E_q)$. The integration over $E_a$ can 
be easily performed in this case and, combined with the $\lambda^{-\frac12}(1,x_a,x_b)$ 
prefactor, gives $|\vec q|$. In the same approximation the last line of 
\eqref{eqn: scattering reaction density simplified} does not depend on the distribution 
functions and gives so-called `reduced cross section': 
\begin{align}
\label{eqn:reduced cross section definition}
\hat \sigma(s)\equiv \frac{1}{8\pi}\int\limits_0^{2\pi} 
	\frac{d\varphi_{ai}}{2\pi}
	\int\limits_{t^-}^{t^+} \frac{dt}{s}
	\EffAmplitude{}{ab\leftrightarrow ij} \dend
\end{align}
Using \eqn\eqref{eqn:reduced cross section definition} and integrating over $E_q$ we recover 
the usual expression for the scattering reaction density:
\begin{align}
	\label{eqn:scattering reaction density canonical}
	\langle\gamma^{ab}_{ij}\rangle & \approx \frac{T}{64\pi^4}\int\limits_{s_{min}}^\infty ds 
	\sqrt{s}K_1\left(\frac{\sqrt{s}}{T}\right) \hat{\sigma}(s)\dend
\end{align}
To take the quantum-statistical effects into account we need to express energies of 
initial and final states in terms of the integration variables. By energy conservation 
$E_b=E_q-E_a$. Therefore $E_a$ and $E_b$ as well as the related momenta $|\vec p_a|$ and 
$|\vec p_b|$ are completely fixed by the second and third integration variables.
Next we consider the final states. By energy conservation $E_j=E_q-E_i$. It remains 
to express $E_i$ in terms of the integration variables. Let us choose the coordinates 
such that $\vec p_a$ points along the $x$-axis and $\vec p_b$ lies in the $xy$-plane. 
Then the momentum transfer $\vec q$ also lies in 
the same plane. Its components are given by
$\vec q=|\vec q|(\cos\theta_{aq},\sin\theta_{aq},0)$ where:
\begin{align}
\cos\theta_{aq}=\frac{2E_a E_q-s+m_b^2-m_a^2}{2|\vec q||\vec p_a|}\dend
\end{align}
The components of 
$\vec p_i$ can be written in the form $\vec p_i=|\vec p_i|(\cos\theta_{ai}, 
\sin\theta_{ai}\cos\varphi_{ai},\sin\theta_{ai}\sin\varphi_{ai})$. Then the 
angle between the vectors $\vec p_i$ and $\vec q$ is given by 
\begin{align}
\cos\theta_{qi}&=\cos\theta_{aq}\cos\theta_{ai}+\sin\theta_{aq}\sin\theta_{ai}\cos\varphi_{ai} \dend
\end{align}
Using energy-momentum conservation we can express $E_i$ in terms of this 
angle and the integration variables:
\begin{align}
\label{eqn: Ei and Theta_qi}
&E_i=\frac12\frac{s}{s+\vec q^2\sin^2\theta_{qi}}
\left[E_q \bigl(1+x_i-x_j)\right.\nonumber\\
&+|\vec q|\cos\theta_{qi}\bigl(\lambda(1,x_i,x_j)-
4x_i \vec q^2/s \sin^2\theta_{qi}\bigr)^\frac12\bigr]\dend
\end{align}
Note that for $4m_\lepton^2 \vec q^2> \lambda(s,m_\lepton^2,m_\higgs^2)$ the
difference under the square root in \eqn\eqref{eqn: Ei and Theta_qi} can become 
negative for some angles. This means that such scattering angles are forbidden
kinematically and should not be integrated over. 

Since \eqn\eqref{eqn: Ei and Theta_qi} implicitly depends on $\theta_{ai}$ it is 
convenient to use this angle as an integration variable instead of $t$. 
The integration measure in \eqn\eqref{eqn: scattering reaction density simplified}
is then modified according to 
$dt\rightarrow 2|\vec p_a||\vec p_i| \sin\theta_{ai}\, d\theta_{ai}$.

To calculate the scattering amplitudes we need the three Mandelstam variables. 
$s$ is an integration variable. $t$ is given by
\begin{align}
\label{eqn: t and Theta_ai}
t=m_a^2+m_i^2-2E_a E_i + 2|\vec p_a||\vec p_i|\cos\theta_{ai}\dend
\end{align}
The remaining one, $u$, can be inferred from the Mandelstam relation $s+t+u=\sum m^2$.

\section{\label{GeneralizedOptTh}Generalized optical theorem and thermal cutting rules}

The generalized optical theorem is a consequence of the unitarity of the
$S$-matrix and can be seen as a consistency condition for the amplitudes 
to ensure conservation of probability.
It can also be seen as a consequence of the Cutkosky cutting rules \cite{cutkosky:429,Eden:2002,Bellac:1992} 
for the computation of the discontinuities of Feynman diagrams. As such it can be applied to 
unstable particles at any given order of perturbation theory. We may write it as
\begin{align}
	-i\Big[&\amp{a}{b}(\{k_i\},\{p_i\})-\amp{b}{a}^*(\{p_i\},\{k_i\})\Big]=\nonumber\\
	&=\sum_{i}\left( \prod_{i_l} \int {\lorentzd{}{i_l}}\right)(2\pi)^4\deltafour{\sum_j k_j -\sum_{i_l} q_{i_l}}\nonumber\\
	&\times\amp{a}{i}(\{k_i\},\{q_{i_l}\})\amp{b}{i}^*(\{p_i\},\{q_{i_l}\})\dend
	\label{eqn:generalized optical theorem}
\end{align}
The amplitudes $\amp{a}{b}$ include all contributing diagrams (at a given order of perturbation 
theory) and the sum on the right-hand side is over all possible real intermediate states $i$ 
which contribute to $\amp{a}{b}$. The generalized optical theorem can be exploited to see explicitly 
why the RIS subtraction works. 
\begin{figure}[!ht]
	\centering
	\includegraphics{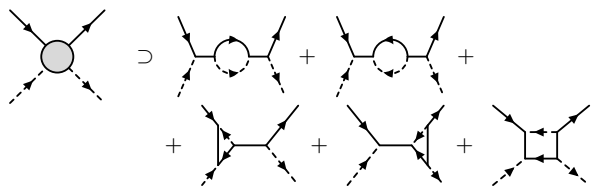}
	\caption{One-loop contributions to $\lepton\higgs\rightarrow\lepton\higgs$ scattering 
	at $\order{\yu^4}$. The cuts through the internal lepton and Higgs lines yield the 
	$s\times s$, $s\times t$ and $t\times t$-contributions to $\lepton\higgs\rightarrow\bar\lepton\bar\higgs$. 
	The cuts through single internal Majorana lines yield the interference terms which 
	contribute to the \CP-violating parameter. \label{fig:gneral diagram bb-bb}}
\end{figure}
To this end, we apply it to the forward scattering processes 
$\lepton\higgs\rightarrow \lepton\higgs$ and $\bar{\lepton}\bar{\higgs}
\rightarrow\bar{\lepton}\bar{\higgs}$, see \fig\figref{gneral diagram bb-bb}, above the 
energy thresholds $s>\m{\majneutrino_1}^2$ (the contribution by $\majneutrino_2$ real 
intermediate states etc.~can be addressed analogously) and $s>(\m{\lepton} +\m{\higgs})^2$. 

\begin{figure}[t]
	\centering
	\includegraphics{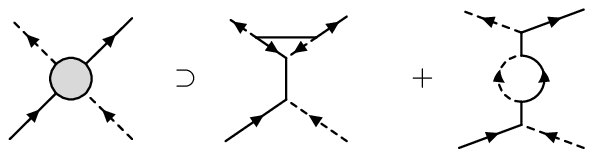}
	\caption{One-loop contributions to $\lepton\bar{\higgs}\rightarrow\lepton\bar{\higgs}$ 
	scattering at $\order{\yu^4}$. Due to medium effects the particles in the loop can be 
	on-shell.\label{fig:diagram bb-bb contribution}}
\end{figure}

\begin{figure}[t]
	\centering
	\includegraphics{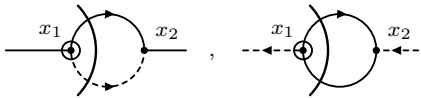}
	\caption{\label{rate through thermal cutting} The thermal width can be obtained 
	using (causal) finite temperature cutting rules.}
\end{figure}

\begin{figure}[t]
	\centering
	\includegraphics{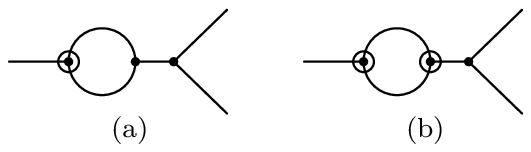}
	\\
	\includegraphics{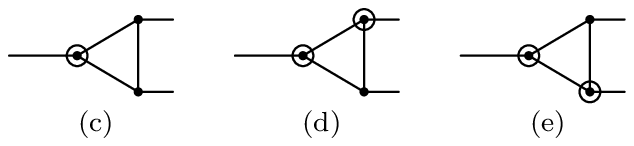}
	\includegraphics{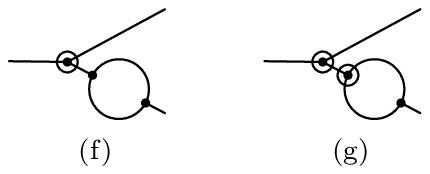}
	\caption{Circlings contributing to the self-energy (a),(b) and vertex (c-e) \CP-violating 
	parameter for Majorana neutrino decay and for Higgs decay (f),(g),(c-e). The contributions 
	by graphs (b),(f) vanish since $\majneutrino_i$ and $\majneutrino_j$ cannot be on-shell 
	simultaneously for $i\not=j$. Contributions by graphs (c-e) are suppressed if the cut is 
	through an internal Majorana neutrino line.\label{fig:selfenergy loop corrections imaginary part causal products}}
\end{figure}

We include all possible graphs up to $\order{\yu^4}$. Furthermore, we sum \eqn\eqref{eqn:generalized optical theorem} 
over all internal degrees of freedom of initial and 
final states and absorb these in the `effective amplitudes' defined in \sect\ref{ConventionalApproach}. 
We get for the process involving particles
\begin{align}
	2\Im & \big\{\sum_{\text{dof's}}\amp{\lepton\higgs}{\lepton\higgs}\big\}
	=\int {\lorentzd{\majneutrino_1}{\momq}}(2\pi)^4\deltafour{\momk+\momp-\momq}\EffAmplitude{}{\lepton\higgs\rightarrow\majneutrino_1} \nonumber\\
	+&\int {\lorentzd{\lepton}{\momq}\lorentzd{\higgs}{\momr}}(2\pi)^4\deltafour{\momk+\momp-\momq-\momr}\EffAmplitude{}{\lepton\higgs\rightarrow\bar{\lepton}\bar{\higgs}}\nonumber\\
	+&\int {\lorentzd{\lepton}{\momq}\lorentzd{\higgs}{\momr}}(2\pi)^4\deltafour{\momk+\momp-\momq-\momr}\EffAmplitude{}{\lepton\higgs\rightarrow{\lepton}{\higgs}}\dend
	\label{eqn:generalized optical theorem applied at order four 1}
\end{align}
The amplitudes squared on the right-hand side contain the relevant graphs including the vertex and 
self-energy contributions, see \fig\ref{treevertexself}, whose interference terms lead to 
{\CP}-violation in the particle decay.
In the same way one finds for $\bar{\lepton}\bar{\higgs}\leftrightarrow\bar{\lepton}\bar{\higgs}$:
\begin{align}
	2\Im & \big\{\sum_{\text{dof's}}\amp{\bar{\lepton}\bar{\higgs}}{\bar{\lepton}\bar{\higgs}}\big\}
	=\int {\lorentzd{\majneutrino_1}{\momq}}(2\pi)^4\deltafour{\momk+\momp-\momq}\EffAmplitude{}{\bar{\lepton}\bar{\higgs}\rightarrow\majneutrino_1} \nonumber\\
	+&\int {\lorentzd{\lepton}{\momq}\lorentzd{\higgs}{\momr}}(2\pi)^4\deltafour{\momk+\momp-\momq-\momr}\EffAmplitude{}{\bar{\lepton}\bar{\higgs}\rightarrow\lepton\higgs}\nonumber\\
	+&\int {\lorentzd{\lepton}{\momq}\lorentzd{\higgs}{\momr}}(2\pi)^4\deltafour{\momk+\momp-\momq-\momr}\EffAmplitude{}{\bar{\lepton}\bar{\higgs}\rightarrow\bar{\lepton}\bar{\higgs}}\dend
	\label{eqn:generalized optical theorem applied at order four 2}
\end{align}
As a consequence of {\CPT} we have $\sum\amp{\lepton\higgs}{\lepton\higgs}
=\sum\amp{\bar{\lepton}\bar{\higgs}}{\bar{\lepton}\bar{\higgs}}$. Therefore the difference of the 
left-hand sides of \eqns\eqref{eqn:generalized optical theorem applied at order four 1} and 
\eqref{eqn:generalized optical theorem applied at order four 2} as well as that of the third 
terms on the right-hand-sides vanish. Subtracting the right-hand sides we obtain
\begin{align}
	& \int {\lorentzd{\lepton}{\momq}\lorentzd{\higgs}{\momr}}(2\pi)^4\deltafour{\momk+\momp-\momq-\momr}
	\big[ \EffAmplitude{}{\lepton\higgs\rightarrow\bar{\lepton}\bar{\higgs}} 
	- \EffAmplitude{}{\bar{\lepton}\bar{\higgs}\rightarrow\lepton\higgs} \big]=\nonumber\\
	& -\int {\lorentzd{\majneutrino_1}{\momq}}(2\pi)^4\deltafour{\momk+\momp-\momq}
	\big[ \EffAmplitude{}{\lepton\higgs\rightarrow\majneutrino_1} - 
	\EffAmplitude{}{\bar{\lepton}\bar{\higgs}\rightarrow\majneutrino_1}\big],
	\label{eqn:generalized optical theorem applied at order four difference}
\end{align}
as a requirement for a consistent approximation of the amplitudes compatible with unitarity and {\CPT}. 
It is obvious that this cannot be satisfied if the scattering amplitudes are in tree-level approximation 
meanwhile the decay amplitudes violate {\CP}. We show in \sect\ref{RISQuantStat} how 
\eqn\eqref{eqn:generalized optical theorem applied at order four difference} can be satisfied by replacing 
the two-body scattering amplitudes $\Xi \rightarrow \Xi'$. The solution for $\Xi'$ amounts to subtracting 
the real intermediate state contributions from $\Xi$.
In order to obtain an equivalent result for the Higgs decay at high temperature (i.e.~for $\m{\higgs}>\m{\lepton}+M_i$) 
we need to consider different processes since the amplitude $\ampabs{\bar{\higgs}}{\lepton\majneutrino_i}$ 
cannot be obtained as cut of the graphs in \fig\figref{gneral diagram bb-bb}. One could try to draw and to 
cut graphs for $\lepton\majneutrino_i\rightarrow\lepton\majneutrino_i$ scattering, but the obtained cuts 
are lepton number conserving and would drop out in the difference $Y_{L}=Y_{\lepton}-Y_{\bar{\lepton}}$. 
Instead the relevant contributions appear as `thermal cuts' of the $t$-channel contributions to 
$\lepton\bar{\higgs}\rightarrow\lepton\bar{\higgs}$ depicted in \fig\figref{diagram bb-bb contribution}. 
Since they exist only at finite density we need to use finite temperature `circling rules', 
which can be seen as a generalization of \eqn\eqref{eqn:generalized optical theorem}, to compute 
the imaginary part of causal $n$-point functions in the real time formalism \cite{Kobes1991prd43}:
\begin{align}
	\label{eqn:imaginary part of causal products}
	2\Im\big\{i^{-1} \mathcal{F}_{R/A}^{(\alpha)} (x_1,\ldots & ,x_\alpha ,\ldots ,x_n;z_j)\big\} = \\
	\mp\sum_{x_i}^{\text{not all}}\sum_{z_j}^{} \Im\Big\{&i^{-1}F_> (x_1,\ldots,\underline{x}_\alpha ,\ldots , x_n;z_j) - \nonumber\\
	- & i^{-1}F_< (x_1,\ldots ,\underline{x}_\alpha ,\ldots , x_n;z_j)\Big\}\nonumber\kend
\end{align}
where `not all' means that not all $x_i$ should be circled at the same time. It was shown in \cite{Garny:2010nj} that causal $n$-point functions are the ones relevant for the computation of {\CP}-violating parameters. The vertex $x_\alpha$ 
with largest or smallest time is always circled. We can use this equation together with the circling 
rules given in \cite{Kobes1991prd43,Garny:2010nj} to compute the imaginary parts of the graphs in 
\fig\figref{diagram bb-bb contribution} and the \CPT-conjugated process 
$\bar{\lepton}{\higgs}\rightarrow\bar{\lepton}{\higgs}$.
Taking the difference of both we obtain, 
similar to \eqn\eqref{eqn:generalized optical theorem applied at order four difference}:
\begin{align}
	&\int {\lorentzd{\lepton}{\momq}\lorentzd{\higgs}{\momr}}(2\pi)^4\deltafour{\momk+\momp-\momq-\momr}\big[ \EffAmplitude{}{\lepton\higgs\rightarrow\bar{\lepton}\bar{\higgs}} 
	- \EffAmplitude{}{\bar{\lepton}\bar{\higgs}\rightarrow\lepton\higgs} \big]=\nonumber\\
	&-\int {\lorentzd{\higgs}{\momq}}(2\pi)^4\deltafour{\momk+\momp-\momq}\big[ \EffAmplitude{}{\lepton\majneutrino_1\rightarrow\bar{\higgs}} 
	- \EffAmplitude{}{\bar{\lepton\majneutrino_1}\rightarrow\higgs}
	\big]\dend
\end{align}
We can also use \eqn\eqref{eqn:imaginary part of causal products} to compute the thermal widths 
which cutoff the $s$- and $t$-channel resonances by cutting the self-energy graphs as shown 
in \fig\ref{rate through thermal cutting}.
Furthermore, the thermal \CP-violating parameters can be obtained using thermal cutting rules, 
see e.g.~\fig\figref{selfenergy loop corrections imaginary part causal products} 
and \cite{Garny:2010nj}. Altogether, the concept of RIS subtraction can be generalized to include 
quantum-statistical effects using thermal quantum field theory in the real time formalism and 
a complete set of reaction densities can be computed. Note however that inconsistencies are 
inherent out of equilibrium and arise e.g.~at higher order in the expansion performed in 
\sect\ref{RateEquations}. Similar computations where performed in 
\cite{Kiessig:2010pr,Kiessig:2011fw} in the imaginary time formalism of thermal quantum field theory.

\section{\label{RateEquationsApp}Rate equations}

In this appendix we present some detailed intermediate steps in the derivation of 
rate equations with quantum-statistical terms. To obtain these we need to assume that the system is close to thermal equilibrium. Under certain conditions it is then possible to reduce Boltzmann-like equations to a set of rate equations for the systems evolution. Assuming that the detailed conditions given in 
\sect\ref{RateEquations} are fulfilled and using \eqn\eqref{eqn:general decay collision term}, we get for the integrated Boltzmann equations with quantum-statistical terms:
\begin{widetext}
	\begin{align}
		\label{eqn:quantum corrected Boltzmann equation appendix}
		\int \dpi{\majneutrino_i\lepton\higgs}{}{\mommaj\momlep\momhig}{} \bigl[\pm\EffAmplitude{}{\majneutrino_i \rightleftarrows\lepton\higgs}\F{\lepton\higgs}{\majneutrino_i}{\momlep\momhig}{\mommaj} - &\EffAmplitude{}{\majneutrino_i \rightleftarrows\bar\lepton\bar\higgs}\F{\bar{\lepton}\bar{\higgs}}{\majneutrino_i}{\momlep\momhig}{\mommaj}\bigr] \\ = 
		\int \dpi{\majneutrino_i\lepton\higgs}{}{\mommaj\momlep\momhig}{} 
		(2\pi)^4\,\deltafour{\mommaj-\momlep-\momhig} \Big\{ \pm & \EffAmplitude{}{\majneutrino_i\rightleftarrows\lepton\higgs}
		(1-\f{\lepton}{\momlep})(1+\f{\higgs}{\momhig})\big[ \frac{\f{\majneutrino_i}{\mommaj}
		-\feq{\majneutrino_i}{\mommaj}}{\qstatff{\majneutrino_i}{\mommaj}\feq{\majneutrino_i}{\mommaj}}
		- (e^{+\frac{\mu_{\lepton}+\mu_{\higgs}}{T}}-1)\big]\nonumber\\
		- & \EffAmplitude{}{\majneutrino_i\rightleftarrows\bar\lepton\bar\higgs}
		(1-\f{\bar{\lepton}}{\momlep})(1+\f{\bar{\higgs}}{\momhig})
		\big[ \frac{\f{\majneutrino_i}{\mommaj}
		-\feq{\majneutrino_i}{\mommaj}}{\qstatff{\majneutrino_i}{\mommaj}\feq{\majneutrino_i}{\mommaj}} 
		- (e^{-\frac{\mu_{\lepton}+\mu_{\higgs}}{T}}-1)\big]\Big\} 
		\qstatff{\majneutrino_i}{\mommaj}
		\frac{\feq{\majneutrino_i}{\mommaj}}{\qstatfeqf{\majneutrino_i}{\mommaj}}\dend\nonumber
	\end{align}
\end{widetext}
This expression is still exact with respect to deviations from equilibrium and, taking into account results before \eqn\eqref{eqn:linear expansion of product of quantum-statistical terms}, it has an obvious expansion in $\mu/T$. We can see that there will be contributions proportional to $\Delta\f{\majneutrino_i}{\mommaj}=({\f{\majneutrino_i}{\mommaj}-\feq{\majneutrino_i}{\mommaj}})$, to $\mu/T$ and proportional to $\Delta\f{\majneutrino_i}{}\mu/T$ at linear order.
The coefficients of these terms were introduced in \sect\ref{RateEquations} (\eqns\eqref{eqn:reaction density decays} and \eqref{eqn:reaction density washout}) and dubbed reaction densities. We find for the $\order{\mu/T}$ expansion of each of the two terms in \eqn\eqref{eqn:quantum corrected Boltzmann equation appendix}, i.e.~for the general collision term of $a b\leftrightarrow N$ in \eqn\eqref{eqn:general decay collision term}:
\begin{align}
	\label{eqn:general decay collision term expanded}
	&\int \dpi{N a b}{}{\mommaj\momlep\momhig}{}\EffAmplitude{}{a b\leftrightarrow N}\F{a b}{N}{\momlep\momhig}{\mommaj}\approx\bigg< \frac{\EffAmplitude{}{a b\leftrightarrow N}}{\EffAmplitude{}{N}}\frac{\Delta\f{N}{\mommaj}}{\feq{N}{}}\gamma^D_{N}\bigg>\nonumber\\
	&-\bigg< \frac{\EffAmplitude{}{a b\leftrightarrow N}}{\EffAmplitude{}{N}}\frac{\Delta\f{N}{\mommaj}}{\feq{N}{}}(\xi_a\frac{a}{T}\feq{a}{}+\xi_b\frac{b}{T}\feq{b}{}
	)\gamma^D_{N}\bigg>\nonumber\\
	&-\frac{\mu_a +\mu_b}{T}\bigg< \frac{\EffAmplitude{}{a b\leftrightarrow N}}{\EffAmplitude{}{N}}\frac{\qstatf{N}{}}{\qstatfeq{N}{}}\gamma^W_{N}\bigg>\kend
\end{align}
where we defined the decay reaction densities as 
\begin{align}
	\label{eqn:reaction density decays washout appendix}
	\langle X \gamma^D_{N}\rangle \equiv & \int \dpi{{a}{N}{b}}{}{{\momlep}{\mommaj}{\momhig}}{}\, 
	(2\pi)^4 \deltafour{\mommaj-\momhig-\momlep} X\EffAmplitude{}{N} \feq{N}{q} f_{a b}\kend\nonumber\\
	\langle X\gamma^W_{N}\rangle \equiv & \big< X\qstatfeqf{N}{}
	\gamma^D_{N}\big>\kend
\end{align}
with $f_{a b}\equiv(1-\xi_a\feq{a}{\momlep}-\xi_b\feq{b}{\momhig})$.
For a general process $aN\leftrightarrow b$ (where we allow again for deviations from equilibrium in $\f{N}{}$) we find in the same way:
\begin{align}
	\label{eqn:linear order expansion Higgs decay term}
	&\int \dpi{N a b}{}{\mommaj\momlep\momhig}{}\EffAmplitude{}{a N\leftrightarrow b}\F{a N}{b}{\momlep\momhig}{\mommaj}\approx \bigg< -\frac{\EffAmplitude{}{aN\leftrightarrow b}}{\EffAmplitude{}{b}}\frac{\Delta\f{N}{\mommaj}}{\feq{N}{}}\gamma^D_{b}\bigg>\nonumber\\
	&-\bigg< \frac{\EffAmplitude{}{aN\leftrightarrow b}}{\EffAmplitude{}{b}}\frac{\Delta\f{N}{\mommaj}}{\feq{N}{}}(-\frac{\mu_a}{T}(1-\xi_a\feq{a}{})+\xi_b\frac{\mu_b}{T}\feq{b}{}
	)\gamma^D_{b}\bigg>\nonumber\\
	&-\frac{\mu_a -\mu_b}{T}\bigg< \frac{\EffAmplitude{}{aN\leftrightarrow b}}{\EffAmplitude{}{b}}\frac{\qstatf{N}{}}{\qstatfeq{N}{}}\gamma^W_{b}\bigg>\kend
\end{align}
where
\begin{align}
	\label{eqn:reaction density decays washout Higgs appendix}
	\langle X \gamma^D_{b}\rangle \equiv & \int \dpi{{a}{N}{b}}{}{{\momlep}{\mommaj}{\momhig}}{}\, 
	(2\pi)^4 \deltafour{\momlep+\mommaj-\momhig} X\EffAmplitude{}{b} \feq{N}{q} f_{a b}\kend\nonumber\\
	\langle X\gamma^W_{b}\rangle \equiv & \big< X\qstatfeqf{N}{}
	\gamma^D_{N}\big>\kend
\end{align}
with, now, $f_{a b}\equiv(\feq{a}{\momlep}+\xi_N\feq{b}{\momhig})$. Note that these reaction densities 
will tend to zero in the zero temperature limit because of their dependence on the distribution functions.
Using the general result \eqref{eqn:general decay collision term expanded} we obtain the contributions to 
\eqn\eqref{eqn:quantum corrected Boltzmann equation appendix}, and therefore to \eqn\eqref{divlepCPpred}, 
proportional to $\Delta\f{\majneutrino_i}{\mommaj}$,
\begin{align}
	\label{eqn:quantum corrected Boltzmann equation Delta f}
	& \frac{s\hubbleratem }{z} \frac{d Y_{Li\atop\majneutrino_i}}{dz}\Big|_{\Delta\f{\majneutrino_i}{}} 
	 = \bigg< \frac{\pm\EffAmplitude{}{\majneutrino_i \rightleftarrows \lepton\higgs}-\EffAmplitude{}{\majneutrino_i \rightleftarrows \bar{\lepton}\bar{\higgs}}}{\EffAmplitude{}{\majneutrino_i}}\frac{\Delta\f{\majneutrino_i}{\mommaj}}{\feq{\majneutrino_i}{}}\gamma^D_{\majneutrino_i}\bigg>\kend
\end{align}
contributions proportional to $\Delta\f{\majneutrino_i}{\mommaj}\cdot {\frac{\mu_\lepton}{T}}$,
\begin{align}
	\label{eqn:quantum corrected Boltzmann equation mu by T times Delta f}
	& \frac{s\hubbleratem }{z} \frac{d Y_{Li\atop\majneutrino_i}}{dz} 
	\Big|_{\Delta\f{\majneutrino_i}{}{\frac{\mu_\lepton}{T}}}= \\
	&\frac{\mu_\lepton}{T}\bigg<\frac{\pm\EffAmplitude{}{\majneutrino_i \rightleftarrows \lepton\higgs}
	+\EffAmplitude{}{\majneutrino_i \rightleftarrows 
	\bar{\lepton}\bar{\higgs}}}{{\EffAmplitude{}{\majneutrino_i}}}(c_{\higgs\lepton}\feq{\higgs}{\momhig}
	-\feq{\lepton}{\momlep}) \frac{\Delta\f{\majneutrino_i}{\mommaj}}{\feq{\majneutrino_i}{}}\gamma^D_{\majneutrino_i}\bigg>\nonumber
\end{align}
or just proportional to $\mu_\lepton/T$:
\begin{align}
	\label{eqn:quantum corrected Boltzmann equation mu by T}
	& \frac{s\hubbleratem }{z} \frac{d Y_{Li\atop\majneutrino_i}}{dz}\Big|_{\frac{\mu_\lepton}{T}} 	= \\
	& -\frac{\mu_\lepton}{T}(1+c_{\higgs\lepton})\bigg<\frac{\pm\EffAmplitude{}{\majneutrino_i 
	\rightleftarrows \lepton\higgs}+\EffAmplitude{}{\majneutrino_i 
	\rightleftarrows \bar{\lepton}\bar{\higgs}}}{{\EffAmplitude{}{\majneutrino_i}}}
	\frac{(1-\f{\majneutrino_i}{})}{(1-\feq{\majneutrino_i}{})}\gamma^W_{\majneutrino_i}\bigg>\dend\nonumber
\end{align}
This leads immediately to the result in \eqn\eqref{eqn:quantum corrected Boltzmann equation Li result}. 
Similarly, we get using \eqn\eqref{eqn:linear order expansion Higgs decay term} for the contributions 
\eqref{divlepCPHiggspred} by (anti-)Higgs decay to the rate equations:
\begin{align}
	\label{eqn:quantum corrected Boltzmann equation Higgs decay all contributions}
	&{\frac{s\hubbleratem }{z}}\frac{d Y_{Li\atop\majneutrino_i}}{dz}\Big|_{\Delta\f{\majneutrino_i}{}} = -\bigg< \frac{\EffAmplitude{}{\bar\higgs \rightarrow \lepton\majneutrino_i}
	\mp\EffAmplitude{}{\higgs \rightarrow \bar{\lepton}\majneutrino_i}}{\EffAmplitude{}{\higgs}}\frac{\Delta\f{\majneutrino_i}{\mommaj}}{\feq{\majneutrino_i}{}}\gamma^D_{\higgs}\bigg>\kend\nonumber\\
	&{\frac{s\hubbleratem }{z}}\frac{d Y_{Li\atop\majneutrino_i}}{dz} 
	\Big|_{\Delta\f{\majneutrino_i}{}{\frac{\mu_\lepton}{T}}} = \nonumber\\
	&-\frac{\mu_\lepton}{T}\bigg<\frac{\EffAmplitude{}{\bar\higgs \rightarrow 
	\lepton\majneutrino_i}\mp\EffAmplitude{}{\higgs \rightarrow 
	\bar{\lepton}\majneutrino_i}}{{\EffAmplitude{}{\higgs}}}(c_{\higgs\lepton}\feq{\higgs}{\momhig}
	-\qstatfeqf{\lepton}{\momlep}) \frac{\Delta\f{\majneutrino_i}{\mommaj}}{\feq{\majneutrino_i}{}}
	\gamma^D_{\higgs}\bigg>\kend\nonumber\\
	&{\frac{s\hubbleratem }{z}}\frac{d Y_{Li\atop\majneutrino_i}}{dz}\Big|_{\frac{\mu_\lepton}{T}} 
	= \nonumber\\
	& -\frac{\mu_\lepton}{T}(1+c_{\higgs\lepton})\bigg<\frac{\EffAmplitude{}{\bar\higgs \rightarrow \lepton\majneutrino_i}
	\pm\EffAmplitude{}{\higgs \rightarrow 
	\bar\lepton\majneutrino_i}}{{\EffAmplitude{}{\higgs}}}\frac{(1-
	\f{\majneutrino_i}{})}{(1-\feq{\majneutrino_i}{})}\gamma^W_{\higgs}\bigg>\dend
\end{align}
In addition we get for the 'extra' terms in \eqns\eqref{eqn:rate divlep extra term} and \eqref{eqn:rate Majorana extra term}:
\begin{widetext}
\begin{align}
	\label{eqn:quantum corrected Boltzmann equation decay extra term}
	\frac{d Y_{Li\atop\majneutrino_i}}{dz}\Big|_{D,\text{extra}}&\propto 
	(1\pm 1)\bigg<\frac{\EffAmplitude{}{\majneutrino_i \rightleftarrows \lepton\higgs} -\EffAmplitude{}{\lepton\higgs \rightleftarrows \majneutrino_i}}{{\EffAmplitude{}{\majneutrino_i}}}\qstatff{\majneutrino_i}{}\gamma^D_{\majneutrino_i}\bigg>+(1\mp 1)\frac{\mu_\lepton}{T}(1+c_{\higgs\lepton})\bigg<\frac{\EffAmplitude{}{\majneutrino_i \rightleftarrows \lepton\higgs} -\EffAmplitude{}{\lepton\higgs \rightleftarrows \majneutrino_i}}{{\EffAmplitude{}{\majneutrino_i}}}\qstatff{\majneutrino_i}{}\gamma^D_{\majneutrino_i}\bigg>\nonumber \\
	&+(1\mp 1)\frac{\mu_\lepton}{T}\bigg<\frac{\EffAmplitude{}{\majneutrino_i \rightleftarrows \lepton\higgs} -\EffAmplitude{}{\lepton\higgs \rightleftarrows \majneutrino_i}}{{\EffAmplitude{}{\majneutrino_i}}}(c_{\higgs\lepton}\feq{\higgs}{}-\feq{\lepton}{})\qstatff{\majneutrino_i}{}\gamma^D_{\majneutrino_i}\bigg>\dend
\end{align}
\end{widetext}
This means that for ${d Y_{Li}}/{dz}$ only the first term and for ${d Y_{\majneutrino_i}}/{dz}$ the last two terms contribute. The terms proportional to the difference of $\EffAmplitude{}{\majneutrino_i \rightleftarrows \lepton\higgs}$ and $\EffAmplitude{}{\majneutrino_i \rightleftarrows \bar\lepton\bar\higgs}$ lead to contributions proportional to the \CP-violating parameter.

To reduce the two-body scattering terms in \eqn\eqref{divlepCPscatt} we can use the same methods to find for the general expression, 
\begin{align}
	\label{eqn:general scattering collision term expanded 0}
	\int&\dpi{ab}{ij}{\momk\momp}{\momq\momr}\,(2\pi)^4\delta (\momk +\momp -\momq -\momr)\EffAmplitude{}{ab\leftrightarrow ij}\\
	\times &\big[\qstatf{a}{\momk}\qstatf{b}{\momp}\f{i}{\momq}\f{j}{\momr}\pm\qstatf{i}{\momq}\qstatf{j}{\momr}\f{a}{\momk}\f{b}{\momp}\big]\kend\nonumber
\end{align}
the $\order{\mu/T}$ expansion:
\begin{align}
	\label{eqn:general scattering collision term expanded}
	&\int\dpi{ab}{ij}{\momk\momp}{\momq\momr}\,(2\pi)^4\delta (\momk +\momp -\momq -\momr)\EffAmplitude{}{ab\leftrightarrow ij}\nonumber\\
	\times &\qstatf{a}{\momk}\qstatf{b}{\momp}\f{i}{\momq}\f{j}{\momr}\big[(1\pm 1)\pm (e^{\frac{\mu_a +\mu_b -\mu_i -\mu_j}{T}}-1)\big]\nonumber\\
	=&\frac{\mu_c +\mu_d \pm (\mu_a +\mu_b)}{T}\langle\gamma^{ab}_{ij}\rangle + (1\pm 1)\langle\gamma^{ab}_{ij}\rangle\nonumber\\
	- &(1\pm 1)\langle (\xi_a\frac{\mu_a}{T}\feq{a}{}+\xi_b\frac{\mu_b}{T}\feq{b}{}-\xi_i\frac{\mu_i}{T}\feq{i}{}-\xi_j\frac{\mu_j}{T}\feq{j}{})\gamma^{ab}_{ij}\rangle\kend\nonumber
\end{align}
where the two-body scattering reaction density was introduced in \eqn\eqref{ReactDensExact}.
The last expression results after first order expansion in $\mu/T$.

\section{\label{SelfEnergy}2PI effective action and self-energies}

In this appendix we derive one-loop contribution to the self-energy of the Majorana field as well 
as one and two-loop contributions to the self-energy of leptons. 

\subsection{\label{2PIAction}2PI effective action}

The 2PI effective action is defined as a functional of the one- and two-point functions consisting 
of an infinite sum of all 2PI vacuum diagrams \cite{PhysRevD.10.2428}. In practice, its expansion 
can be characterized in terms of the number of loops appearing in each diagram:
\begin{align}
	i\Gamma_{2\unit{PI}}[\pLept{}{},\pMaj{}{},\pHiggs{}{}]=\sum_{n} i\Gamma_{2\unit{PI}}^{(n)}[\pLept{}{},\pMaj{}{},\pHiggs{}{}]\dend
\end{align}
The two lowest order contributions, $i\Gamma_{2\unit{PI}}^{(2)}$ and $i\Gamma_{2\unit{PI}}^{(3)}$, 
relevant for leptogenesis are shown in \fig\ref{fig:2PI contributions}. Their contributions to the 
2PI action read
\begin{subequations}
\label{2PIContrib}
\begin{align}
	\label{G2}
	i\Gamma_{2\unit{PI}}^{(2)}=& - \int_{\cal C} d^4u\ d^4w \ \Tr\big[h P_R\pMaj{}{}(u,w)P_L\ h^{\dagger}\nonumber\\ 
	&\hspace{60pt}\ \times \ \pLept{}{}(w,u)\epsilon \pHiggs{*}{}(u,w) \epsilon \big]\kend\hphantom{aa}\\
	\label{G3}
	i\Gamma_{2\unit{PI}}^{(3)}=& \frac{1}{2}\int_{\cal C} d^4u\ d^4w\ d^4\eta\ d^4\xi \ \Tr\big[h P_R
	\pMaj{}{}(u,w)\nonumber\\
	& \times CP_R h^T \pLept{T}{}(\eta,w)\epsilon \pHiggs{*}{}(w,\xi)\epsilon h^* P_LC \nonumber\\
	& \times \pMaj{}{}(\eta,\xi)P_L h^{\dagger} \pLept{}{}(\xi,u)\ \epsilon \pHiggs{*}{}(u,\eta) \epsilon \big]\dend
\end{align}
\end{subequations}
In \eqn\eqref{2PIContrib} the trace is taken over flavor, Dirac and $SU(2)_L$ indices whereas the 
transposition only acts in flavor and Dirac space. 

\subsection{\label{LeptSelfEn}Lepton self-energies}

By functional differentiation of the 2PI effective action with respect to the 
two-point function we obtain the corresponding self-energy which enters the 
Schwinger-Dyson equation:
\begin{align}
	\label{defSE}
	\sLept{(n-1)}{\alpha \beta}(x,y) = -i\frac{\delta \Gamma^{(n)}_{2\unit{PI}}[\pLept{}{}]}{\delta \pLept{\, T}{\beta \alpha}(y,x)}\dend
\end{align}
Here, flavor indices are shown explicitly whereas the $SU(2)_L$ and Dirac structure 
is embodied implicitly in matrix notation. The resulting self-energy is given by a combination 
of the Majorana, lepton and Higgs propagators. Since we do not consider the flavor effects 
and the early Universe was in an $SU(2)_L$-symmetric state:
\begin{subequations}
\begin{align}
	\pHiggs{}{ab}(x,y) &= \delta_{ab}\pHiggs{}{}(x,y)\kend\;\; \pLept{\alpha \beta}{ab}(x,y) = \delta_{ab}\delta^{\alpha\beta}\pLept{}{}(x,y)\dend
\end{align}
\end{subequations}
Since $\epsilon^4 = -\epsilon^2 = \mathbf{1}$ the lepton self-energy also becomes 
diagonal: $\sLept{\alpha \beta}{ab}=\delta_{ab}\sLept{\alpha\beta}{}$. Furthermore, 
in the unflavored approximation it is convenient to sum over lepton flavors: 
$\sLept{}{}\equiv \sum_\alpha \sLept{\alpha\alpha}{}$. Then the one- and two-loop 
order contributions to the lepton self-energy, see \fig\ref{fig:2PI contributions}, 
take the form:
\begin{subequations}
\begin{align}
	\label{SE1L}
	&\sLept{(1)}{}(x,y) = - (\yu^{\dagger}\yu)_{ji}
	P_R \pMaj{ij}{}(x,y)P_L\pHiggs{}{}(y,x)\kend\\
	\nonumber
	\label{SE2L}
	&\sLept{(2)}{}(x,y) = - (\yu^\dagger \yu)_{ij}(\yu^\dagger \yu)_{lk}
	\int_{\cal C} d^4w d^4\eta P_R \pMaj{jk}{}(x,w)C \nonumber\\
	&\times P_R \pLept{T}{}(\eta,w)P_L C \pMaj{li}{}(\eta,y)
	P_L \pHiggs{}{}(y,w) \pHiggs{}{}(\eta,x)\dend
\end{align}
\end{subequations}
Eventually, it is the Wightman components that we are interested in since they enter the gain- 
and loss terms on the right-hand side of \eqn\eqref{MasterEquationWigner}. Therefore, we insert 
the usual decomposition of the propagators $G\in \{\pHiggs{}{}, \pLept{}{}, \pMaj{}{}\}$ into 
the spectral and statistical parts, \eqn\eqref{FRhoDecomposition}, into the self-energies 
\eqref{SE1L} and \eqref{SE2L}. A formal decomposition of the self-energy in analogy to 
\eqref{FRhoDecomposition} allows us to identify its spectral and statistical part and define 
the Wightman components in coordinate space as $\sLept{}{\gtrless} = \sLept{}{F} \mp \textstyle 
\frac{i}{2} \sLept{}{\rho}$. For the one-loop self-energy \eqref{SE1L} they read
\begin{align}
 \label{SE1loop}
 \sLept{(1)}{\gtrless}(x,y) = 
 -(\yu^\dagger \yu)_{ji} P_R \pMaj{ij}{\gtrless}(x,y)P_L\pHiggs{}{\lessgtr}(y,x)\dend
\end{align}
In the case of the two-loop contribution, \eqn\eqref{SE2L}, the computation becomes slightly elaborate. 
The complication is due to the appearance 
of 32 different terms after inserting the decomposition \eqref{FRhoDecomposition} for each of the five propagators 
into \eqn\eqref{SE2L} as well as due to the two remaining integrations over the internal space-time 
arguments $w$ and $\eta$. The decomposition makes the path-ordering explicit and allows us to 
convert the integration along the CTP into an integration along the positive branch. The 32 terms 
contain different combinations of the $\sign$-functions. These can be rewritten by using relations 
given in Appendix C of \cite{Garny:2009rv}. After some simple but lengthy algebra we obtain for 
the Wightman components:
\begin{widetext}
\begin{align}
	\label{SeWC}
	\sLept{(2)}{\gtrless}&(x,y) = (\yu^\dagger \yu)_{ij}(\yu^\dagger \yu)_{lk}
	\int d\omega \int d\eta \nonumber\\
	&\times \bigl[ P_R\pMaj{jk}{R}(x,\omega)CP_R\pLept{T}{F}(\eta,\omega)P_LC\pMaj{li}{\gtrless}(\eta,y)P_L\ \pHiggs{}{\lessgtr}(y,\omega)\pHiggs{}{A}(\eta,x)\nonumber\\
	&\; \, +P_R\pMaj{jk}{F}(x,\omega)CP_R\pLept{T}{R}(\eta,\omega)P_LC\pMaj{li}{\gtrless}(\eta,y)P_L\ \pHiggs{}{\lessgtr}(y,\omega)\pHiggs{}{A}(\eta,x)\nonumber\\
	&\; \, +P_R\pMaj{jk}{R}(x,\omega)CP_R\pLept{T}{A}(\eta,\omega)P_LC\pMaj{li}{\gtrless}(\eta,y)P_L\ \pHiggs{}{\lessgtr}(y,\omega)\pHiggs{}{F}(\eta,x)\nonumber\\
	&\; \, +P_R\pMaj{jk}{\gtrless}(x,\omega)CP_R\pLept{T}{R}(\eta,\omega)P_LC\pMaj{li}{A}(\eta,y)P_L\ \pHiggs{}{F}(y,\omega)\pHiggs{}{\lessgtr}(\eta,x)\nonumber\\
	&\; \, +P_R\pMaj{jk}{\gtrless}(x,\omega)CP_R\pLept{T}{A}(\eta,\omega)P_LC\pMaj{li}{F}(\eta,y)P_L\ \pHiggs{}{R}(y,\omega)\pHiggs{}{\lessgtr}(\eta,x)\nonumber\\
	&\; \, +P_R\pMaj{jk}{\gtrless}(x,\omega)CP_R\pLept{T}{F}(\eta,\omega)P_LC\pMaj{li}{A}(\eta,y)P_L\
	\pHiggs{}{R}(y,\omega)\pHiggs{}{\lessgtr}(\eta,x)\nonumber\\
	&\; \, +P_R\pMaj{jk}{R}(x,\omega)CP_R\pLept{T}{\lessgtr}(\eta,\omega)P_LC\pMaj{li}{A}(\eta,y)P_L\ \pHiggs{}{\lessgtr}(y,\omega)\pHiggs{}{\lessgtr}(\eta,x)\nonumber\\
	&\; \, +P_R\pMaj{jk}{\gtrless}(x,\omega)CP_R\pLept{T}{\gtrless}(\eta,\omega)P_LC\pMaj{li}{\gtrless}(\eta,y)P_L\
	\pHiggs{}{R}(y,\omega)\pHiggs{}{A}(\eta,x)\bigr]\dend
\end{align}
\end{widetext}
Specific approximations will allow us to interpret both expressions, \eqn\eqref{SE1loop} and 
\eqn\eqref{SeWC}, as describing decay, inverse decay and scattering processes of quasiparticles 
in the medium. 
 
The expressions for the one- and two-loop self-energies given by \eqns\eqref{SE1loop} and 
\eqref{SeWC} depend explicitly on two coordinates in four dimensional space-time. However, 
the self-energies which govern the gain- and loss term on the right-hand side of 
\eqn\eqref{MasterEquationWigner} are expressed in terms of phase space coordinates. 
Let us therefore exchange the pair of space-time arguments $(x,y)$ for an equivalent 
set of center and relative coordinates, $X\equiv (x+y)/2$ and $s\equiv x-y$. In contrast 
to thermal equilibrium, the out of equilibrium propagators depend not only on the relative 
coordinate $s$ but also on the center coordinate $X$. Performing a so-called Wigner 
transformation \cite{Calzetta:1986cq}, i.e. a Fourier transformation with respect to 
the relative coordinate $s$, we can trade the latter for a momentum space variable:
\begin{subequations}
\label{WT}
\noeqref{WT1a,WT1b}
\begin{align}
	\label{WT1a}
	G_{F}(X_{uv},p) & = \int d^4 s_{uv} e^{ips_{uv}}G_{F}(X_{uv},s_{uv})\kend\\
	\label{WT1b}
	G_{\rho}(X_{uv},p) &= -i \int d^4 s_{uv} e^{ips_{uv}}G_{\rho}(X_{uv},s_{uv})\kend\hphantom{aa}
\end{align}
\end{subequations}
where we have used $X_{uv}=(u+v)/2$ and $s_{uv}=u-v$, $(u,v) \in \{x,y,w,\eta\}$ 
according to the various combinations appearing in \eqns\eqref{SE1loop} and \eqref{SeWC}. 
Note that the factor $-i$ in the definition \eqref{WT1b} is conventional and makes the 
Wigner transform of the spectral propagator a hermitian matrix. Definitions of the 
Wigner transforms of the advanced and retarded propagators coincide with that for 
the statistical propagator. 

The Wigner transform of \eqn\eqref{SE1loop} is obtained straightforwardly:
\begin{align}
 \label{SEOneLoopWigner}
 \sLept{(1)}{\gtrless}(t,\momlep)=&-(\yu^\dagger \yu)_{ji}\int\lorentzdd{\momhig} 
 \lorentzdd{\mommaj}(2\pi)^4 \delta(\mommaj-\momhig-\momlep)\nonumber\\
	 &\times P_R\pMaj{ij}{\gtrless}(t,\mommaj)
	 P_L\pHiggs{}{\lessgtr}(t,\momhig)\dend
\end{align}
Note that motivated by the homogeneity and isotropy of the early Universe we only 
indicate time-dependence of the propagators and self-energy, $t\equiv X^0_{xy}$. 
To obtain the Wigner transform 
of \eqn\eqref{SeWC} we will use an additional approximation: each of the Wigner transforms 
of the propagators we replace by $G(X_{uv},p) \rightarrow G(X,p)$. 
This means that we neglect the variations of $X_{uv}$ from the center coordinate $X=X_{xy}$ 
at which the self-energy is evaluated. Technically, it corresponds to a gradient expansion 
to lowest order and therefore disregards all memory effects. This can be compared to the 
`Sto\ss{}zahlansatz' within the usual approach to the Boltzmann equation. It is convenient
to represent the resulting expression as a sum of three terms:
\begin{align}
	\label{SE3p}
	\sLept{(2)}{\gtrless} = \sLept{(2.1)}{\gtrless} + 
	\sLept{(2.2)}{\gtrless} + \sLept{(2.3)}{\gtrless}\dend
\end{align}
The first term on the right-hand side of \eqn\eqref{SE3p} corresponds to the Wigner transform of 
lines one to three and four to six in \eqn\eqref{SeWC}:
\begin{align}
	\label{SED}
	&\sLept{(2.1)}{\gtrless}(t,\momlep) = \int \lorentzdd{\mommaj} \lorentzdd{\momhig}\
	(2\pi)^4\delta(\momlep-\mommaj+\momhig) \\
	&\times\big[ (\yu^\dagger \yu)_{ij}(\yu^\dagger \yu)_{lk}
	\varLambda_{jk}(t,q,k)P_L C \pMaj{li}{\gtrless}(t,\mommaj) 
	P_L\pHiggs{}{\lessgtr}(t,\momhig)\nonumber\\
	&+ (\yu^\dagger \yu)_{ji} (\yu^\dagger \yu)_{kl}
	P_R \pMaj{il}{\gtrless}(t,\mommaj) C P_R V_{kj}(t,\mommaj,\momlep)
	\pHiggs{}{\lessgtr}(t,\momhig)\big]\nonumber \dend
\end{align}
where we have introduced two functions containing loop corrections:
\begin{align}
	\label{LC1}
	&\varLambda_{jk}(t,\mommaj,\momhig) \equiv \int \lorentzdd{k_1} \lorentzdd{k_2} \lorentzdd{k_3}\nonumber\\
	& \times (2\pi)^4\delta(\mommaj+k_1+k_2)\ (2\pi)^4 \delta(\momhig+k_2-k_3)\nonumber\\
	& \times \bigl[ P_R \pMaj{jk}{R}(t,-k_3) C P_R \pLept{T}{F}(t,k_2)\pHiggs{}{A}(t,k_1)\nonumber\\
	& +\ P_R \pMaj{jk}{F}(t,-k_3) C P_R \pLept{T}{R}(t,k_2) \pHiggs{}{A}(t,k_1)\nonumber\\
	& +\ P_R \pMaj{jk}{R}(t,-k_3) C P_R \pLept{T}{A}(t,k_2) \pHiggs{}{F}(t,k_1) \bigr]\kend
\end{align}
and $V_{kj}(t,q,k)\equiv P\, \varLambda^{\dagger}_{kj}(t,q,k)\, P$. As we will see,
\eqref{SED} describes \CP-violating decay of the heavy Majorana neutrino. 
The second term on the right-hand side of \eqn\eqref{SE3p} is given by the Wigner transform of 
the seventh line of \eqn\eqref{SeWC} and describes lepton number violating scattering processes: 
\begin{align}
	\label{SeSca}
	&\sLept{(2.2)}{\gtrless}(t,\momlep) = \int \lorentzdd{\momlep_2} 
	\lorentzdd{\momhig_1} d\Pi_{\momhig_2}(2\pi)^4\delta(\momlep+\momhig_1-\momlep_2-\momhig_2)\nonumber\\
	&\times (\yu^\dagger \yu)_{ij}(\yu^\dagger \yu)_{lk}
	\bigl[P_R\pMaj{jk}{R}(t,\momlep_2+\momhig_2) CP_R\pLept{T}{\lessgtr}(t,-\momlep_2)P_L\nonumber\\
	& \times C\pMaj{li}{A}(t,\momlep_2-\momhig_1)P_L \pHiggs{}{\lessgtr}(t,\momhig_1)\pHiggs{}{\lessgtr}(t,-\momhig_2)\bigr]\dend
\end{align}
Finally the last term in \eqn\eqref{SE3p} corresponds to the last line of \eqn\eqref{SeWC},
\begin{align}
	\label{SeLC}
	&\sLept{(2.3)}{\gtrless}(t,\momlep) = \int d\Pi_{\momlep_2} d\Pi_{\momhig_1} 
	d\Pi_{\momhig_2}(2\pi)^4\delta(\momlep+\mommaj_1-\momlep_2-\mommaj_2)\nonumber\\
	&\times (\yu^\dagger \yu)_{ij} (\yu^\dagger \yu)_{lk}
	\bigl[P_R\pMaj{jk}{\gtrless}(t,-\mommaj_1) CP_R
	\pLept{T}{\gtrless}(t,\momlep_2)P_LC \nonumber\\
	&\times \pMaj{li}{\gtrless}(t,\mommaj_2)P_L 
	\pHiggs{}{A}(t,-\mommaj_2-\momlep_2)\pHiggs{}{R}(t,\mommaj_1-\momlep_2)\bigr]\kend
\end{align}
and can be identified with lepton number conserving processes which do not contribute to
generation of the lepton asymmetry.

\subsection{\label{MajoranaSelfEn}Majorana self-energy}

Differentiating \eqn\eqref{G2} with respect to the 
two-point function of the Majorana neutrino and using definitions of the 
{\CP}-conjugate two-point functions we obtain for the Majorana self-energy
\begin{align}
	\label{SigmaMajorana}
	\sMaj{ij}{}(x,y)=&-g_w\bigl[ (\yu^\dagger \yu)_{ij}\, P_L \pLept{}{} (x,y)P_R \, \pHiggs{}{}(x,y) \\
	& + (\yu^\dagger \yu)^*_{ij}\, P_R P \pLeptcp{}{}(\bar x,\bar y)P P_L\,
	\pHiggscp{}{}(\bar x,\bar y)\,\bigr]\kend\nonumber
\end{align}
where we have assumed the $SU(2)_L$ symmetry of the medium and neglected flavor effects.
The factor $g_w=2$ in \eqn\eqref{SigmaMajorana} comes from the summation over the $SU(2)_L$ indices. 
The {\CP} conjugate self-energy differs from \eqn\eqref{SigmaMajorana} only in the propagators 
replaced by their {\CP} conjugate counterparts and the couplings replaced by their complex conjugates.

From \eqn\eqref{SigmaMajorana} we can read off the Wightman components of the self-energy:
\begin{align}
	\label{eqn: Majorana Wightman self-energy} 
	\sMaj{ij}{\gtrless}(x,y)&=-g_w\bigl[
	(\yu^\dagger\yu)_{ij} P_L \pLept{}{\gtrless}(x,y)P_R\pHiggs{}{\gtrless}(x,y)\nonumber\\
	&+(\yu^\dagger \yu)^*_{ij} P_R P \pLeptcp{}{\gtrless}(\bar x,\bar y)\, 
	P P_L \pHiggscp{}{\gtrless}(\bar x,\bar y) \bigr]\dend
\end{align}
Its {\CP}-conjugate can be obtained by complex conjugating the couplings and replacing
the two-point functions by their {\CP}-conjugates. To calculate amplitudes of the scattering 
processes we will need its Wigner transform: 
\begin{align}
	\label{eqn: Majorana self-energy}
	\sMaj{ij}{\gtrless}(t,\mommaj) = & -g_w \int \lorentzdd{\momhig} 
	\lorentzdd{\momlep}(2\pi)^4\delta(\mommaj-\momlep-\momhig) \nonumber\\ 
	& \times \bigl[ \, (\yu^\dagger \yu)_{ij} P_L 
	\pLept{}{\gtrless} (t,\momlep)P_R 
	\pHiggs{}{\gtrless}(t,\momlep) \nonumber \\ 
	& + (\yu^\dagger \yu)_{ji}P_R P 
	\pLeptcp{}{\gtrless}(t,\bar{\momlep}) P P_L 
	\pHiggscp{}{\gtrless} (t,\bar{\momhig}) \bigr] \dend\hphantom{aaaa}
\end{align}
From \eqn\eqref{eqn: Majorana self-energy} we can deduce the Wigner transform of the corresponding
spectral self-energy:
\begin{align}
	\label{SigmaRhohCPSymmInit}
	\sMaj{ij}{\rho}(t,\mommaj)=&-{\frac{g_w}{16\pi}}
	\bigl[(h^\dagger h)_{ij}P_L \sMaj{}{\rho}(t,\mommaj)P_R\nonumber\\
	& + (h^\dagger h)^*_{ij} P_R \sMajcp{}{\rho}(t,\bar{\mommaj})P_L\bigr]\kend
\end{align}
where we have introduced 
\begin{align*}
	\sMaj{}{\rho}(t,\mommaj)&\equiv 
	16\pi{\textstyle\int}\lorentzdd{\momhig} \lorentzdd{\momlep} (2\pi)^4\deltafour{\mommaj-\momhig-\momlep}\nonumber\\
	& \times \bigl[\pHiggs{}{F}(t,\momhig)\pLept{}{\rho}(t, \momlep)+
	\pHiggs{}{\rho}(t,\momhig)\pLept{}{F}(t, \momlep)\bigr]\kend\\
	\sMaj{}{\rho}(t,\mommaj)& \equiv 
	16\pi{\textstyle\int}\lorentzdd{\momhig} \lorentzdd{\momlep} (2\pi)^4\deltafour{\mommaj-\momhig-\momlep}\nonumber\\
	& \times P\bigl[\pHiggscp{}{F}(t,\momhig)\pLeptcp{}{\rho}(t,\momlep)
	+\pHiggscp{}{\rho}(t,\momhig) \pLeptcp{}{F}(t,\momlep)\bigr]P\kend
\end{align*}
and $\bar \mommaj\equiv (\mommaj^0,-\vec{\mommaj}\,)$. 
In the quasiparticle approximation the Wigner trans\-forms of the two-point functions
of leptons and the Higgs are given by \eqns\eqref{KBLepton}--\eqref{leptonQPspectralfunc} 
and \eqns\eqref{KBHiggs}--\eqref{QPforHiggs} respectively.
In a {\CP}-symmetric medium, which the early Universe was to a very good approximation, 
$\pHiggscp{}{\rho}(t, \momhig)=\pHiggs{}{\rho}(t,\momhig)$ and $\pHiggscp{}{F}(t,\momhig)=
\pHiggs{}{F}(t,\momhig)$. 
The homogeneity and isotropy of the early Universe furthermore imply, that there is 
no dependence on the momentum direction and the spatial central coordinate so that 
$\pHiggscp{}{\rho}(t,\bar \momhig)=\pHiggs{}{\rho}(t,\momhig)$ 
and $\pHiggscp{}{F}(t, \bar \momhig)=\pHiggs{}{F}(t,\momhig)$. Just like for 
scalars, in a {\CP}-symmetric medium the fermion two-point functions are related by 
$\pLeptcp{}{\rho}(t, \momlep)= \pLept{}{\rho}(t,\momlep)$ 
and $\pLeptcp{}{F}(t,\momlep)=\pLept{}{F}(t,\momlep)$. As 
for the $\momlep\rightarrow \bar \momlep$ transformation,
the terms in the lepton propagators which carry spinor structure are not invariant 
under it:
\begin{align}
	\label{ppbar}
	P_L\,\slashed{\bar \momlep}\, P_R=P_L P\,\slashed{\momlep}\,P P_R= P P_R\,\slashed{\momlep}\,P_L P\dend
\end{align}
Since $P^2=\mathbf{1}$, \eqn\eqref{ppbar} implies that in a homogeneous, isotropic and 
{\CP}-symmetric medium $\sMajcp{}{}$ and $\sMaj{}{}$ are left and right 
projections of the same `vector' integral $L_{\rho}$:
\begin{align}
	\label{SigmaRhohCPSymm}
	\sMaj{ij}{\rho}&=-{\frac{g_w}{16\pi}}\bigl[(h^\dagger h)_{ij}P_L+(h^\dagger h)^*_{ij} P_R\,\bigr] L_{\rho}\dend
\end{align}
Explicit form of $L_\rho$ depends on the kinematic regime and is presented below.

To evaluate the decay amplitudes as well as amplitudes of the $s$-channel scattering processes we 
need to evaluate it for positive $q^2=M^2$ and $\mommaj^2=s$. If $\mommaj^0$ 
is also positive then $L_\rho$ takes the form:
\begin{align}
	\label{SigmaRhoVectMedium}
	L_{\rho}(t,\mommaj)=16\pi \int & \lorentzd{\higgs}{\momhig} 
	\lorentzd{\lepton}{\momlep} (2\pi)^4 \deltafour{\mommaj-\momhig-\momlep}\,\slashed{\momlep}\nonumber\\
	&\times \,\bigl[1+f_\higgs(t,\momhig)-f_\lepton(t,\momlep)\bigr]\dend
\end{align}
To obtain \eqn\eqref{SigmaRhoVectMedium} we have used the quasiparticle approximation for 
the two-point functions and integrated over the zeroth components of the momenta. Integrating 
out the energy-momentum conserving delta-function we obtain for its Lorentz components: 
\begin{subequations}
	\label{SigmaVTherm}
	\begin{align}
		L^0_\rho&=\frac{2T}{y}I_{1}\left(y_0,y\right)\kend\\
		\vec L_\rho&=\frac{\vec\mommaj}{|\vec \mommaj|}
		\frac{2T}{y^2}\bigl[\,y_0 I_1\left(y_0,y\right)\nonumber\\
		&-{\textstyle\frac12}(y_0^2-y^2)(1+x_\lepton-x_\higgs)I_0\left(y_0,y\right)\bigr]\kend
	\end{align}
\end{subequations}
where $y_0\equiv q^0/T$ and $y\equiv |\vec \mommaj|/T$. The integral functions $I_n$ 
are defined by 
\begin{align*}
	I_n(y_0,y) \equiv \int\limits_{z^{-}}^{z^{+}}dz\, z^n
	\left(1+\frac1{e^{y_0-z}-1}-\frac1{e^z+1}\right)\kend
\end{align*}
where, in complete analogy with \eqn\eqref{eqn: decay integration limits},
the integration limits are given by 
\begin{align}
\label{eqn: integration limits In}
z^\pm =
{\textstyle\frac12}\bigl[\,y_0\bigl(1+x_\lepton-x_\higgs\bigr)
\pm y \lambda^\frac12(1,x_\lepton,x_\higgs)\bigr]\dend
\end{align}
For positive $q^2$ and negative $\mommaj_0$ the components of $L_\rho$ are related
to the ones above by $L^0_\rho(-\mommaj^0,\vec \mommaj\,)=L^0_\rho(\mommaj^0,\vec \mommaj\,)$ 
and $\vec L_\rho(-\mommaj^0,\vec \mommaj\,)=-\vec L_\rho(\mommaj^0,\vec \mommaj\,)$ respectively.

To evaluate amplitudes of the $t$- and $u$-channel processes we also need 
to calculate $L_\rho$ for negative square of the momentum transfer. In this case 
momentum-energy
conservation ensures that $\momlep^0$ and $\momhig^0$ cannot be positive or negative 
simultaneously. If they have different signs then, assuming homogeneity and isotropy 
of the medium and using relations \eqref{oddp0}, we find
\begin{align}
	\label{SigmaRhoVectMediumTUchan}
	L_{\rho}(t,\mommaj)&=16\pi{\textstyle\int} \lorentzd{\higgs}{\momhig} 
	\lorentzd{\lepton}{\momlep} (2\pi)^4 \,\slashed{\momlep}\nonumber\\
	\times\bigl\{ &\deltafour{\mommaj+\momhig-\momlep}[f_{\bar\higgs}(t,\momhig)
	+f_\lepton(t,\momlep)]\nonumber\\
	+ &\deltafour{\mommaj-\momhig+\momlep}[f_\higgs(t,\momhig)
	+f_{\bar\lepton}(t,\momlep)]\bigr\}\dend 
\end{align}
\eqn\eqref{SigmaRhoVectMediumTUchan} implies that for negative square of the 
momentum transfer $L_\rho$ vanishes in vacuum. Although it is in principle possible 
to retain the thermal masses of leptons and the Higgs in the calculation, the 
resulting expressions are quite lengthy in this case. Neglecting the thermal masses 
we obtain for the Lorentz components of $L_\rho$ in this regime:
\begin{subequations}
\label{Lrhonegmom}
\begin{align}
L_\rho^0&=\frac{2T}{y}\sum_\pm I_1^\pm(y_0,y)\kend \\
\vec L_\rho&=\frac{\vec q}{|\vec q|}\frac{2 T}{y^2}\sum_\pm\bigl[\,y_0 I_1^\pm(y_0,y)-{\textstyle\frac12}(y_0^2-y^2)I_0^\pm(y_0,y)\bigr]\kend
\end{align}
\end{subequations}
where the integral functions are given by
\begin{align*}
I_n^\pm(y_0,y)\equiv \int\limits_{\frac12(y\pm y_0)}^\infty dz\, z^n
\left(\frac{1}{e^z+1}+\frac{1}{e^{z\mp y_0}-1}\right)\dend
\end{align*}
Note that in this regime $y>y_0$ and therefore the lower integration limit is positive.

To compute the scattering amplitude we need to calculate the product $q L_\rho$.
For the $s$-channel we find:
\begin{align}
	qL_\rho&=q^2(1+x_\lepton-x_\higgs)\,y^{-1}I_0(y_0,y)\kend
\end{align} 
whereas the corresponding expression for the $t$- and $u$-channels reads
\begin{align}
\label{tuLrho}
qL_\rho&=q^2\, y^{-1}\sum_\pm I_0^\pm(y_0,y)\dend
\end{align}
At low temperatures \eqn\eqref{tuLrho} is exponentially small and vanishes in the vacuum limit.

To analyze the Higgs decay we need to evaluate the spectral loop 
integral in a region of the phase space where the effective Higgs 
mass exceeds the sum of the Majorana and lepton masses. Using 
properties of the distribution functions under the $\momlep^0 \rightarrow
-\momlep^0$ transformation we find after some algebra from \eqn\eqref{LRhoHiggs}:
\begin{align*}
	L_\rho(t,\mommaj)&= 16\pi\int \dpi{{\higgs}{\lepton}}{}{{\momhig}{\momleploop}}{} 
	(2\pi)^4 \delta(\mommaj+\momlep-\momhig)\,\slashed{\momlep}\notag \bigl[\,\feq{\higgs}{\momhig}+\feq{\lepton}{\momlep}\,\bigr].
\end{align*}
Just like in \eqn\eqref{SigmaRhoVectMedium}, the integration is over the (on-shell) momenta 
of the Higgs and lepton and the Majorana momentum serves as a constraint. Note that in 
this case we are interested only in the on-shell Majorana momenta and therefore $q^2=M^2$. 
After integrating out the delta-function we obtain a result similar to \eqn\eqref{SigmaVTherm}:
	\begin{align}
		L^0_\rho&=\frac{2T}{y}J_1\left(y_0,y\right)\kend\\
		\vec L_\rho&=\frac{\vec\mommaj}{|\vec \mommaj|}\frac{2 T}{y^2} 
		\bigl[\,y_0J_1\left(y_0,y\right)-{\textstyle\frac12}(x_\higgs-x_\lepton-1)J_0\left(y_0,y\right)
		\bigr]\kend
	\end{align}
where the integral function is defined as:
\begin{align*}
	J_n(y^0,y) \equiv \int\limits^{z^+}_{z^{-}}dz\, z^n
	\left(\frac1{e^z+1}+\frac1{e^{z+y^0}-1}\right)\dend
\end{align*}
The integration limits are given by an expression similar to \eqn\eqref{eqn: integration limits In}
but with $1+x_\lepton-x_\higgs$ replaced by $x_\higgs-x_\lepton-1$. When the effective Higgs 
mass approaches the kinematic limit, $m_\higgs=m_\lepton+M$ the upper integration 
limit approaches the lower one, and the integral vanishes. 

\section{\label{NumericalParameters} Numerical parameters}

\begin{figure}[!ht]
	\begin{center}
	\includegraphics[width=\columnwidth]{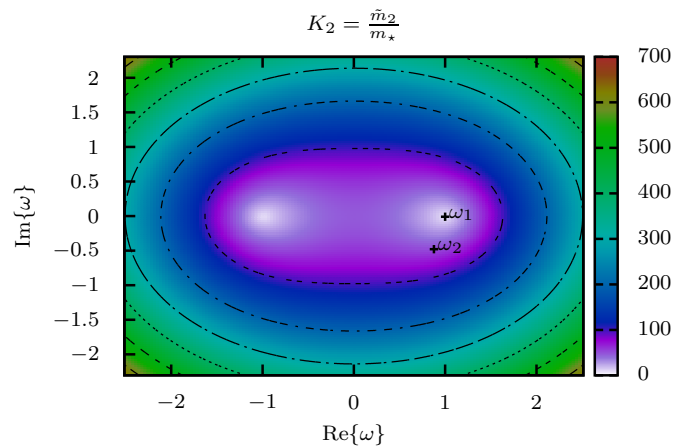}
	\end{center}
	\caption{\label{m2tildeVsReomegaImomega} Washout parameter for $\majneutrino_2$ decay in the complex $\omega$-plane. The values for the benchmark points are $K_2(\omega_1)=9.22$ and $K_2(\omega_2)=53.38$. $K_2$ has a minimum value of $K_2(-1)=K_2(1)=8.28173$. }
\end{figure}

To perform the quantitative analysis we need to specify the Yukawa couplings. For simplicity 
we focus on the case of a very heavy third Majorana neutrino $M_3\gg M_2> M_1$ (MSM). In 
this limit the Yukawa couplings can be expressed in terms of the observed active neutrino masses 
and mixing angles and only one complex additional free parameter $\omega$. 
\begin{figure}[!ht]
	\begin{center}
	\includegraphics[width=\columnwidth]{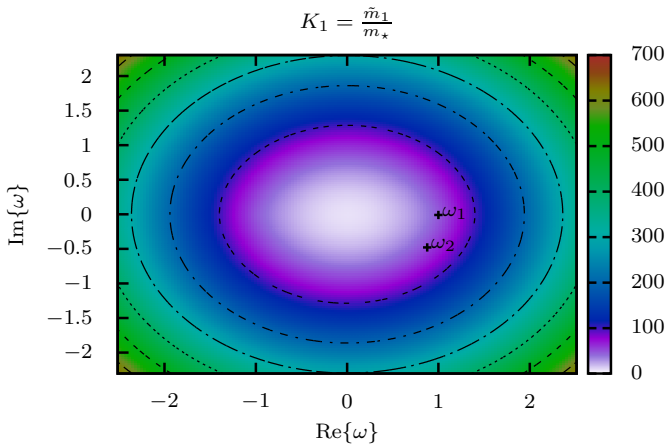}
	\end{center}
	\caption{\label{m1tildeVsReomegaImomega} Washout parameter for $\majneutrino_1$ decay in the complex $\omega$-plane. The values for the benchmark points are $K_1(\omega_1)=47.20$ and $K_1(\omega_2)=54.97$. $K_1$ has a minimum value of $K_1(0)=8.28$.}
\end{figure}
In the Casas-Ibarra 
parameterization \cite{Ibarra:2003up,Antusch:2011nz,Blanchet:2006dq,Guo:2006qa} the Yukawa 
couplings are given by
\begin{subequations}
\label{mtildevsomega}
\begin{align}
(\yu^\dagger \yu)_{11} & = \frac{M_1}{v^2}(m_2\abs{1-\omega^2}+m_3\abs{\omega^2}) \kend\\
(\yu^\dagger \yu)_{22} & = \frac{M_2}{v^2}(m_3\abs{1-\omega^2}+m_2\abs{\omega^2})\kend\\
\label{tildem12}
(\yu^\dagger \yu)_{12} &=\frac{\sqrt{M_1M_2}}{v^2}(m_2\, \omega \sqrt{1-\omega^2}\,^*-m_3\,\omega^*\sqrt{1-\omega^2})\kend
\end{align}
\end{subequations}
where $v\approx 174$ GeV is the Higgs vev and we have assumed normal hierarchy. 
In this case the physical neutrino masses are given by 
\begin{align*}
m_2&=(\Delta m^2_{sol})^\frac12\approx 8.71\cdot 10^{-12} \quad {\rm GeV}\kend\\
m_3&=(\Delta m^2_{sol}+\Delta m^2_{atm})^\frac12\approx 5.0\cdot 10^{-11}\quad {\rm GeV}\dend
\end{align*}
For illustration we choose the benchmark points $\omega=\exp(-0.01\cdot I)$ and $\omega=\exp(-0.5\cdot I)$ 
denoted by BM1 and BM2 in the plots. As masses of the right-handed neutrinos we choose $M_1=10^9$ GeV 
and $M_2=\sqrt{10} M_1$ respectively. This choice of parameters is such that effects related to 
the resonant enhancement will be unimportant but contributions from both heavy Majorana neutrinos 
can be relevant. Note however that there are lower bounds on the washout parameters 
$K_i\equiv \Gamma_i/\hubblerate\big|_{T=M_1} =\tilde{m}_i/m_{\star}$ (with `equilibrium neutrino 
mass' $m_{\star}={16 \pi^{5/2} \sqrt{g^{*}_{SM}}v^2 /\big(3 \sqrt{5}{M_{Pl}}\big)} $), 
see \figs\ref{m1tildeVsReomegaImomega} and \ref{m2tildeVsReomegaImomega}. 
The freeze-out of the asymmetry will therefore typically occur late (i.e.~$T\ll M_i$) which 
renders medium effects in general small. However for the qualitative issues discussed in this paper
our preference is to specify a consistent set of parameters for which we can discuss the generation 
of the lepton asymmetry in terms of two heavy Majorana neutrinos. This is of course not a 
general restriction for the employed techniques. 

\end{appendix}

\newpage

\bibliographystyle{apsrev}


\end{document}